\documentclass[conference]{IEEEtran}
\usepackage{cite}
\usepackage{amsmath,amssymb,amsfonts}
\usepackage{algorithm, algpseudocode}
\usepackage{graphicx}
\usepackage{textcomp}
\usepackage{xcolor}
\usepackage{fancyhdr}
\usepackage{hyperref}
\usepackage{physics}
\usepackage{multirow}
\usepackage{booktabs}
\usepackage{subfig}
\usepackage[export]{adjustbox}

\def\BibTeX{{\rm B\kern-.05em{\sc i\kern-.025em b}\kern-.08em
    T\kern-.1667em\lower.7ex\hbox{E}\kern-.125emX}}

\newcommand{\swap} {{SWAP }}
\newcommand{\ucl} {{2QAN}}
\newcommand{\tket} {{t$\ket{\textup{ket}}$}}
\newcommand{\h} {{Hamiltonian}}

\title{2QAN: A quantum compiler for 2-local qubit Hamiltonian simulation algorithms }
\author{
    \IEEEauthorblockN{Lingling Lao, Dan E. Browne}
    \IEEEauthorblockA{\textit{Department of Physics and Astronomy, University College London, London WC1E 6BT, United Kingdom}}
}

\begin{document}
\maketitle
\pagestyle{plain}

\begin{abstract}
Simulating quantum systems is one of the most important potential applications of quantum computers.
The high-level circuit defining the simulation needs to be compiled into one that complies with hardware limitations such as qubit architecture (connectivity) and instruction (gate) set.
General-purpose quantum compilers work at the gate level and have little knowledge of the mathematical properties of quantum applications, missing further optimization opportunities.
Existing application-specific compilers only apply advanced optimizations in the scheduling procedure and are restricted to the CNOT or CZ gate set.
In this work, we develop a compiler, named \ucl, to optimize quantum circuits for 2-local qubit Hamiltonian simulation problems, a framework which includes the important quantum approximate optimization algorithm (QAOA).  
In particular, we exploit the flexibility of permuting different operators in the Hamiltonian (no matter whether they commute) and propose permutation-aware techniques for qubit routing, gate optimization and scheduling to minimize compilation overhead. \ucl~can target different qubit topologies and different hardware gate sets. Compilation results on four applications (up to 50 qubits) and three quantum computers (namely, Google Sycamore, IBMQ Montreal and Rigetti Aspen) show that \ucl~outperforms state-of-the-art general-purpose compilers and application-specific compilers. Specifically, \ucl~can reduce the number of inserted SWAP gates by 11.5X, reduce overhead in hardware gate count by 68.5X, and reduce overhead in circuit depth by 21X. 
Experimental results on the Montreal device demonstrate that benchmarks compiled by \ucl~achieve the highest fidelity. 
\end{abstract}


\section{Introduction}

Quantum simulation has broad applications in quantum many-body physics \cite{raeisi2012quantum}, quantum chemistry \cite{poulin2014trotter}, and quantum field theory \cite{jordan2012quantum} and has been demonstrated in different quantum technologies \cite{hempel2018quantum, Arute2020fh} (see \cite{georgescu2014quantum} for a detailed review). The problem of Hamiltonian simulation can grow exponentially with respect to the system size, becoming intractable by classical computers for large systems. Quantum computers can be used to efficiently solve this problem, as first proposed by Richard Feynman \cite{feynman1982simulating}. The main challenge is to find an efficient circuit that asymptotically approximates the time evolution of a Hamiltonian. Different approaches have been proposed for efficient Hamiltonian simulation, including product formulas \cite{trotter1959product, lloyd1996universal,suzuki1991general}, Taylor series \cite{berry2015simulating}, quantum walk \cite{berry2012black}, and quantum signal processing \cite{low2017optimal}. The product formula has a straightforward implementation and good performance in practice \cite{babbush2015chemical, childs2018toward}. Given a Hamiltonian that is decomposed as a sum of Hermitian terms ($H=\sum_{j=1}^{L}h_jH_j$), the product formula approximates the exponential of this \h~as a product of exponentials of individual terms and each individual exponential can be efficiently implemented by a quantum circuit. The approximation errors are caused by the anti-commuting terms in the Hamiltonian. 
To achieve a desired precision, the product is divided into many repetitions of small time steps, the circuit therefore has a periodic structure.

\begin{figure*}[ht!]
    \centering
    \subfloat[Google Sycamore (54 qubits) \cite{arute2019quantum}]{
    \includegraphics[scale=0.7]{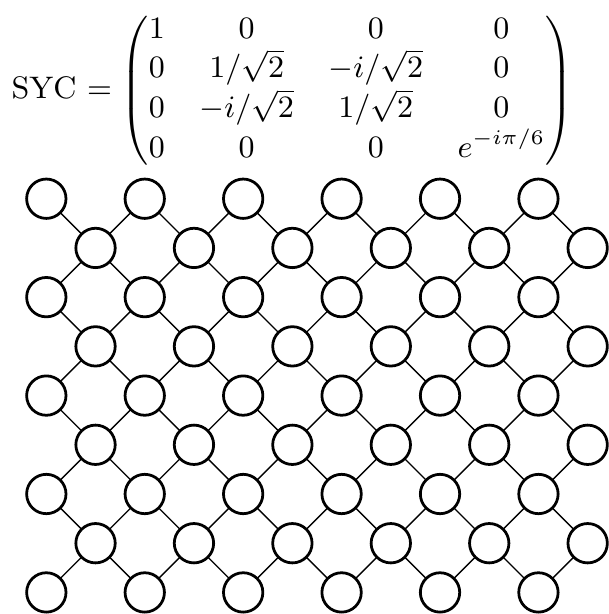}
    \label{fig:arch_sycamore}
    }\hspace{5mm}
    \subfloat[IBM Montreal (27 qubits) \cite{ibm_devices}]{
    \includegraphics[scale=0.5]{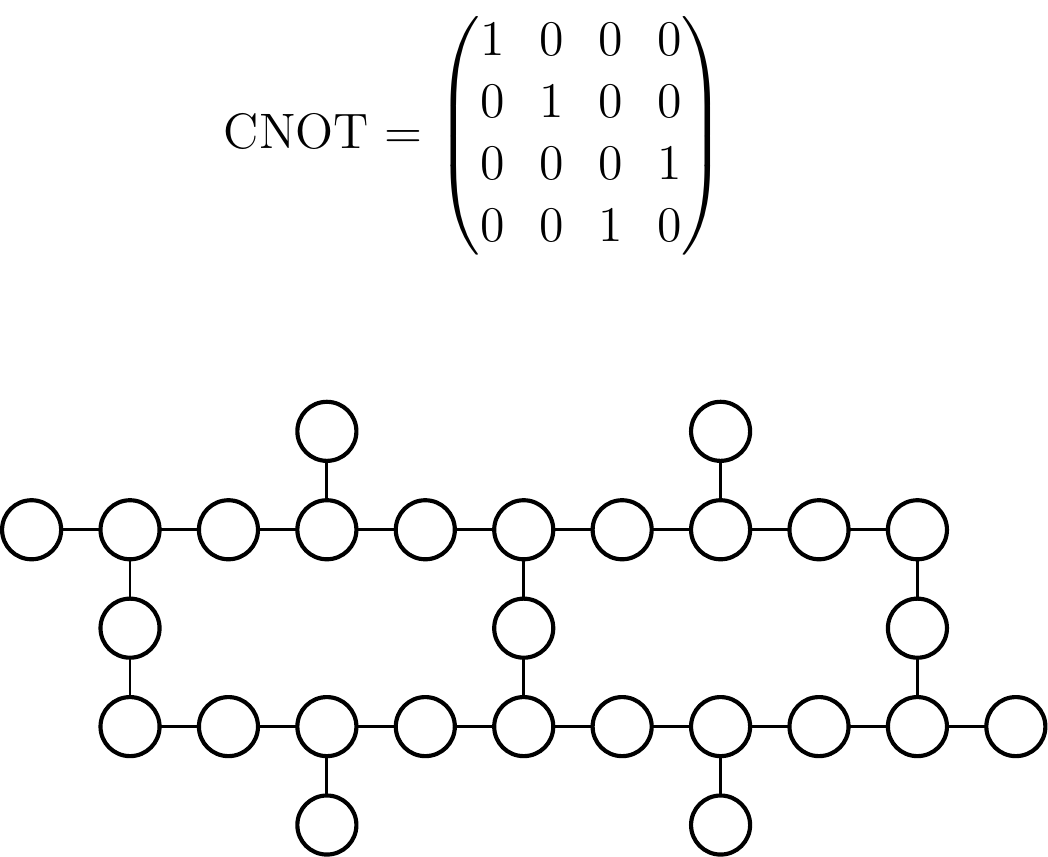}
    \label{fig:arch_monstreal}    
    }\hspace{5mm}
    \subfloat[Rigetti Aspen (16 qubits) \cite{rigetti}]{
    \includegraphics[scale=0.5]{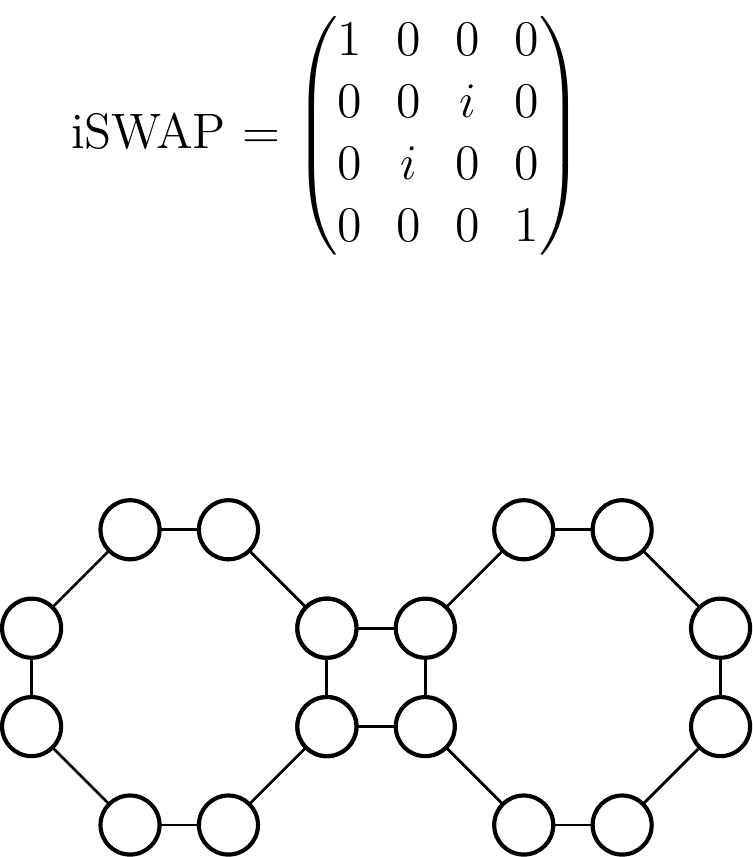}
    \label{fig:arch_aspen}    
    }
    \caption{Device topologies and hardware two-qubit gates of different quantum computers. Nodes represent qubits and edges represent the connectivity between qubit pairs. All three devices allow arbitrary single-qubit rotations.}
    \label{fig:devices}
\end{figure*}

This high-level circuit representation is typically hardware-agnostic and needs to be decomposed into the instruction set supported by the underlying quantum hardware.
In noisy intermediate-scale quantum (NISQ) computers \cite{preskill2018quantum}, the universal instruction set is normally composed of arbitrary single-qubit rotations and one or a few two-qubit gates (e.g., the SYC, CNOT, iSWAP gates in quantum processors from Google, IBM, Rigetti respectively in Figure~\ref{fig:devices}).
Furthermore, these quantum computers only allow two-qubit gates on restricted qubit pairs, i.e., limited qubit connectivity. Compilation techniques are required to map circuit qubits onto hardware qubits and insert SWAP gates to move qubits to be neighbours, increasing circuit size in terms of gate count and circuit depth. 
Two-qubit gates have much higher error rates than single-qubit rotations and qubits have short coherence time \cite{arute2019quantum}. Therefore, it is critical to minimize compilation overhead for high-fidelity circuit implementation.

\begin{figure}[tbh!]
    \centering
    \includegraphics[width=0.49\textwidth]{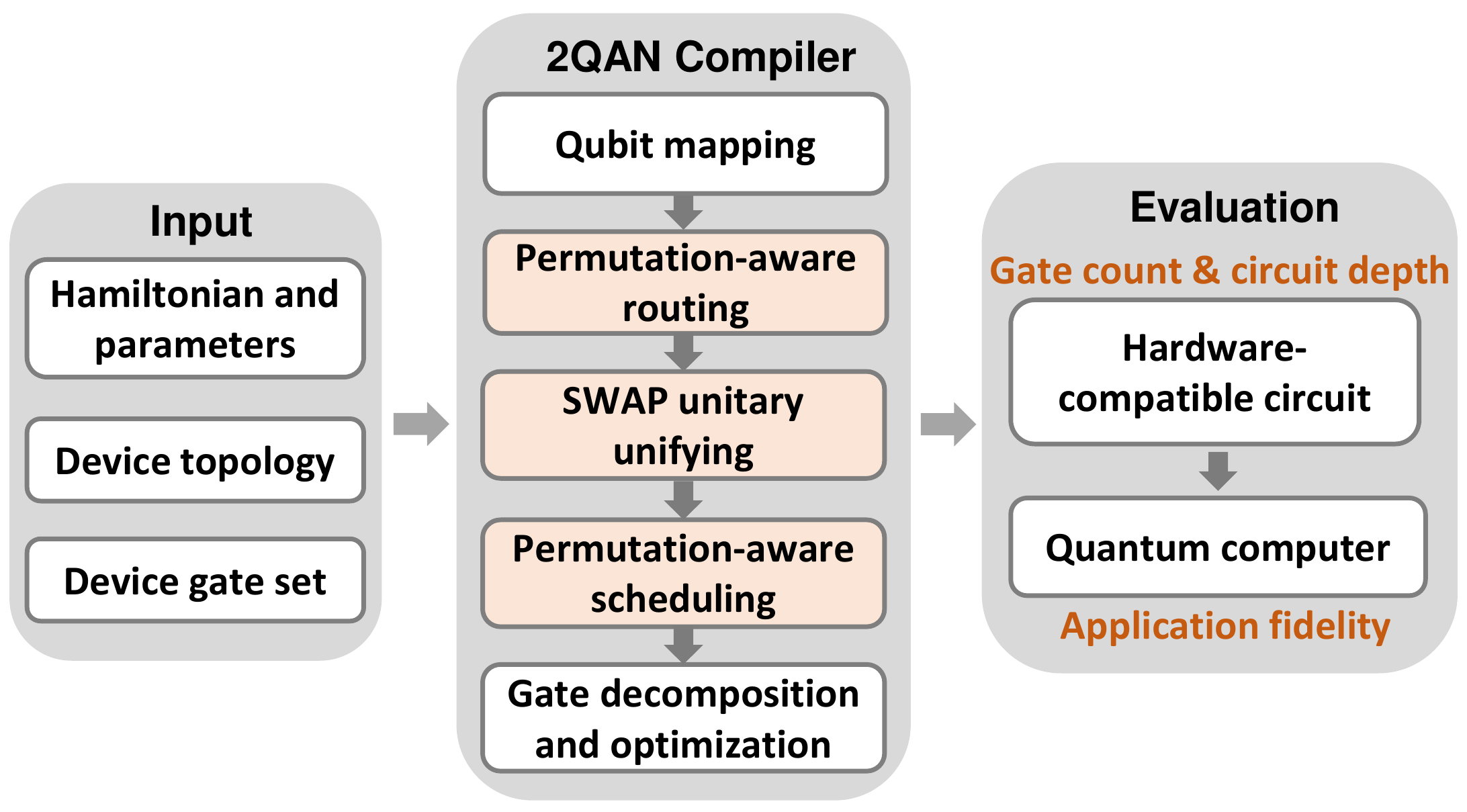}
    \caption{Overview of the \ucl~compiler. The highlighted compilation passes exploit the application-level semantics and cannot be replaced by general-purpose compilation techniques. Other passes are implemented by existing tools.}
    \label{fig:framework}
\end{figure}

Many approaches have been proposed for compiling quantum circuits onto NISQ computers \cite{zulehner2018efficient,cowtan2019qubit,lao21qmap,childs2019circuit,li2019tackling,murali2019noise,tket,qiskit}. These compilers operate at the gate level and are designed for general circuits with little knowledge regarding the mathematical properties of target applications. 
Exploiting the synergy between applications and compilation techniques can provide more optimizations on the implementation circuit, which will further improve application performance. 
For the quantum simulation problems, one application-specific property is the flexible ordering of different terms in the \h. That is, one could permute the exponentials of these terms without losing computational accuracy. This is, however, difficult or even impossible to be recognized at the gate level, especially when many of these operators anti-commute with each other (changing the order of anti-commuting gates violates program semantics). 
Several application-specific compilers have also been developed \cite{dallaire2016quantum,venturelli2018compiling,shi2019optimized,gokhale2019partial, alam2020circuit, alam2020efficient, alam2020noise,li2021software, li2021paulihedral}, but their optimization techniques are performed on one particular compilation pass and are limited to the CNOT or CZ gate set.

In this work, we identify the flexibility in \h~operator permutation and exploit it in the circuit compilation procedure. In particular, we propose permutation-aware qubit routing, gate scheduling, and gate optimization techniques to efficiently compile circuits for 2-local qubit \h~simulation problems. The quantum approximate optimization algorithm (QAOA) \cite{farhi2014quantum} also has Hamiltonians in this form. The developed compiler, named \textbf{\ucl}\footnote{Pronounced "toucan". We propose the name 2QAN since this compiler targets \textbf{2}-local \textbf{Q}ubit Hamiltonian simulation, is hardware \textbf{A}dapted and designed for \textbf{N}ISQ quantum devices.}, can target different qubit topologies and different gate sets.
Evaluation results on three quantum computers show that \ucl~can significantly reduce gate count and circuit depth compared to state-of-the-art quantum compilers. Furthermore, we also experimentally demonstrate the advantages of \ucl~on the IBMQ Montreal device. The evaluation toolflow is presented in Figure \ref{fig:framework}. 

The main contributions of this work are:
\begin{itemize}
    \item We discover an optimization opportunity for compiling quantum simulation problems on NISQ computers, that is, the order of different terms in a \h~is flexible and a quantum compiler can permute their exponentials (operators in the product formula) to minimize compilation overhead.
    \item We exploit this flexibility and develop \ucl, a quantum compiler for efficiently compiling 2-local qubit \h~simulation problems on NISQ computers. 
    We first propose a permutation-aware qubit routing heuristic to minimize the number of inserted SWAP gates. Then we implement a unitary unifying pass that combines a SWAP gate with a circuit gate to further reduce gate overhead. Moreover, we design a permutation-aware gate scheduling technique to minimize circuit depth. The routing and scheduling algorithms have quadratic time complexity in the number of gates and is scalable for large systems.
    \item We perform all the proposed permutation-aware compilation passes prior to the gate decomposition pass as shown in Figure \ref{fig:framework}. That is, these optimization algorithms are independent of the underlying hardware two-qubit gates, allowing 2QAN to target different instruction sets. 
    \item We evaluate the proposed compiler by compiling the Heisenberg model, XY model, Ising model, and the QAOA circuits onto three industrial quantum computers, Google Sycamore \cite{arute2019quantum}, IBMQ Montreal \cite{ibm_devices}, Rigetti Aspen \cite{rigetti}. Across all benchmarks and quantum computers, the \ucl~compiler outperforms state-of-the-art quantum compilers, including the general-purpose compilers \tket~\cite{tket} and Qiskit~\cite{qiskit}, and the application-specific compilers, Paulihedral~\cite{li2021paulihedral} and the QAOA compiler~\cite{ alam2020circuit, alam2020efficient, alam2020noise}. For example, \ucl~can reduce the SWAP count, two-qubit gate overhead, and depth overhead by on average 3.6x (9.1x), 4x (10.4x), and 2.1x (3.4x), respectively compared to \tket~(Qiskit). Furthermore, experimental results of running QAOA on the IBMQ Montreal device show that this substantial compilation overhead reduction provides significant improvement in application performance. \ucl~achieves the highest fidelity for all problem sizes and can increase the QAOA layers while the other compilers cannot. 
\end{itemize}

\section{Background}
\label{sec:background}
\subsection{Quantum simulation}
\noindent
\textbf{Product formula:} Consider a system \h~$H$ that is decomposed into a sum of polynomially many Hermitian terms $H_j$, $H=\sum_{j=1}^{L}h_jH_j$, its time evolution can be described by the unitary $U=\textup{exp}(it\sum_{j=1}^{L}h_jH_j)$. The goal is to find an efficient circuit construction for this unitary. One of the leading approaches is the product formula or the Trotter formula \cite{trotter1959product, lloyd1996universal, suzuki1991general},
\begin{equation}
    V(t)=\prod_{j=1}^{L}\textup{exp}(ith_jH_j)
    \label{equ:1order}
\end{equation}
and each individual operator $V_j(t)=\textup{exp}(ith_jH_j)$ can be efficiently implemented by a quantum circuit.
$V(t) =U$ if all the terms commute (i.e., $H_jH_k=H_kH_j$). $(V(t/r))^r$ approximates $U$ for large $r$ if some terms do not commute (which is typically the case in natural physical systems). This algorithm is referred as the first-order approximation. $V(t/r)$ is called one \textit{Trotterization step} and the circuit has $r$ such repetitions.
Reducing $r$ will help save the simulation cost substantially. 
For example, the second-order approximation 
\begin{equation}
    V_2(t)=\prod_{j=1}^{L}\textup{exp}(ith_jH_j/2)\prod_{j=L}^{1}\textup{exp}(ith_jH_j/2)
    \label{equ:2order}
\end{equation}
can reduce $r$ from $O((tL\Lambda)^2/\epsilon)$ (first-order) to $O((tL\Lambda)^{1+\frac{1}{2}}/\epsilon^{\frac{1}{2}})$ \cite{suzuki1991general}. $\Lambda$ is the magnitude of the strongest term and $\epsilon$ is the desired error threshold.

\noindent
\textbf{Operator permutation:}
We note that the order of terms in the \h~can be chosen from any of the $L!$ permutations. This flexibility allows one to optimize the \h~simulation circuit for each Trotter step by \textit{rearranging the $L$ individual operators} ($V_j(t/r)$) in the product formula.
Optimizing each Trotter step will improve the overall circuit construction since a large number ($r$) of such repetitions need to be performed, which will be the focus of this work.
Such rearrangement is not trivial for general-purpose quantum compilers because some of the operators anti-commute (e.g., $\textup{exp}(itX_1X_2)\textup{exp}(itY_2Y_3)\neq\textup{exp}(itY_2Y_3)\textup{exp}(itX_1X_2)$) and reordering them at the gate level will cause logic violations in the compilation.

\noindent
\textbf{2-local qubit \h:}
In this work we consider 2-local qubit Hamiltonians that originate in many physical systems,
\begin{equation}
    H = \sum_{(u,v)\in E}H_{uv} + \sum_{k\in V}H_k . 
    \label{equ:2h}
\end{equation}
$H_{uv}$ are two-qubit Hamiltonian terms and $H_k$ are single-qubit Hamiltonians. The interaction graph of this \h~is represented by $G(V,E)$, $V$ is the set of qubits and $E$ is the set of edges.
The transverse Ising model, XY model, and Heisenberg model all have Hamiltonians in this form, 
\begin{equation}
    H_{\textup{Ising}} = \sum_{(u,v)\in E}\gamma_{uv} Z_{u}Z_{v} + \sum_{k\in V}\beta_{k} X_k,
    \label{equ:ising}
\end{equation}
\begin{equation}
    H_{\textup{XY}} = \sum_{(u,v)\in E}(\alpha_{uv} X_{u}X_{v}+\beta_{uv} Y_{u}Y_{v}),
    \label{equ:xy}
\end{equation}
\begin{equation}
    H_{\textup{Heisenberg}} = \sum_{(u,v)\in E}(\alpha_{uv} X_{u}X_{v}+\beta_{uv} Y_{u}Y_{v}+\gamma_{uv} Z_{u}Z_{v}),
    \label{equ:heisenberg}
\end{equation}
where $X, Y, Z$ are the Pauli operators. 
The computational complexity of computing important properties such as their ground state energy depends on the problem graph. 
Barahona \cite{barahona1982computational} showed that computing the ground state energy of a three-dimensional Ising lattice is NP-hard, whereas polynomial classical algorithms are known for the two-dimensional Ising model in the absence of a magnetic field \cite{onsager1944crystal}. Cubitt and Montanaro \cite{cubitt2016complexity} proved the QMA-completeness of the Heisenberg model and the XY model (QMA is the quantum analogue of NP \cite{kitaev2002classical}).

\begin{figure*}[thb!]
    \centering
    \subfloat[Problem circuit]{
    \includegraphics[scale=0.9]{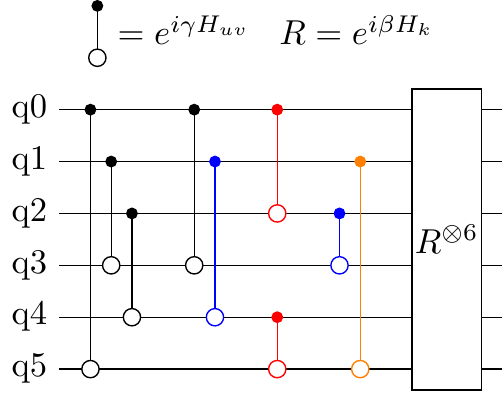}
    \label{fig:eg}
    }\hspace{2mm}
    \subfloat[Generic compiler]{
    \includegraphics[scale=0.9]{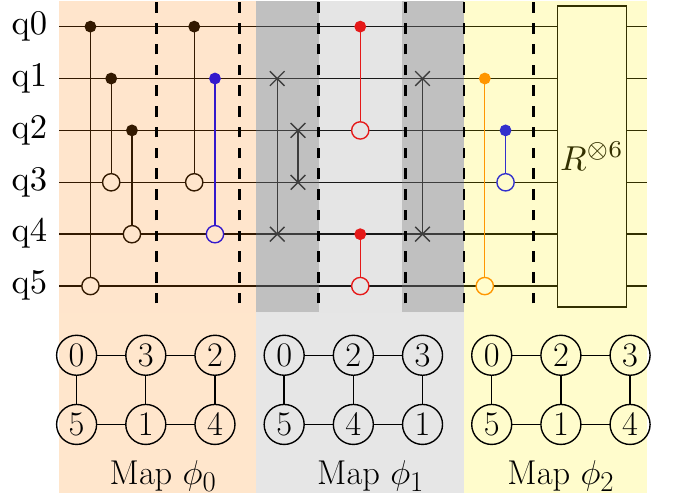}
    \label{fig:eg_tket}    
    }
    \subfloat[Application-specific compiler]{
    \includegraphics[scale=0.9]{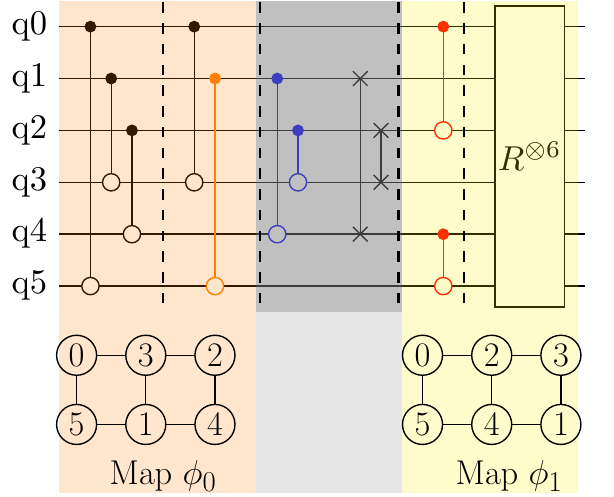}
    \label{fig:eg_ucl}    
    }
    \caption{Examples of compiling a 6-qubit 2-local \h~(Equation \ref{equ:2h}) to a $2\times 3$ grid architecture. (a) The circuit construction of one Trotter step, where the two(single)-qubit operators implement the evolution of two(single)-qubit \h s. These two-qubit operators may not commute. (b-c) Compilation procedure. Top figures show the compiled circuits and bottom figures show the qubit mappings (nodes are qubits and edges represent their connectivity).
    Gates between dashed lines can be performed in parallel. To avoid confusion, SWAPs in this Figure are applied on the corresponding hardware qubits, we draw them on the circuit qubits for better readability.
    (b) Compilation procedure using a generic compiler that obeys the gate dependencies in (a). 
    In total, the compiled circuit has 12 two-qubit gates and a depth of 7. 
    (c) Compilation procedure by using 2QAN compiler that exploits the flexible operator ordering in \h~simulation problems. 
    In total, the compiled circuit has 9 two-qubit gates and a depth of 5 (Each of inserted SWAPs can be merged with a circuit unitary in gray area).}
    \label{fig:egs}
\end{figure*}

\subsection{Circuit compilation}
NISQ computers have hardware limitations such as native gate set and qubit connectivity, making quantum applications not directly executable.
A hardware gate set is typically composed of arbitrary single-qubit rotations and a few two-qubit gates which vary across QC vendors. For example, IBM devices currently have the CNOT as the native two-qubit gate \cite{ibm_devices}, Rigetti implement both CZ and iSWAP gates \cite{rigetti,abrams2019implementation}, and Google support CZ, SYC, $\sqrt{\textup{iSWAP}}$ gates \cite{arute2019quantum,foxen2020continuous}. High-level quantum circuits need to be decomposed into the given gate set by using analytical algorithms \cite{kraus2001optimal, khaneja2001time} or numerical approaches \cite{davis2019heuristics, lao2021isca}.
Furthermore, two-qubit gates can only be executed on the connected (nearest neighbouring, NN) qubits. Movement operations such as SWAP gates need to be inserted for performing non-NN gates, increasing the number of gates and circuit depth.
NISQ computers have limited qubit coherence time and high gate error rates.
It is crucial to minimize circuit sizes for reliable implementation of quantum applications.
To this purpose, quantum compilation techniques, including qubit mapping and routing, gate decomposition and scheduling, have been developed to efficiently transform high-level quantum circuits into hardware-compatible ones. 
For general quantum circuits, the routing and scheduling algorithms assign dependencies between gates to maintain the correctness of program semantics \cite{zulehner2018efficient,cowtan2019qubit,lao21qmap,childs2019circuit,li2019tackling,murali2019noise,tket,qiskit}. These dependencies are typically generated based on the gate order in an input circuit.
These general-purpose quantum compilers work at the gate level and do not consider application-level semantics. Any \textit{application-specific} optimizations that can improve application performance are desirable for NISQ computing. 

\noindent
\textbf{Compilation for \h~simulation:}  
As mentioned previously, when constructing a circuit for each Trotter step, one is free to use any permutation of the operators in the product formula (even they do not commute). 
Such application-level property allows more optimizations for \h~simulation problems but is not exploited by general-purpose quantum compilers.
Figure \ref{fig:egs} shows an example of compiling a 6-qubit \h~circuit to a $2\times 3$ grid architecture. A generic compiler inserts 3 SWAP gates and outputs a 7-depth circuit with 12 two-qubit gates. In contrast, a compiler that considers the flexibility in operator permutation only uses 2 SWAPs. Moreover, these \swap gates can be combined with other gates in the circuit, further decreasing the gate count and circuit depth (the compiled circuit only has 9 two-qubit gates and depth 5).
In this work, we design such an application-specific compiler for 2-local qubit \h~simulation problems to improve application fidelity.

\section{Compilation techniques }
\label{sec:compiler}

In this section, we introduce the proposed compilation techniques for quantum simulation. Figure \ref{fig:framework} shows the overview of our \ucl~compiler.

\subsection{Qubit mapping}
The goal of qubit mapping is to find an optimal qubit initial placement such that the number of qubit moving operations required for implementing all two-qubit gates is minimized.
Similar to the approaches in~\cite{dousti2014squash, bahreini2015minlp,lao2018mapping}, the qubit mapping problem is formulated as a quadratic assignment problem (QAP). That is, the problem of allocating each qubit in the circuit (facility) to one qubit in the device (location) with the cost defined by a function of the distance and interaction times (flow) between circuit qubits. 
Let $n$ be the number of circuit qubits or physical qubits and denote by the set $N=\{1,2,\cdots, n\}$. The objective function is
\begin{equation}\label{equ:qap}
\min_{\phi \in \mathit{S}_n} \sum_{i=1}^{n}\sum_{j=1}^{n}f_{ij}d_{\phi(i)\phi(j)} 
\end{equation}
where $\mathit{S}_n$ is the set of all permutations $\phi$: $N \rightarrow N$, $f_{ij}$ is the interaction times between circuit qubits $i$ and $j$, $d_{\phi(i)\phi(j)}$ is the distance between hardware qubits $\phi(i)$ and $\phi(j)$ and is calculated by using the Floyd-Warshall algorithm.

QAP is a NP-hard problem \cite{qap} and we use the Tabu search heuristic algorithm \cite{glover1989tabu, glover1990tabu} to efficiently find good qubit mappings in this work. 
Other heuristics such as simulated annealing \cite{burkard1984thermodynamically} and greedy randomized adaptive search procedure \cite{li1993greedy} can be also used for solving this problem.
Prior works observed that the QAP formulation of qubit mapping may not work well for general quantum circuits \cite{dousti2014squash, bahreini2015minlp,lao2018mapping}. This is because qubits need to interact in a specific order (gate dependency) and the initial mapping benefits diminish after insertion of SWAPs, e.g., some NN qubits may be moved further apart but will interact later. This is not the case for 2-local \h~simulation problems as any operator that is NN in a qubit map can be scheduled directly regardless of their order in the circuit (see examples in Figure \ref{fig:egs}).

\subsection{Qubit routing}
Generally, not all two-qubit gates are nearest neighbouring in an initial qubit map. Movement operations such as SWAP gates need to be inserted to move non-NN qubits, causing overheads in gate count and circuit depth. Efficient qubit routing techniques are required to minimize compilation overhead for reliable computation.
Different from existing qubit routing algorithms that respect the gate order in the input circuit \cite{cowtan2019qubit,li2019tackling,tannu2019not,ding2020square}, the routing in \ucl~exploits the operator permutation flexibility in a \h. The pseudocode is in Algorithm \ref{alg:routing} and an example is shown in Figure \ref{fig:eg_ucl}.

Given an input circuit that implements one Trotter step of a \h, our routing algorithm starts by searching all the two-qubit gates that are NN in an initial qubit layout found by the mapping algorithm (e.g., there are 7 NN two-qubit gates in Figure \ref{fig:eg_ucl}). These gates are directly mapped and SWAP gates are needed to perform the remaining two-qubit gates (Lines 2-3). For these non-NN gates, it compares their qubit distances in the hardware (the distance matrix in Equation \ref{equ:qap}) and selects the shortest-distance one to route (Line 5). If there are multiple shortest ones, select the first one in this set.
Then the routing algorithm finds all possible SWAP gates that act on one of the qubits of the chosen gate $g$ (Line 6)
It evaluates these SWAPs based on a SWAP selection criteria (will be explained shortly) and selects the best one as the first SWAP for gate $g$ (Line 7).
The qubit map is updated and NN gates are found for the new map and are removed from the un-routed gate set (Lines 8-10). This procedure (Lines 5-10) is repeated until all two-qubit gates are performed. The time complexity of the routing algorithm is $O(m^2n)$, $m$ is the number of two-qubit gates and $n$ is the number of qubits.

\begin{algorithm}
   \begin{algorithmic}[1]
        \Require Un-routed circuit, initial map $\phi_0$, device topology
            \Ensure Routed circuit, a set of qubit maps $\{\phi_i\}$ and a set of NN gates corresponding to each map $\{G_{\phi_{i}}\}$
        \State Initialize the set of qubit maps $\Phi=\{\phi_0\}$
        \State Initialize the set of NN gates for each map, $G=\{G_{\phi_{0}}\}, G_{\phi_{0}} \gets$ all NN two-qubit gates for map $\phi_0$
        \State Initialize $G_{\textup{ur}} \gets$ all un-routed (non-NN) two-qubit gates
        \While{$G_{\textup{ur}} \neq \varnothing $}
            \State Select the gate $g$ $\in G_{\textup{ur}}$ that has shortest distance in $\phi_i$
            \State $S_g\gets$ Find all SWAP gates on qubits in $g$
            \State Select the best SWAP from $S_g$ and add it to $G_{\phi_{i}}$
            \State Update qubit map from $\phi_i$ to $\phi_{i+1}$
            \State $G_{\phi_{i+1}} \gets$Find NN gates in $G_{\textup{ur}}$ for map $\phi_{i+1}$ 
            \State Remove all gates in $G_{\phi_{i+1}}$ from $G_{\textup{ur}}$, add $ G_{\phi_{i+1}}$ to $G$, add $\phi_{i+1}$ to $\Phi$
        \EndWhile
   \end{algorithmic}
   \caption{Permutation-aware routing}
   \label{alg:routing}
\end{algorithm}

\noindent
\textbf{SWAP selection criteria:} The best SWAP gate is evaluated based on three criteria:
\begin{enumerate}
    \item \textit{Least SWAP count:} It will lead to the minimal cost in Equation \ref{equ:qap} (i.e., the minimal number of SWAP gates) for remaining non-NN gates.
    \item \textit{Shortest circuit depth:} It can be most interleaved with previously mapped gates, introducing the least depth overhead.
    \item \textit{Best gate optimization:} It can be merged with a circuit gate, i.e., if there is a circuit gate applied on the same qubits as this SWAP, the compiler will replace these two gates by a single unitary representing their product (more details will be introduced in the next section). 
\end{enumerate}

For our compiler configuration, we will evaluate the best SWAP based on the three criteria in the above priority order. Evaluating these criteria in a different order may further improve the compiler but will not be explored here. When there are multiple best options, a random one will be chosen. 
For example, both \swap(0,3) and \swap(2,3) in map $\phi_0$ of Figure \ref{fig:eg_ucl} can be added for performing the two-qubit gate on pair (0,2) and they have the same cost regarding the first two criteria. \swap(2,3) is selected because it can be combined with a circuit gate that operate on the same qubits.
For implementing the circuit gate on (4,5), both \swap(1,5) and \swap(1,4) are the best, and the compiler randomly selects one.

\subsection{Unitary unifying}
\label{sec:unify}
\begin{figure}[tbh!]
    \centering
    \includegraphics[scale=0.8]{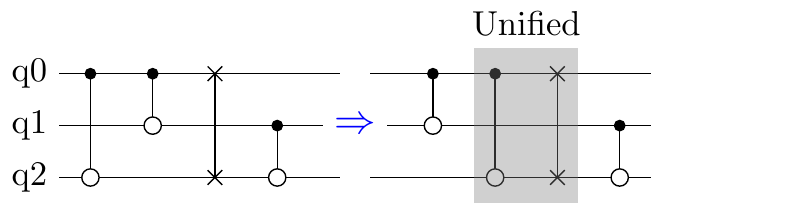}
    \caption{Unitary unifying example. The flexible operator permutation allows the circuit gate on qubits (0,2) to be rescheduled and combined with the SWAP gate.}
    \label{fig:swap}
\end{figure}
\begin{figure*}[tbh!]
    \centering
    \includegraphics[scale=0.72]{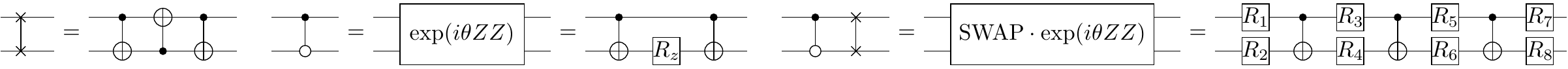}
    \caption{Examples of decomposing the SWAP, the unitary $\textup{exp}(i\theta ZZ)$, and their product into CNOTs and single-qubit rotations.}
    \label{fig:decomposition}
\end{figure*}

\noindent
\textbf{SWAP unitary unifying:}
For each added SWAP gate, the compiler searches over all circuit gates. If there is a circuit gate operating on the same qubit pair as this SWAP, the compiler reschedules the circuit gate to the SWAP gate cycle and unifies them into a single unitary (see the example in Figure \ref{fig:swap}). Such rescheduling is allowed because of the operator permutation flexibility in \h~simulation. The unified unitary (referred as \textit{dressed} SWAP) is the product of the circuit gate unitary and the SWAP unitary. 
This unitary unifying helps further reduce compilation overhead. 
Figure \ref{fig:decomposition} shows the decomposition of a SWAP gate, a circuit gate $\textup{exp}(i\theta ZZ)$, and their unified unitary into CNOT gates and single-qubit rotations.
The unified unitary requires 3 CNOTs while the implementation of these two gates with individual decomposition (in the left circuit in Figure \ref{fig:swap}) would require 5 CNOTs in total.

\noindent
\textbf{Circuit unitary unifying:}
Similarly, we merge all circuit gates that act on the same qubit pair into one single unitary. For example, there are three two-qubit Pauli terms on one qubit pair in the Heisenberg model (Equation \ref{equ:heisenberg}). The exponential of a two-qubit Pauli operator normally requires 2 CNOTs. Implementing each of these three exponentials individually would use 6 CNOTs in total \cite{li2021paulihedral} while the unified unitary only requires 3 CNOTs (Any two-qubit gate can be implemented by at most 3 CNOTs \cite{vatan2004optimal,vidal2004universal}).
To reduce gate count, we pro-process all 2-local \h~simulation circuits by using this circuit unitary unifying prior to other compilation passes (not shown in Figure \ref{fig:framework}). 

\subsection{Gate scheduling}
\noindent
\textbf{Scheduling without dependency:} When connectivity limitations are not considered, the individual operators in one Trotter step (Equation \ref{equ:1order}) can be performed in any order. Graph coloring algorithms \cite{jensen2011graph} can be used to schedule such circuits. In the graph construction, nodes represent operators (gates), and two nodes are connected by an edge if they have common qubits and cannot be scheduled in parallel. We use the default greedy algorithm in NetworkX version 2.5 for our scheduling pass. The circuits without taking into account topology constraints are scheduled by this method and are baseline circuits used for calculating compilation overhead.

\begin{figure}[thb!]
    \centering
    \subfloat[Generic scheduler]{
    \includegraphics[scale=0.9]{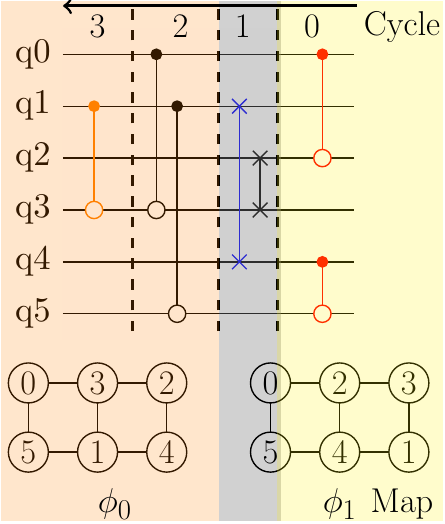}
    \label{fig:schedule1}    
    }\hspace{5mm}
    \subfloat[Hybrid scheduler]{
    \includegraphics[scale=0.9]{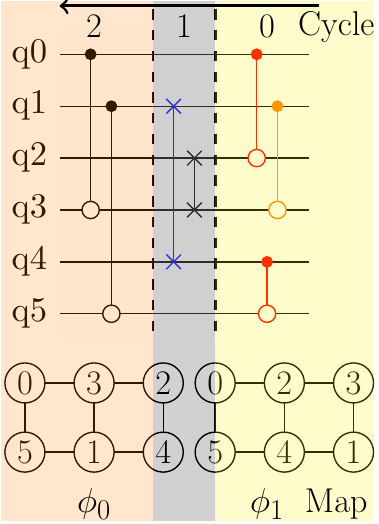}
    \label{fig:schedule2}    
    }
    \caption{ALAP scheduling examples. (a) A generic scheduler respects the gate order provided by the routing pass. The scheduled circuit takes 4 cycles. (b) The hybrid scheduler considers the flexibility of permuting circuit gates and respects the dependency between a SWAP gate and its corresponding circuit gates. 
    The scheduled circuit takes 3 cycles.}
    \label{fig:scheduling}
\end{figure}

\noindent
\textbf{Scheduling with dependency:}
For connectivity-constrained quantum computers, the routing pass in previous section will be applied. The router outputs a list of qubit maps and a set of NN gates corresponding to each map, including both circuit gates and (dressed) SWAP gates.
The order of a (dressed) SWAP gate and a circuit gate cannot be exchanged if the SWAP is inserted to make the circuit gate NN.
One can generate a gate dependency graph based on the gate order after qubit routing and apply conventional scheduling algorithms \cite{cowtan2019qubit,lao21qmap} to minimize circuit depth (decoherence limit). This approach may perform sufficiently well for some circuits such as the one in Figure~\ref{fig:eg_ucl}, but more optimizations can be achieved for other circuits by considering the operator permutation in \h~simulation. In this work, we apply such application-specific optimizations and the pseudocode of the proposed gate scheduling algorithm is presented in Algorithm \ref{alg:scheduling} and an example is shown in Figure \ref{fig:scheduling}. 

\begin{algorithm}
   \begin{algorithmic}[1]
        \Require The set of maps $\Phi=\{\phi_i\}$ and the set of NN gates corresponding to each map $G=\{G_{\phi_i}\}$ 
        \Ensure Scheduled circuit
        \State Use graph coloring algorithm to schedule gates in $G_{\phi_0}$
        \State Initialize scheduling cycle $t=0$ and the set of scheduled gates in this cycle $C_t=\varnothing$
         \State Initialize qubit map for cycle $t$: $M_t \gets$ last map in $\Phi$ (for an ALAP scheduling)
         \State $G_{\textup{us}} \gets$ All un-scheduled gates $\in G$
        \While{$G_{\textup{us}} \neq \varnothing $}
            \For{each circuit gate $g \in G_{\textup{us}}$}
            \State Schedule $g$ in $C_t$ iff it is NN in $M_t$ and its qubits are free
            \EndFor
            \For{each SWAP gate $\in G_{\textup{us}}$}
            \State Schedule it in $C_t$ iff it is NN in $M_t$, its qubits are free, and there is no dependency violation 
            \State $M_{t+1}\gets$ the map after applying this SWAP on $M_t$ 
            \EndFor
        \State $t \gets t+1$ and update $G_{\textup{us}}$ 
        \EndWhile
        \State Reverse the gate sequence in $\{C_0, \cdots, C_t\}$ (ALAP)
   \end{algorithmic}
   \caption{Permutation-aware scheduling }
   \label{alg:scheduling}
\end{algorithm}

\noindent
\textbf{Hybrid scheduling:}
Algorithm \ref{alg:scheduling} is a hybrid of the above two scheduling techniques.
The flexible operator permutation property implies that circuit gates can be scheduled in any qubit map where they are NN.
The NN circuit gates in the initial qubit map do not have dependencies and are first scheduled using the graph coloring algorithm (Line 1). 
For other qubit maps, dependencies between SWAP and circuit gates need to be respected.
The algorithm initializes the circuit cycle $t=0$.
We implement an as-late-as-possible (ALAP) schedule in this work so the algorithm assigns the last qubit map to cycle 0 ($M_0$, Line 2). For cycle $t$, it first finds all circuit gates that can be scheduled (Lines 6-8). A circuit gate can be scheduled at $t$ only if its qubits are NN in map $M_t$ and these qubits are not currently occupied by any other gates. For example, the gate on (1,3) in Figure \ref{fig:schedule2} is scheduled at cycle 0 while it needs to be scheduled at a different cycle in Figure \ref{fig:schedule1} because of the dependency constraint when using a generic scheduler (which can schedule it until its predecessor gate on (1,5) is performed).
Afterwards, the scheduler finds all SWAP gates that can be scheduled at this cycle (Lines 9-12). Except the qubit NN and availability requirements, a SWAP gate can be scheduled at $t$ only if the circuit gates that depends on it have been scheduled (e.g., red gates depend on the two SWAPs in Figure \ref{fig:schedule2}). Once a SWAP is inserted, the qubit map for next cycle will be updated.
The procedure in Lines 6-13 is repeated until all gates are scheduled. Finally, the algorithm reverses the scheduled gate sequence (Line 15).
The time complexity of this hybrid scheduling algorithm scales quadratically with the number of gates.

\section{Experimental setup}

\noindent
\textbf{Benchmarks:} 
We consider the Hamiltonians in a linear array with nearest neighbouring (NN) and next nearest neighbouring (NNN) interactions for the transverse Ising model, XY model, and Heisenberg model, they are noted as \textbf{NNN Ising}, \textbf{NNN XY}, and \textbf{NNN Heisenberg}. The time evolution of a \h~is implemented by using the product formula $(\prod_{j=1}^L\textup{exp}(ih_{j}H_{j}t/r))^r$, $r$ is the number of Trotter steps. 
The coefficients of $H_{j}$ are randomly sampled from $(0,\pi)$. The number of two-qubit operators for NNN Ising, NNN XY, and NNN Heisenberg models in each Trotter step is $2n-3$, $n$ is the number of qubits ranging from 6 to 50 in our evaluation.

In addition, we also use QAOA \cite{farhi2014quantum} for solving the MAX-CUT problems on 3-regular graphs (\textbf{QAOA-REG-3}) as our benchmark.
QAOA is a popular benchmark for testing the performance of quantum computers and has been demonstrated in different quantum processors \cite{lacroix20improving,Harrigan2021quantum}. In particular, the connectivity differs for each problem graph, which is well suited for evaluating compilation techniques. 
QAOA has the same form as the Ising model, 
its problem \h~is $C=\sum_{(u,v)\in E}Z_{u}Z_{v}$ and drive \h~is $B=\sum_{k\in V}X_{k}$. 
The circuit implementation of one-layer QAOA is  
\begin{equation}
    U(\gamma, \beta)=\prod_{(u,v)\in E}\textup{exp}(i\gamma Z_{u}Z_{v})\prod_{k\in V}\textup{exp}(i\beta X_{k}),
    \label{equ:qaoa}
\end{equation}
Different from \h~simulation, these parameters ($\gamma,\beta$) differ in every layer.
In our evaluation, we randomly sample 10 graph instances for each problem size. The operator parameters for each instance are chosen at their theoretically optimal values and are calculated by using tools provided in ReCirq~\cite{recirq}. The number of two-qubit operators in one-layer QAOA-REG-3 is $3n/2$, we consider qubit numbers $n$ from 4 to 22.

\noindent
\textbf{Quantum computers:} We compile these benchmarks onto three quantum computers, \textbf{Google Sycamore \cite{arute2019quantum}, IBMQ Montreal \cite{ibm_devices}, Rigetti Aspen \cite{rigetti}}. As shown in Figure \ref{fig:devices}, Sycamore has a grid architecture with SYC as hardware two-qubit gate, Montreal has a dodecagon lattice with CNOT as native gate, Aspen has connected octagons and iSWAP as native gate. All three devices support arbitrary single-qubit rotations. Experiments on Montreal were performed on 29th October, 2021, the average CNOT error rate was 1.241\%, average read-out error rate was 1.832\%, and average T1=87.75 $us$ and T2=72.65 $us$.

\noindent
\textbf{Quantum compilers:}
We first compare our \textbf{\ucl}~compiler with two state-of-the-art general-purpose compilers, the \textbf{\tket}~compiler version 0.11.0 with the recommended `FullPass' \cite{tket} and the \textbf{Qiskit} compiler version 0.26.2 with optimization level 3 \cite{qiskit}. We then compare \ucl~with two application-specific compilers, the \textbf{QAOA compiler} (IC-QAOA) with default settings \cite{alam2020circuit, alam2020efficient, alam2020noise} and the \textbf{Paulihedral} compiler for quantum simulation \cite{li2021paulihedral}. 
For circuits with larger number of qubits, the default mapping in \tket~may fail to find a qubit initial placement and we use their `LinePlacement' pass instead. 
\tket~and Qiskit have advanced circuit optimizations for the CNOT or CZ gates.
For devices that have different hardware two-qubit gates, we disable the gate decomposition pass in Qiskit and \tket. 
The \ucl~compiler always performs permutation-aware passes prior to gate decomposition.
The mapped circuits that have application-level unitaries 
will need to be decomposed into hardware gate sets.
We apply the `SynthesiseIBM' decomposition pass in \tket~to decompose the mapped circuits by \ucl~for the Montreal device.
We use the analytical method in Cirq~\cite{cirq} to decompose QAOA and Ising unitaries into SYC gates. For other application unitaries and hardware gates, we use the numerical approach developed in \cite{lao2021isca} for finding more efficient decomposition. 
Both Qiskit and \ucl~involve randomization in the mapping procedure, we run their mapping passes 5 times and choose the best results. We also pro-process the input circuits for \tket~and Qiskit by applying the circuit unitary unifying in Section \ref{sec:unify} to reduce compilation overhead.

\noindent
\textbf{Metrics:}
Similar to prior works, we use the \textbf{number of inserted SWAP gates} (smaller is better), the \textbf{number of hardware two-qubit gates} (smaller is better), the \textbf{depth of two-qubit gates} (shorter is better), and the \textbf{depth of all gates} (shorter is better) as metrics to compare the performance of different compilers. We also calculate the increase in gate count and circuit depth (i.e., \textbf{compilation overhead}, less is better) compared to the circuits without considering connectivity constraints (i.e., the baseline implementation). Furthermore, we experimentally evaluate the application performance of QAOA benchmarks on the Montreal device. The performance is measured by the \textbf{normalized cost function} $\left \langle C \right \rangle/C_{\textup{min}}$ (larger is better) \cite{Harrigan2021quantum}. 1 means the perfect result and 0 corresponds to the random guessing result. 

\noindent
\textbf{Implementation:}
We implement \ucl~in Python 3.8.
All compilation in our evaluation was performed on a laptop with an Intel Core i7 processor (2.30GHz and 32GB RAM).

\section{Compiler evaluation}
\label{sec:results}

\begin{figure*}[tbh!]
 \centering
 \subfloat[SWAP count for NNN Heisenberg model]{
     \includegraphics[width=0.33\textwidth]{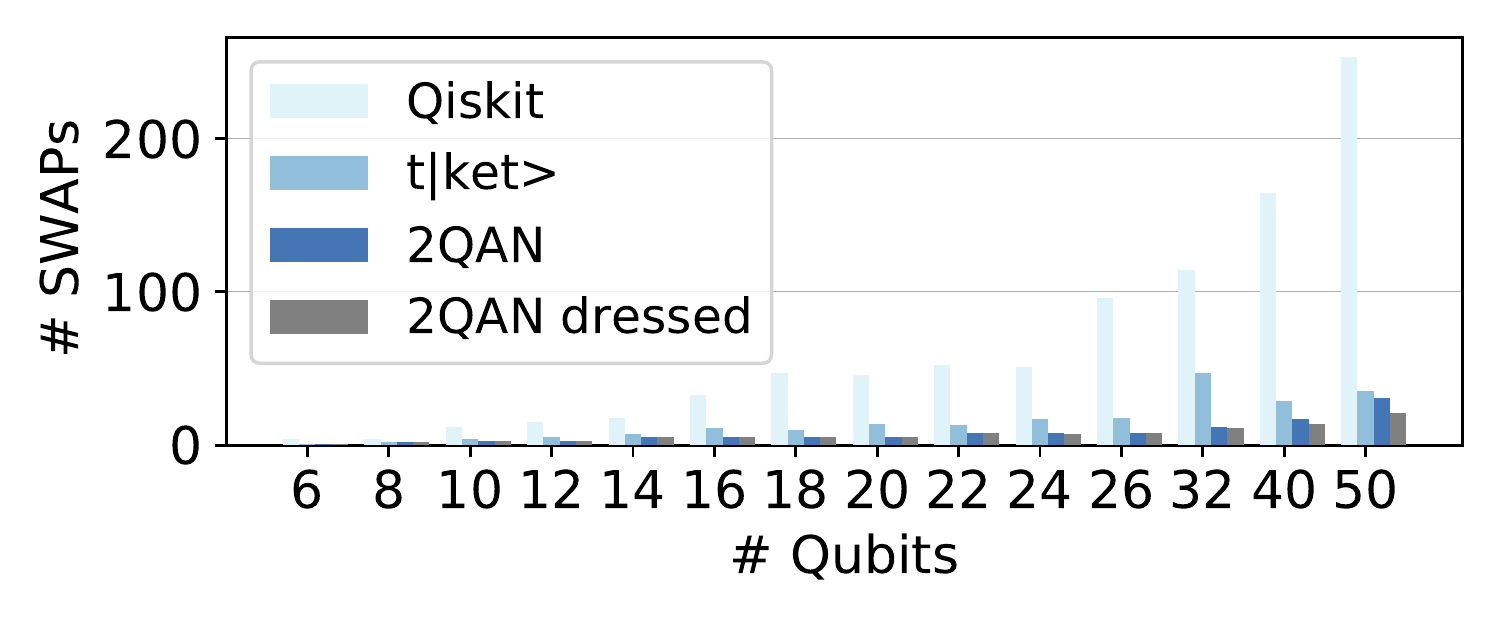}
    \label{fig:grid_hb_swap}
    }
    \subfloat[SYC count for NNN Heisenberg model]{
     \includegraphics[width=0.33\textwidth]{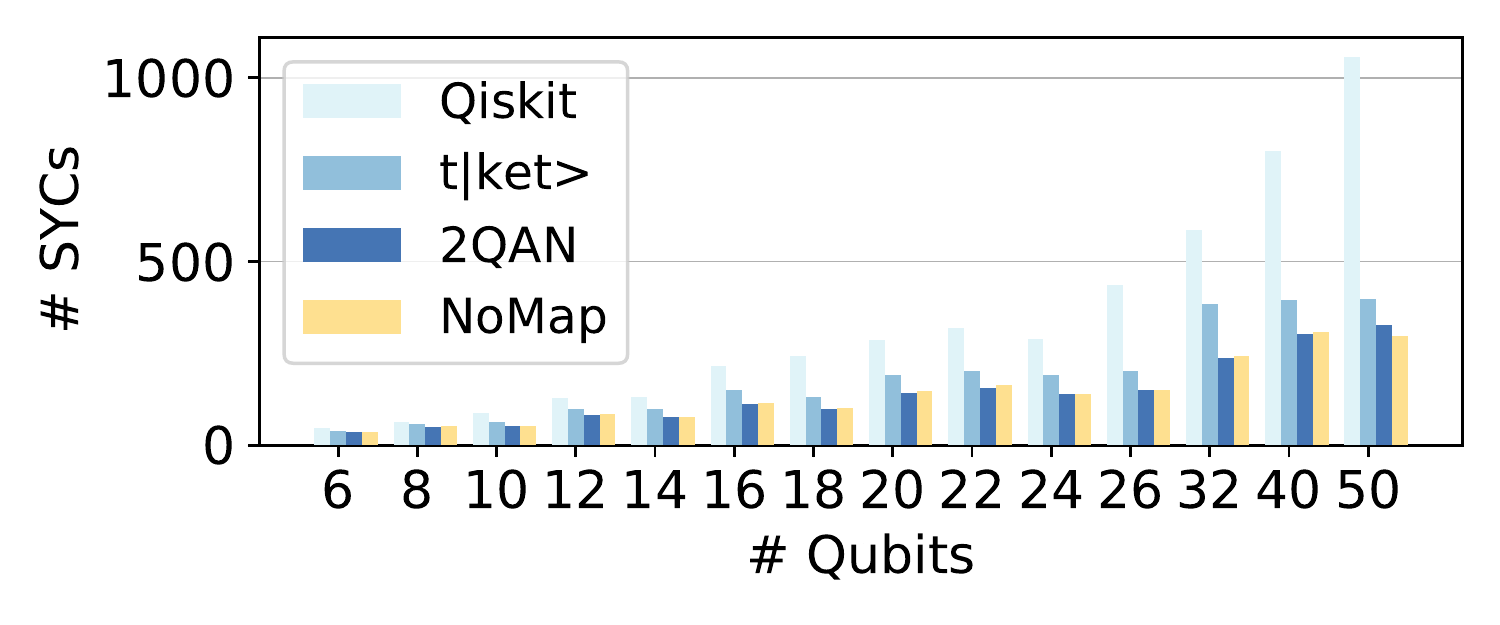}
    \label{fig:grid_hb_sycamore}
    }
    \subfloat[SYC depth for NNN Heisenberg model]{
    \includegraphics[width=0.33\textwidth]{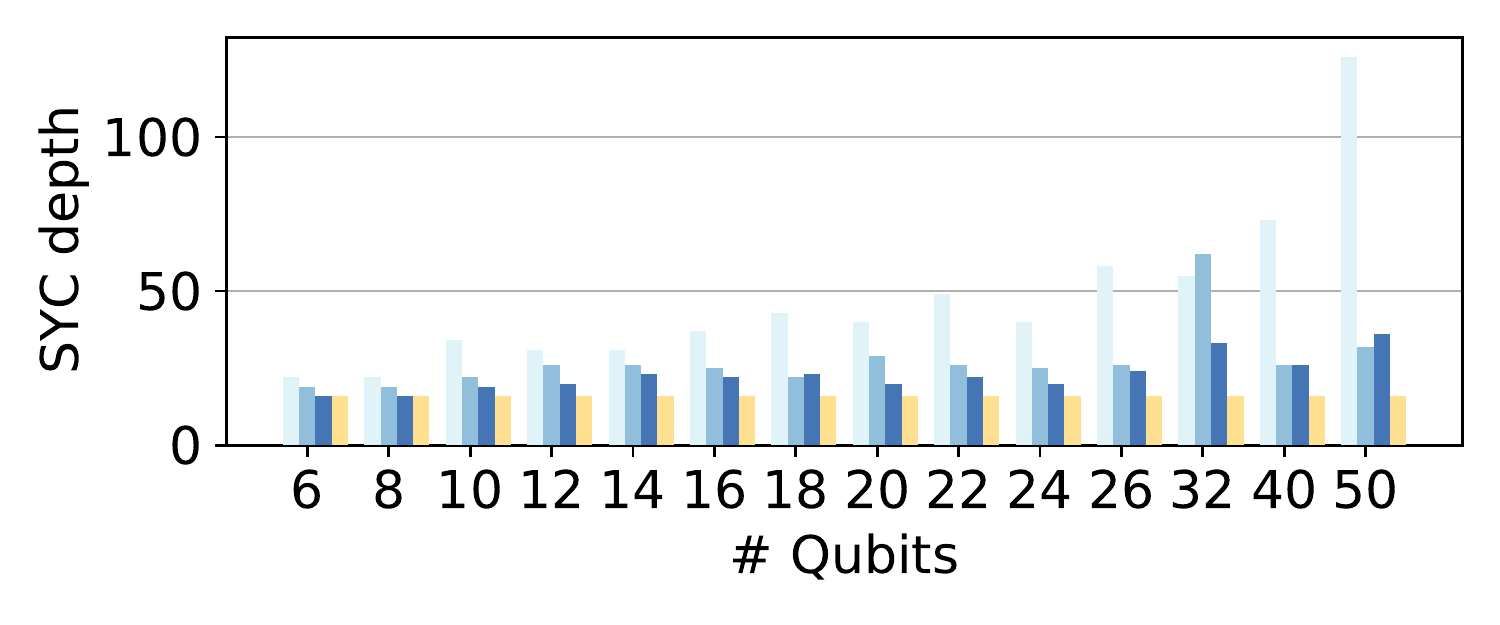}
    \label{fig:grid_hb_sycamore_depth}
    }
    
    \subfloat[SWAP count for NNN XY model]{
     \includegraphics[width=0.33\textwidth]{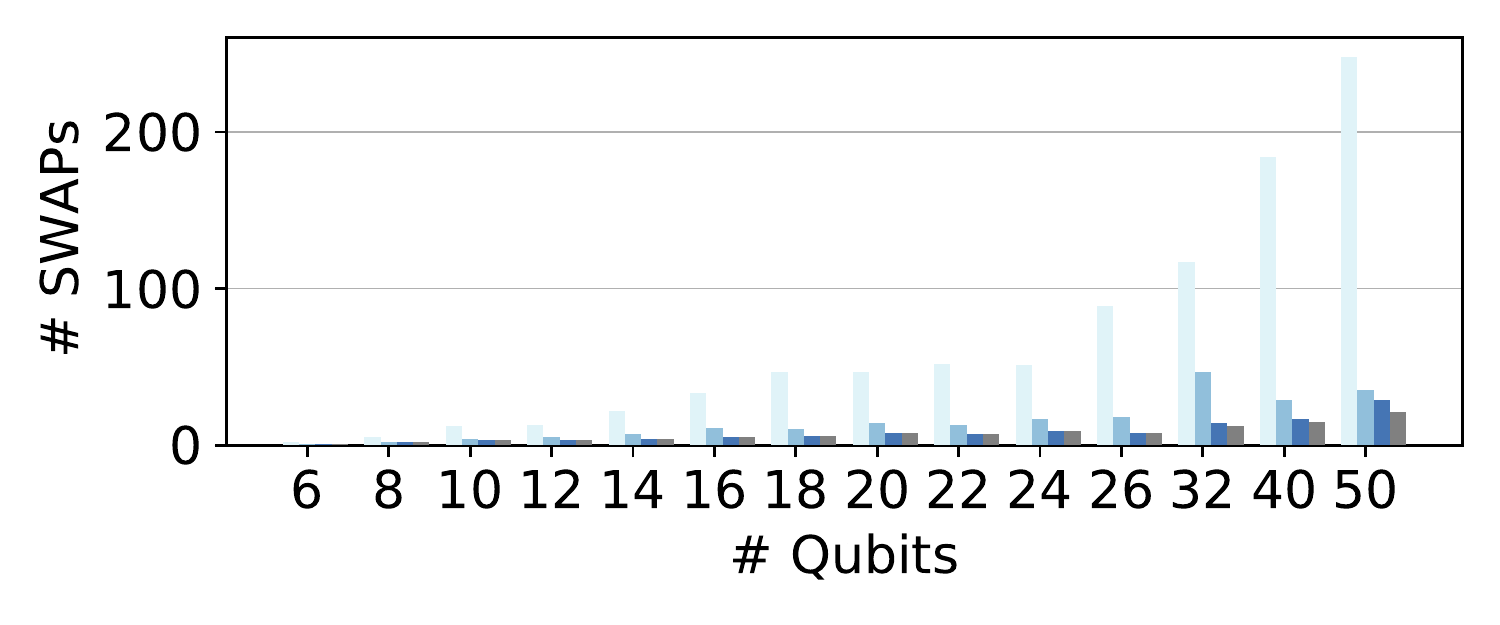}
    \label{fig:grid_xy_swap}
    }
    \subfloat[SYC count for NNN XY model]{
     \includegraphics[width=0.33\textwidth]{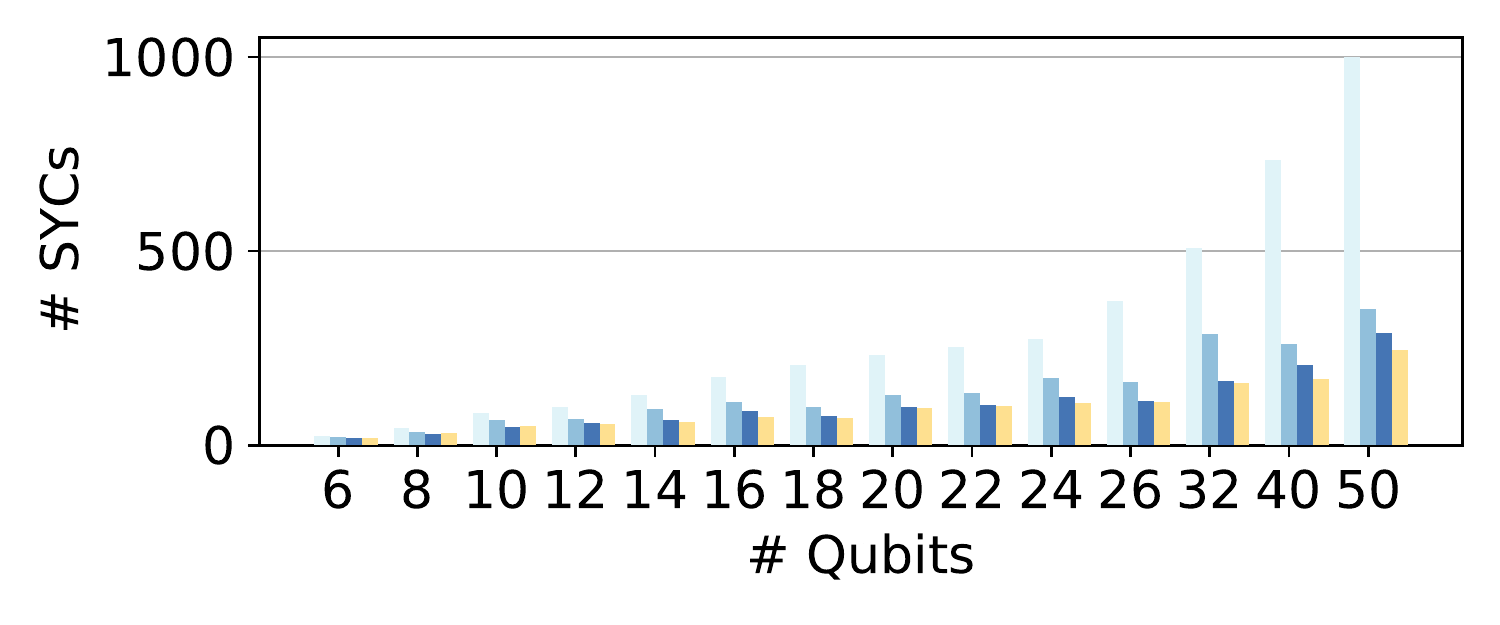}
    \label{fig:grid_xy_sycamore}
    }
    \subfloat[SYC depth for NNN XY model]{
    \includegraphics[width=0.33\textwidth]{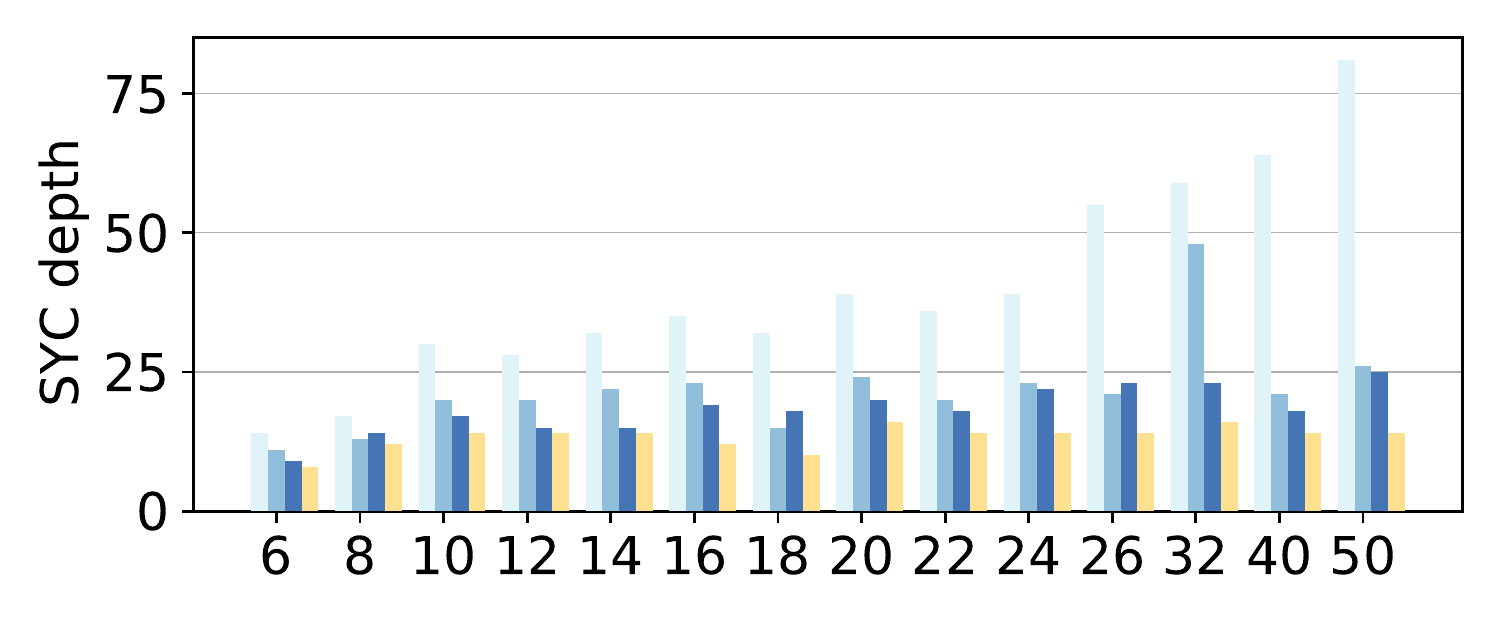}
    \label{fig:grid_xy_sycamore_depth}
    }
    
    \subfloat[SWAP count for NNN Ising model]{
     \includegraphics[width=0.33\textwidth]{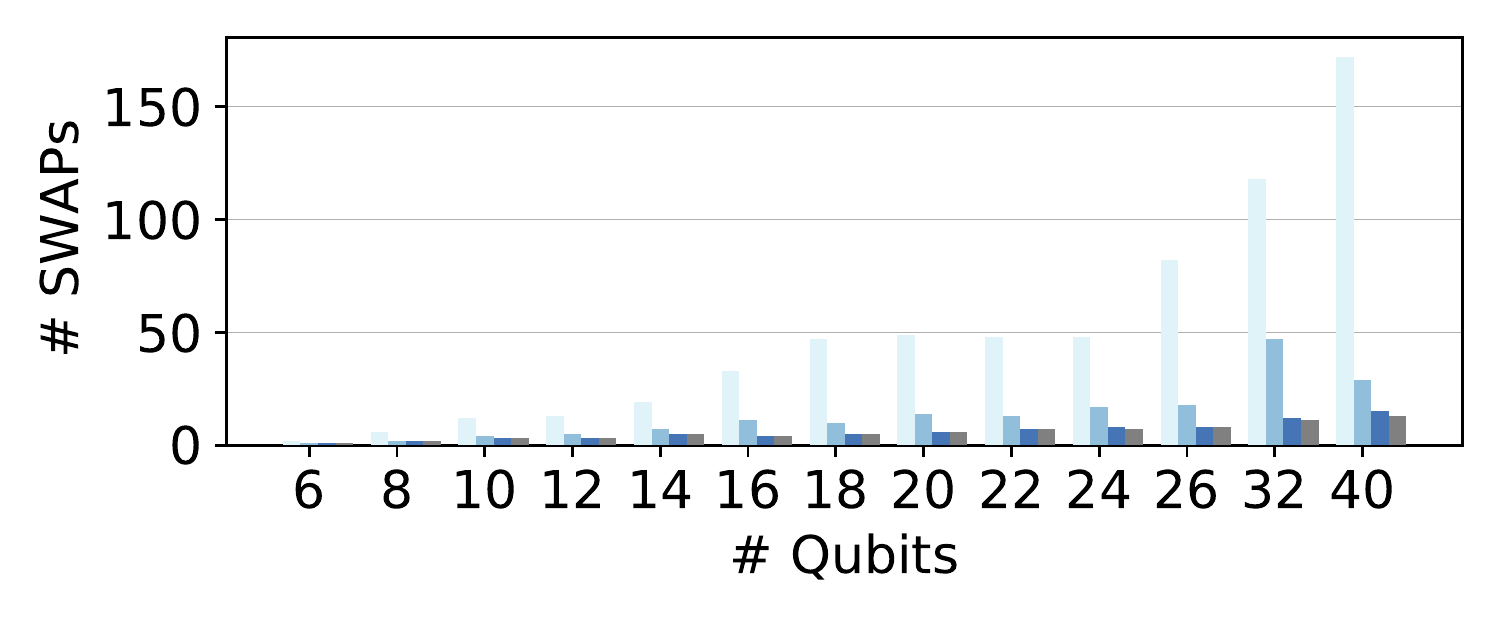}
    \label{fig:grid_ising_swap}
    }
    \subfloat[SYC count for NNN Ising model]{
     \includegraphics[width=0.33\textwidth]{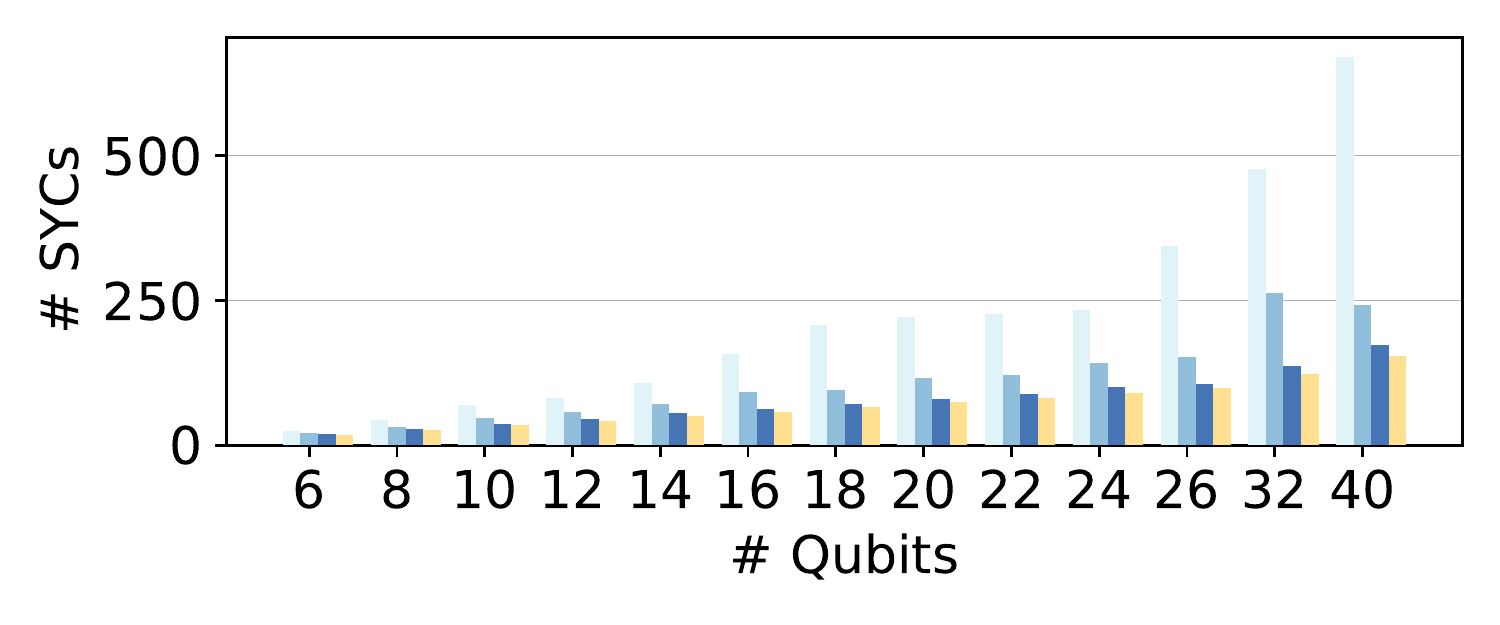}
    \label{fig:grid_ising_sycamore}
    }
    \subfloat[SYC depth for NNN Ising model]{
    \includegraphics[width=0.33\textwidth]{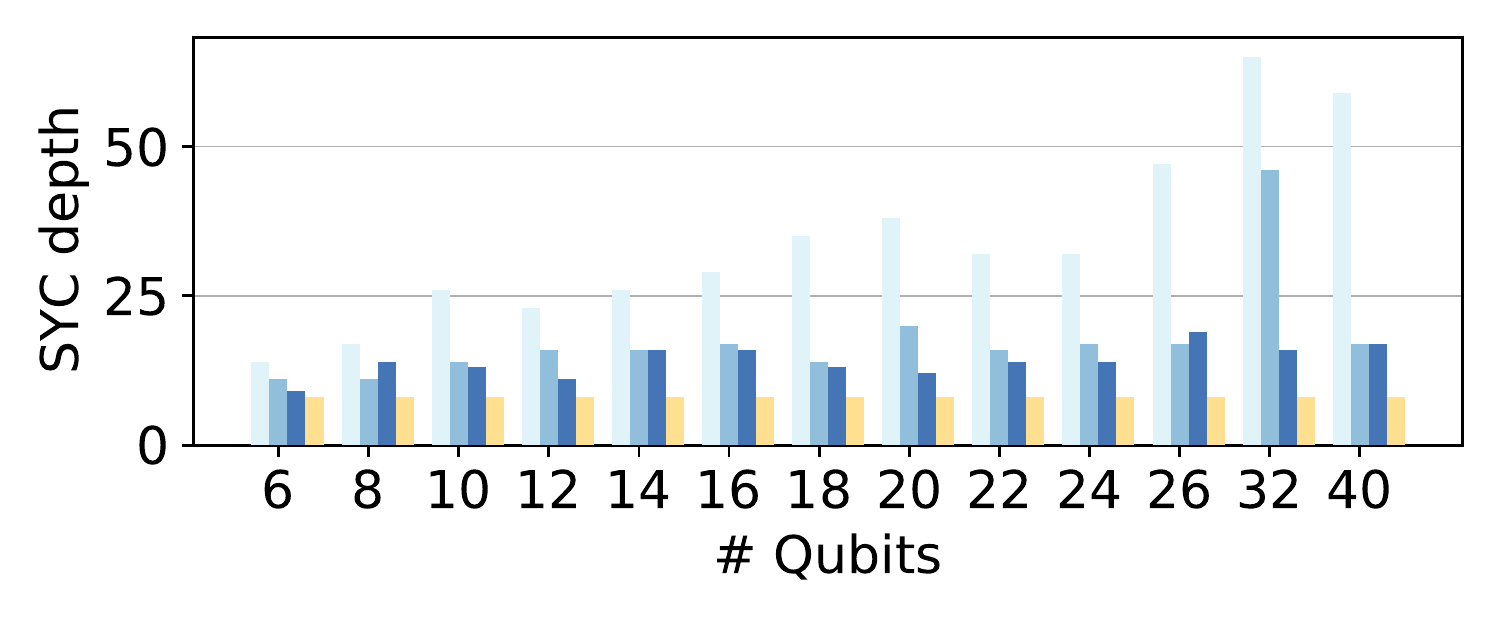}
    \label{fig:grid_ising_sycamore_depth}
    }
    
    \subfloat[SWAP count for QAOA-REG-3]{
     \includegraphics[width=0.33\textwidth]{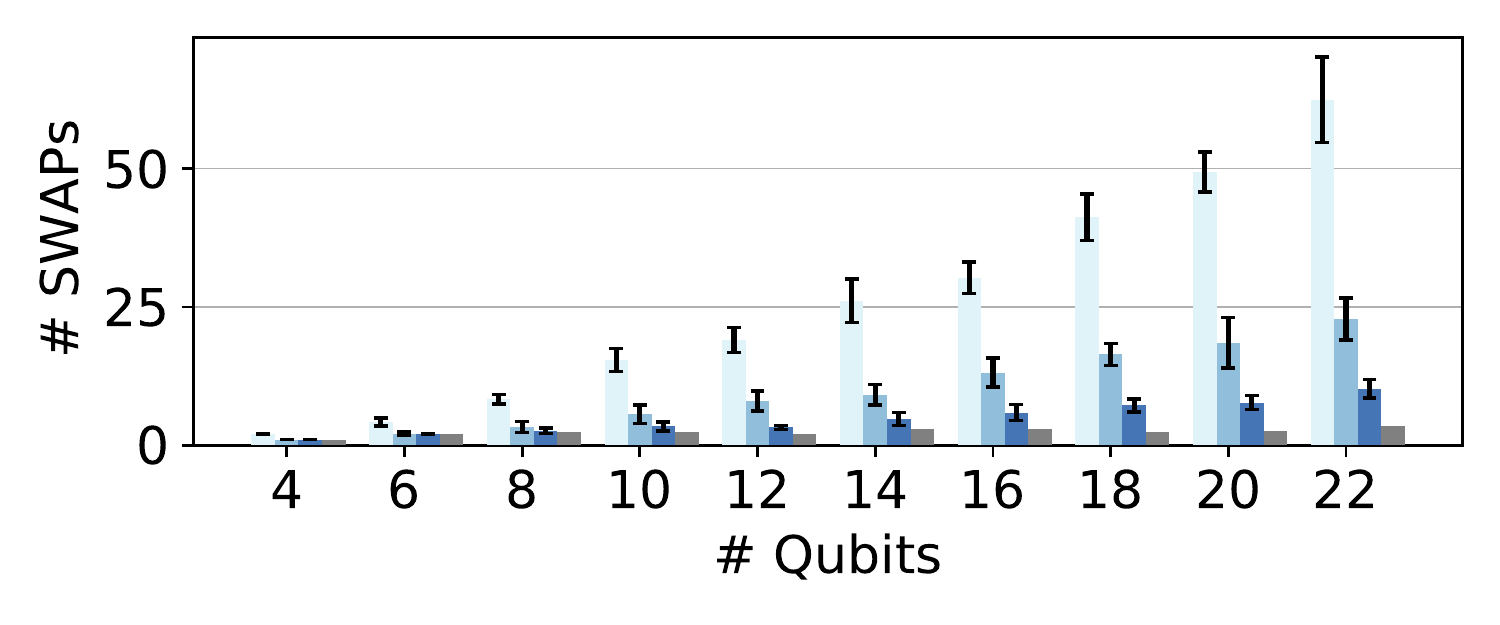}
    \label{fig:grid_qaoa_swap}
    }
    \subfloat[SYC count for QAOA-REG-3]{
     \includegraphics[width=0.33\textwidth]{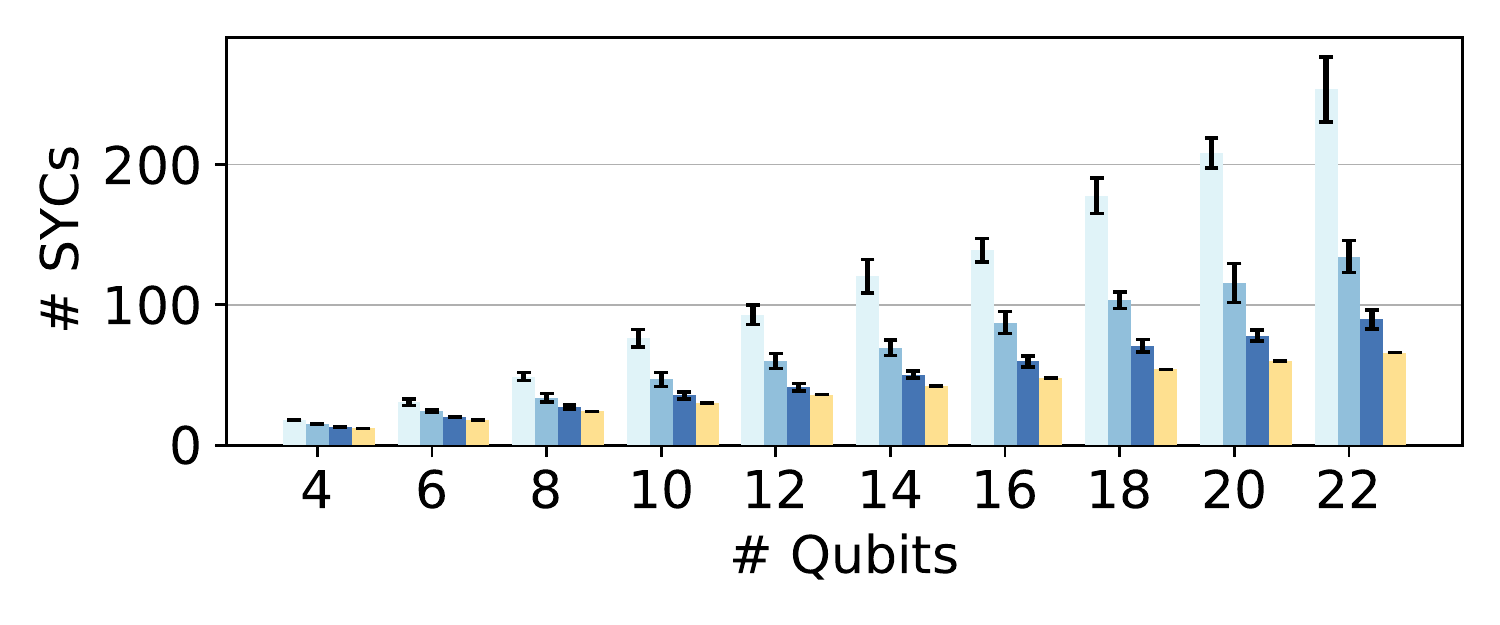}
    \label{fig:grid_qaoa_syc}
    }
    \subfloat[SYC depth for QAOA-REG-3]{
    \includegraphics[width=0.33\textwidth]{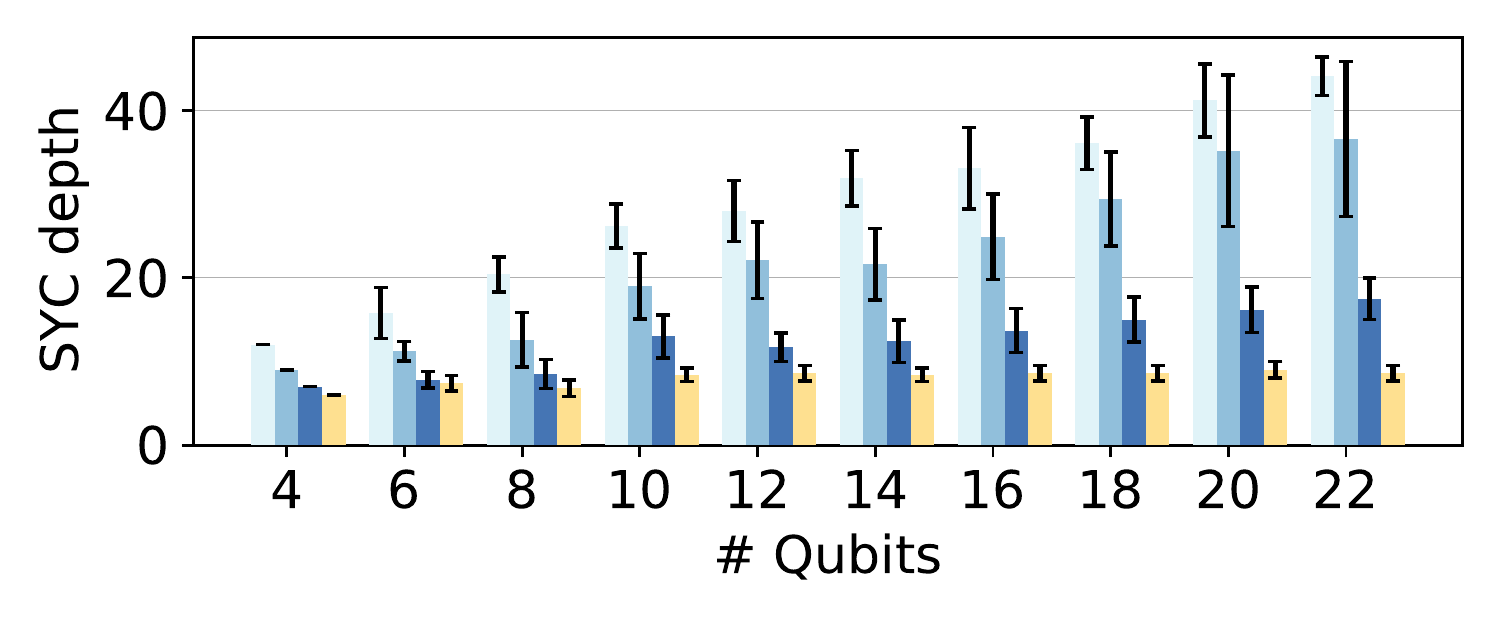}
    \label{fig:grid_qaoa_sycdepth}
    }
\caption{Compilation results of the one-layer NNN Heisenberg model, NNN XY, NNN Ising model, and QAOA-REG-3 on the Google Sycamore device. Each QAOA problem size is averaged over 10 different instances (error bars show the standard deviation) and operator parameters of QAOA circuits were chosen at their theoretically optimal values. `2QAN dressed' shows the number of SWAPs that were merged with circuit gates by \ucl, which helps to reduce hardware gate count. `NoMap' as our baseline, represents the compilation results when assuming all-to-all qubit connectivity. The 2QAN compiler has least compilation overhead (\# SWAPs, \# SYCs, and circuit depth) compared to  \tket \cite{tket}, Qiskit \cite{qiskit}. }
\label{fig:grid_syc}
\end{figure*}

\begin{figure*}[tbh!]
 \centering
    \subfloat[SWAP count for NNN Heisenberg model]{
     \includegraphics[width=0.33\textwidth]{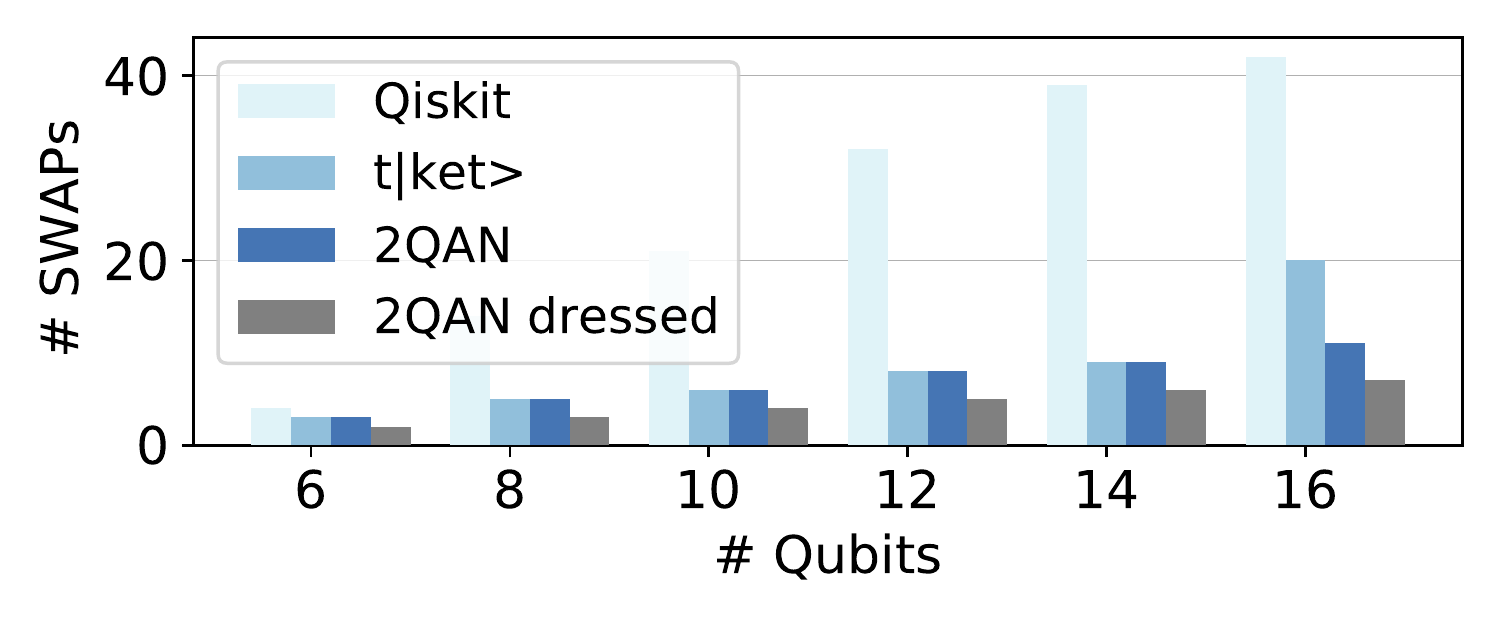}
    \label{fig:aspen_hbxxz_swap}
    }
    \subfloat[iSWAP count for NNN Heisenberg model]{
     \includegraphics[width=0.33\textwidth]{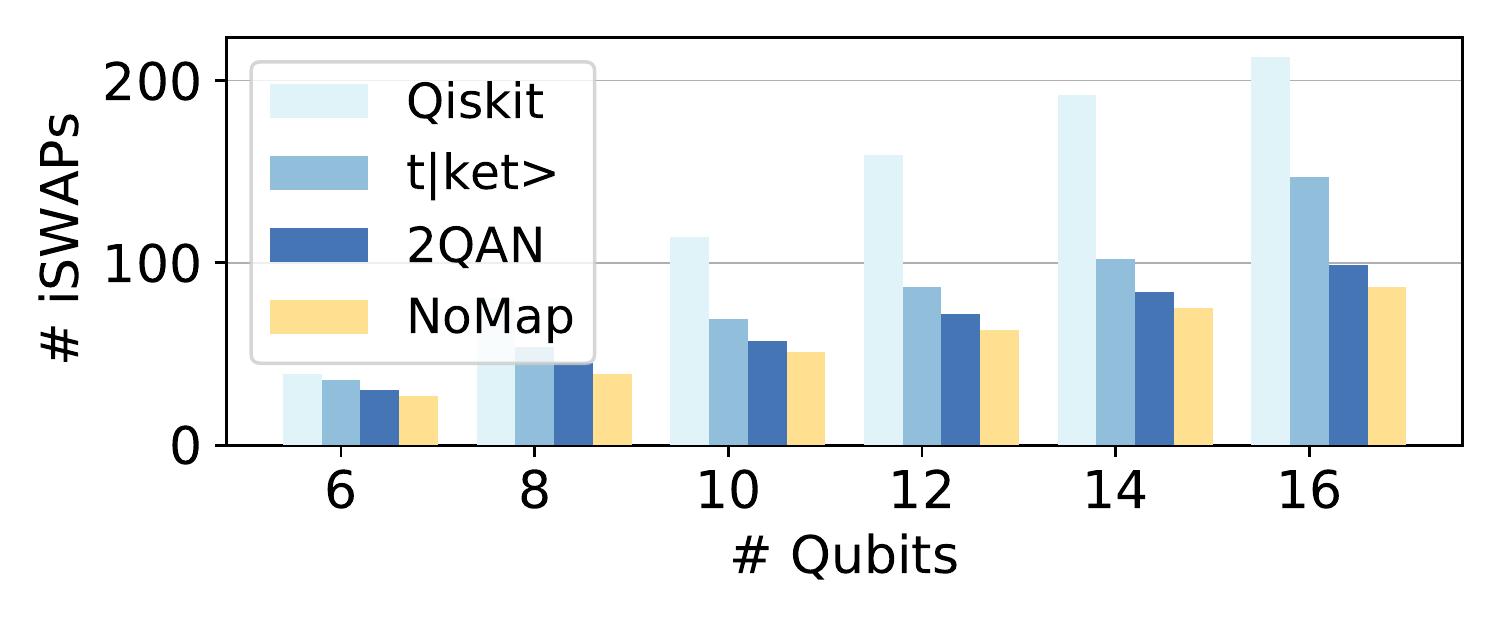}
    \label{fig:aspen_hbxxz_iswap}
    }
    \subfloat[iSWAP depth for NNN Heisenberg model]{
    \includegraphics[width=0.33\textwidth]{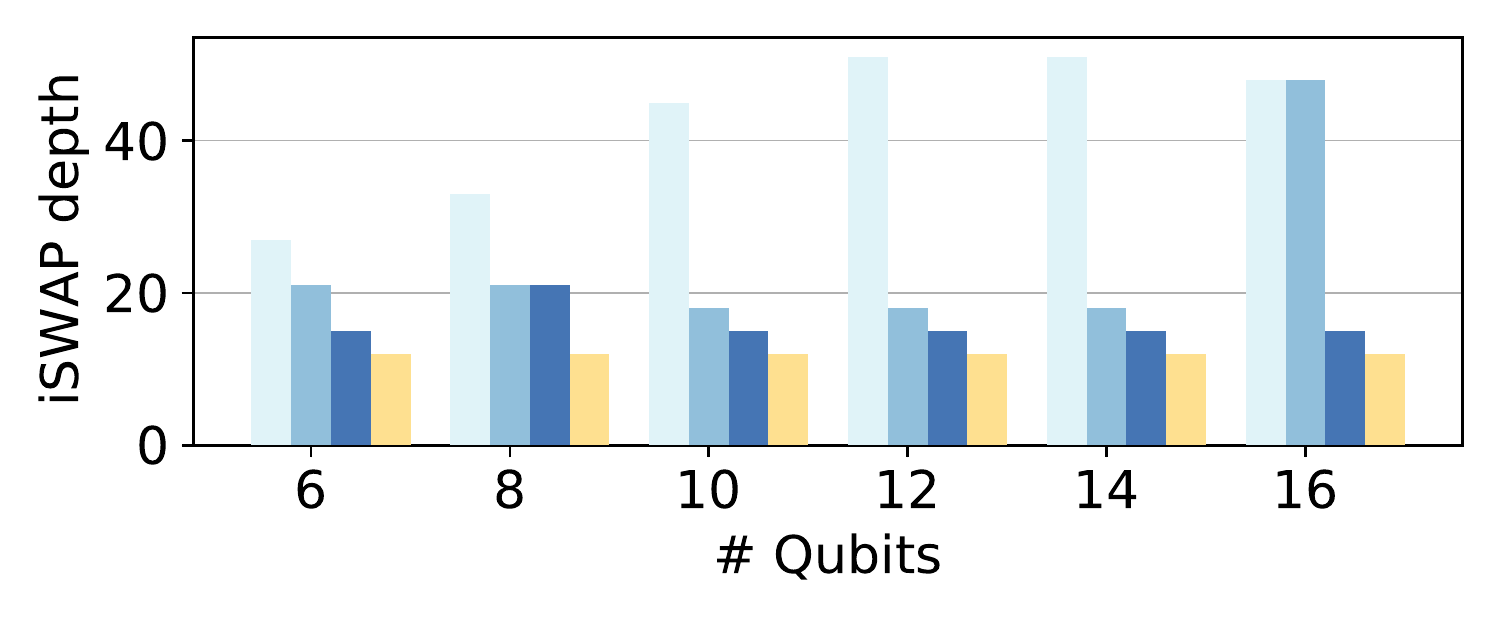}
    \label{fig:aspen_hbxxz_iswap_depth}
    }
    
    \subfloat[SWAP count for NNN XY model]{
     \includegraphics[width=0.33\textwidth]{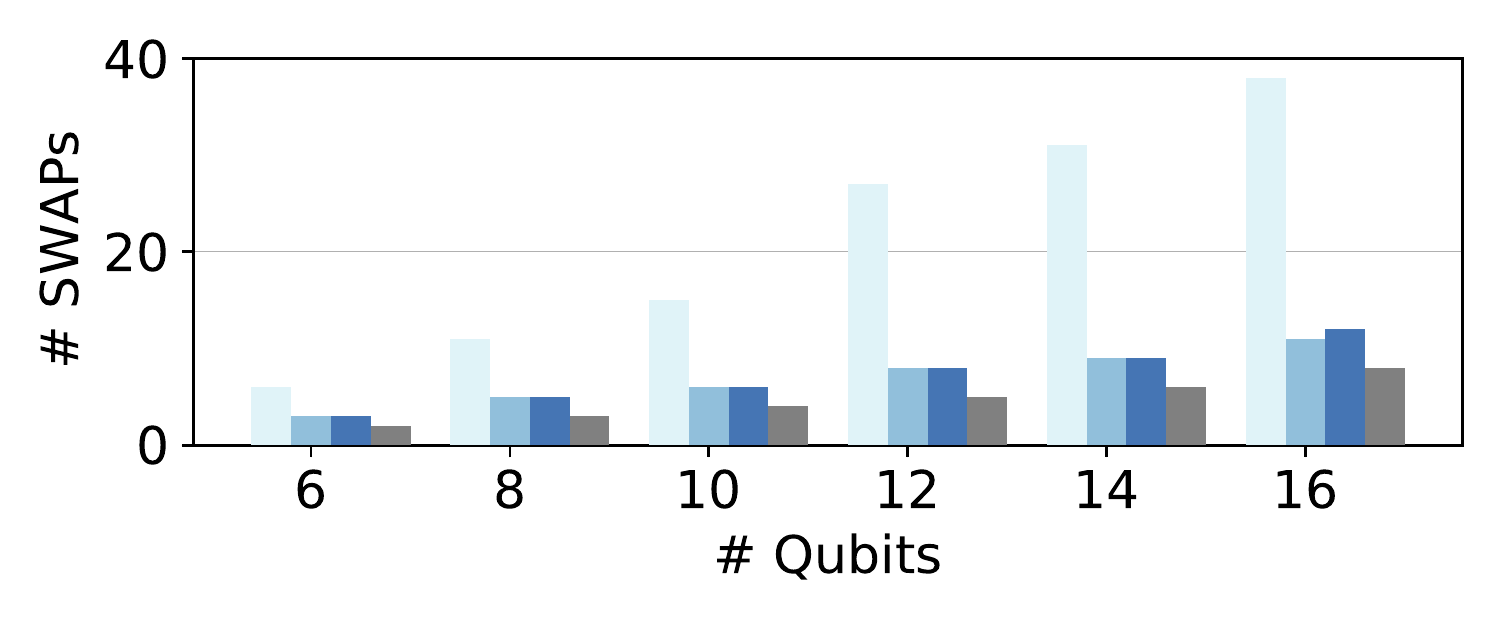}
    \label{fig:aspen_xy_swap}
    }
    \subfloat[iSWAP count for NNN XY model]{
     \includegraphics[width=0.33\textwidth]{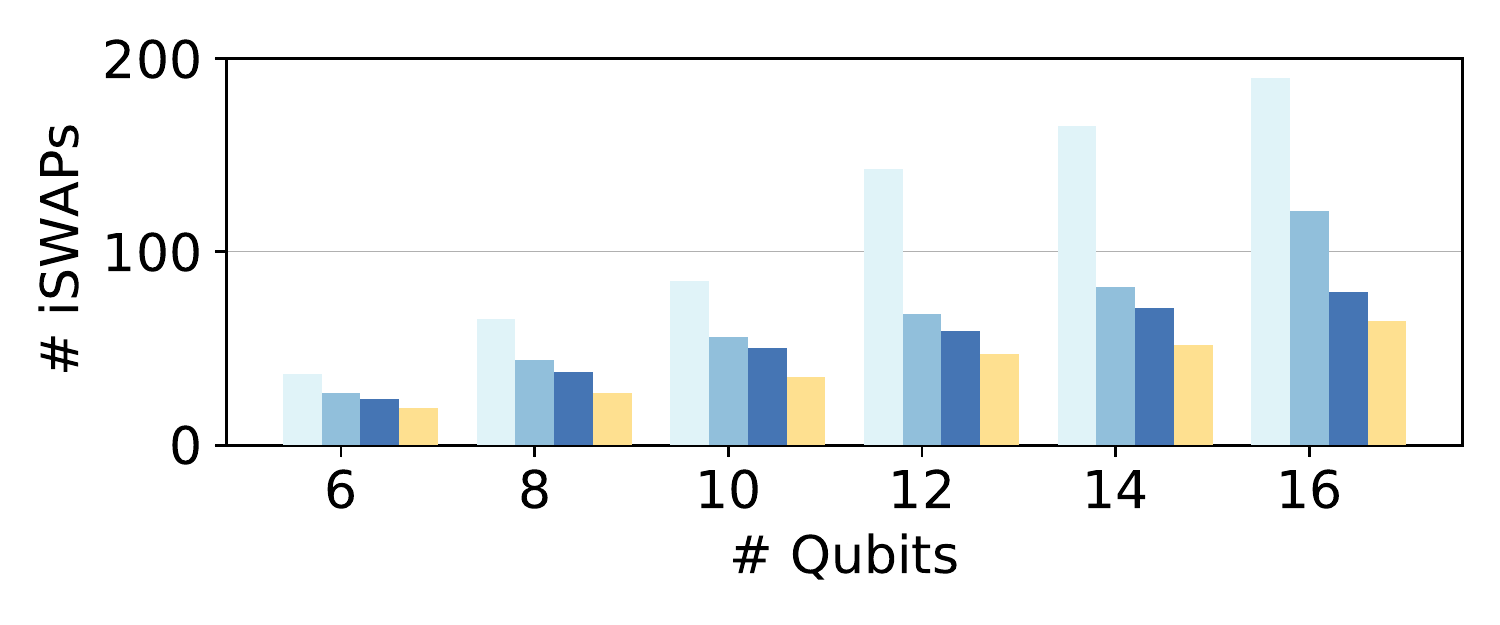}
    \label{fig:aspen_xy_iswap}
    }
    \subfloat[iSWAP depth for NNN XY model]{
    \includegraphics[width=0.33\textwidth]{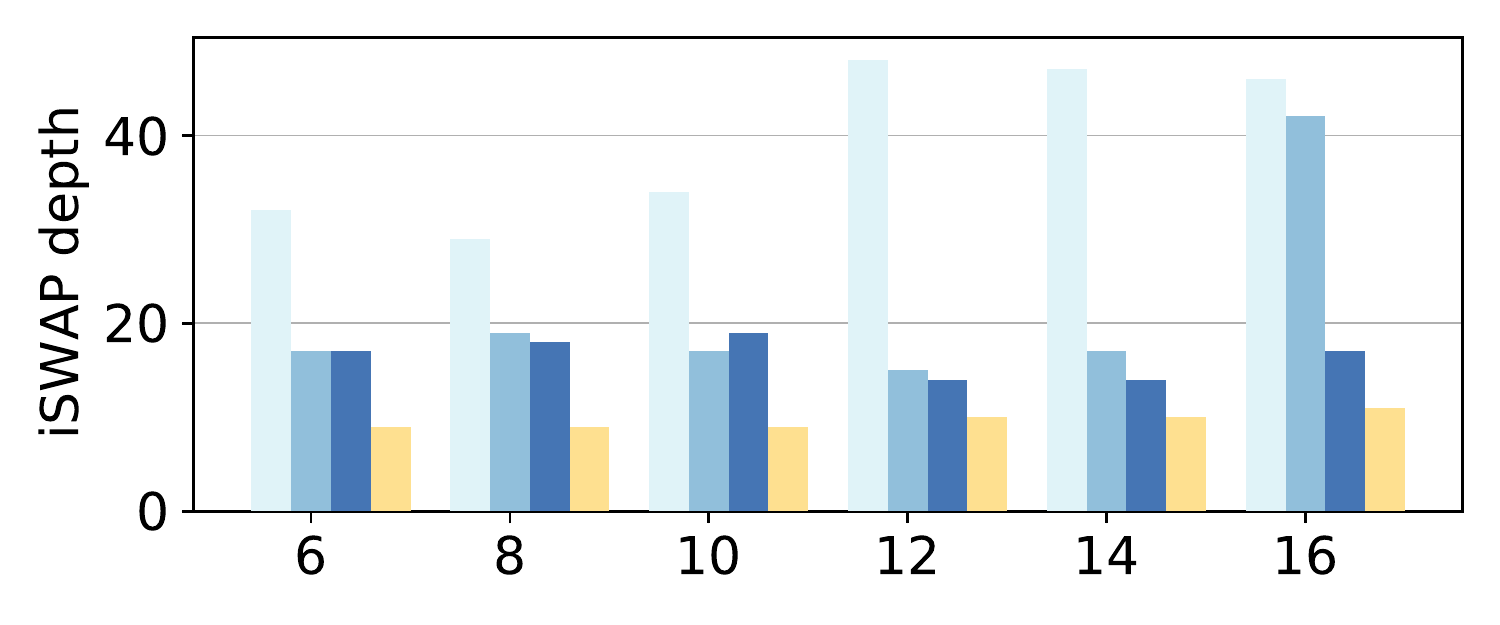}
    \label{fig:aspen_xy_iswap_depth}
    }
    
    \subfloat[SWAP count for NNN Ising model]{
     \includegraphics[width=0.33\textwidth]{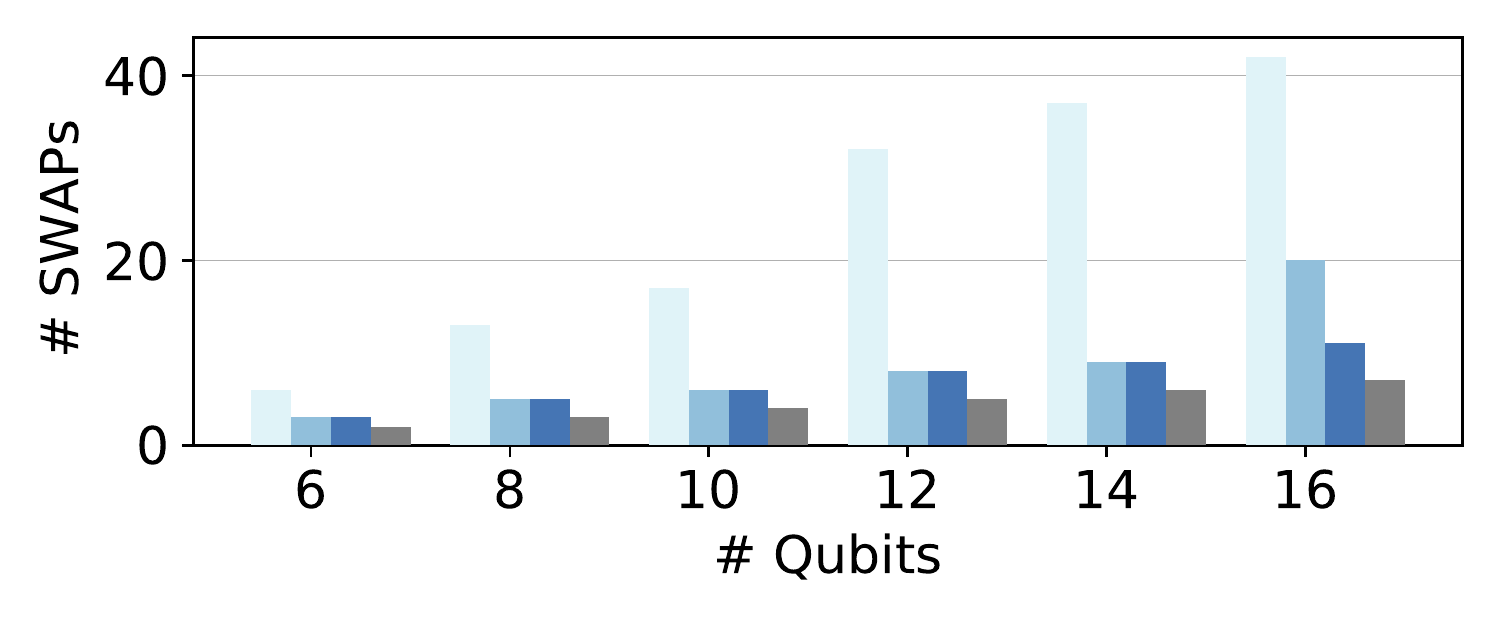}
    \label{fig:aspen_ising_swap}
    }
    \subfloat[iSWAP count for NNN Ising model]{
     \includegraphics[width=0.33\textwidth]{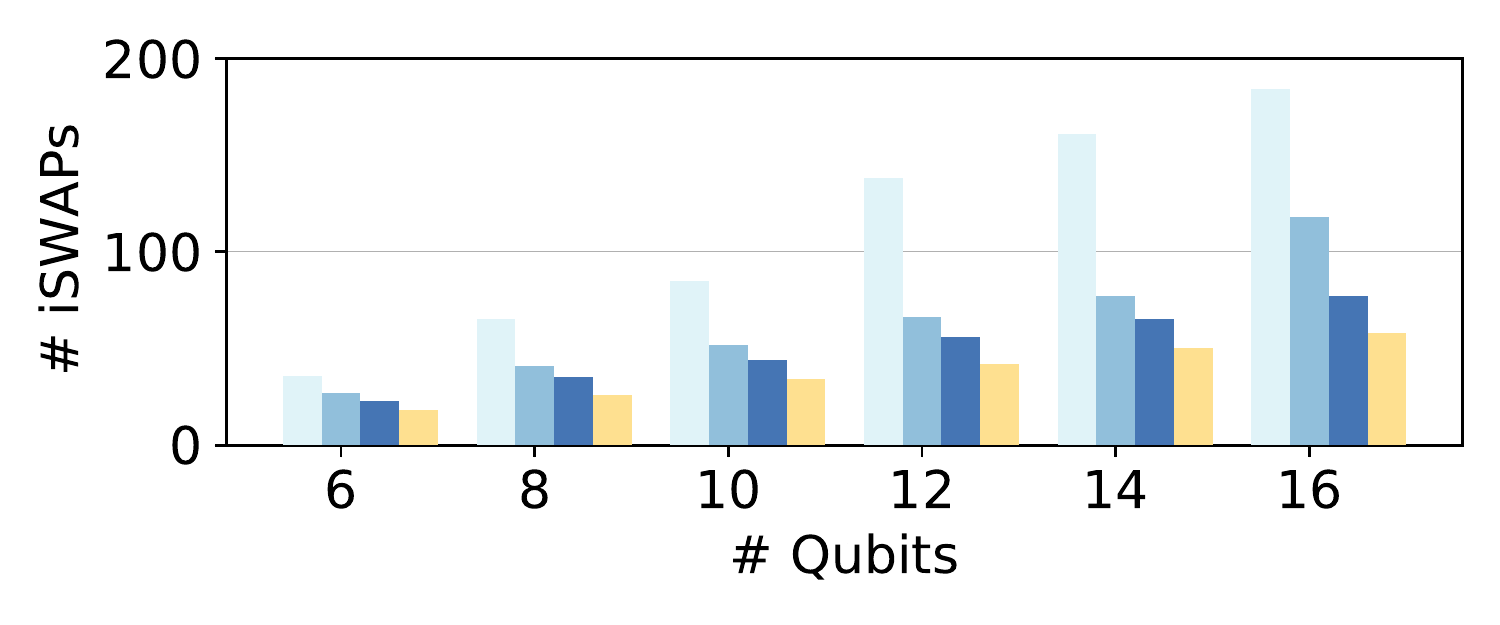}
    \label{fig:aspen_ising_iswap}
    }
    \subfloat[iSWAP depth for NNN Ising model]{
    \includegraphics[width=0.33\textwidth]{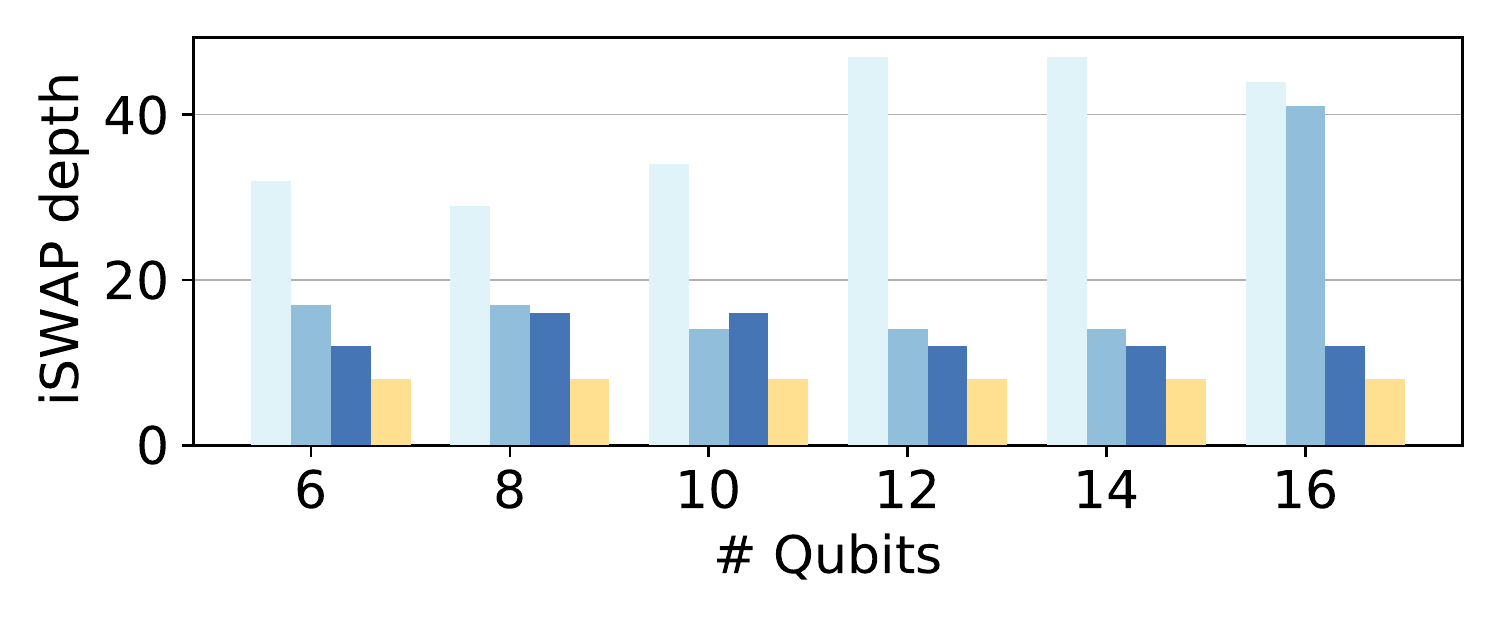}
    \label{fig:aspen_ising_iswap_depth}
    }
    
    \subfloat[SWAP count for QAOA-REG-3]{
     \includegraphics[width=0.33\textwidth]{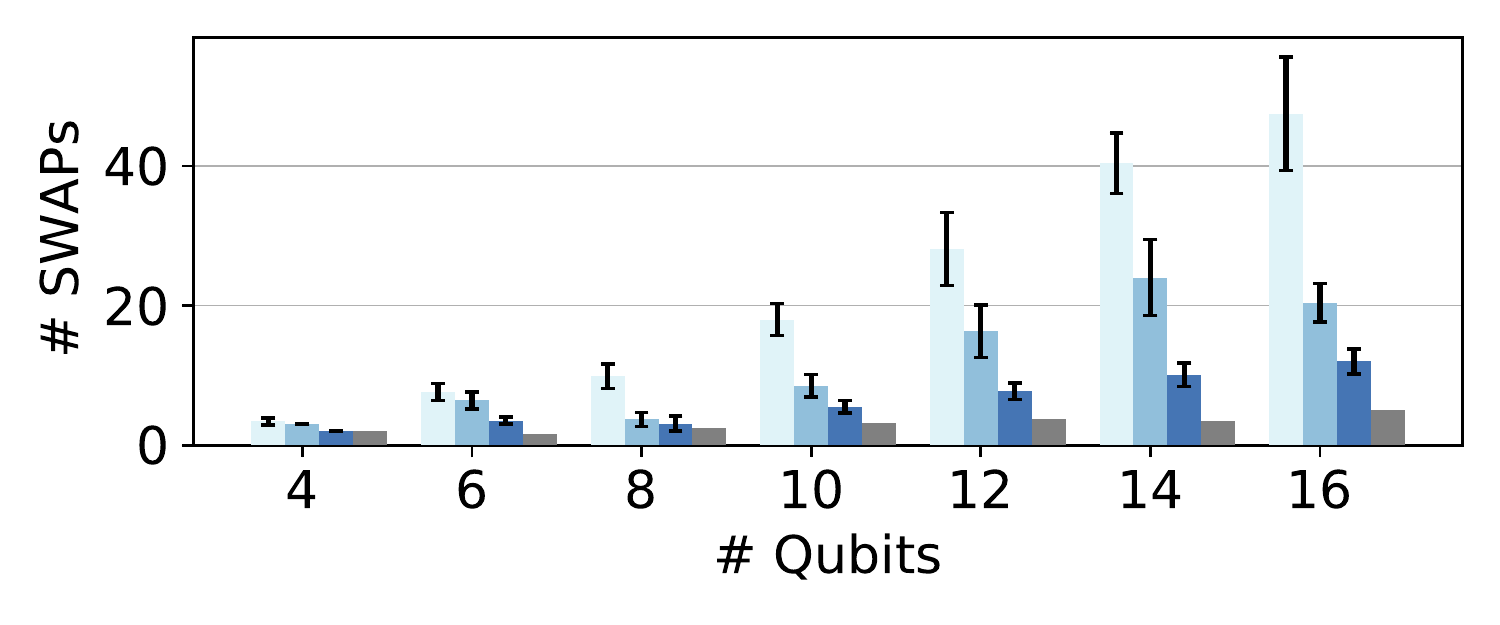}
    \label{fig:aspen_qaoa_swap}
    }
    \subfloat[iSWAP count for QAOA-REG-3]{
     \includegraphics[width=0.33\textwidth]{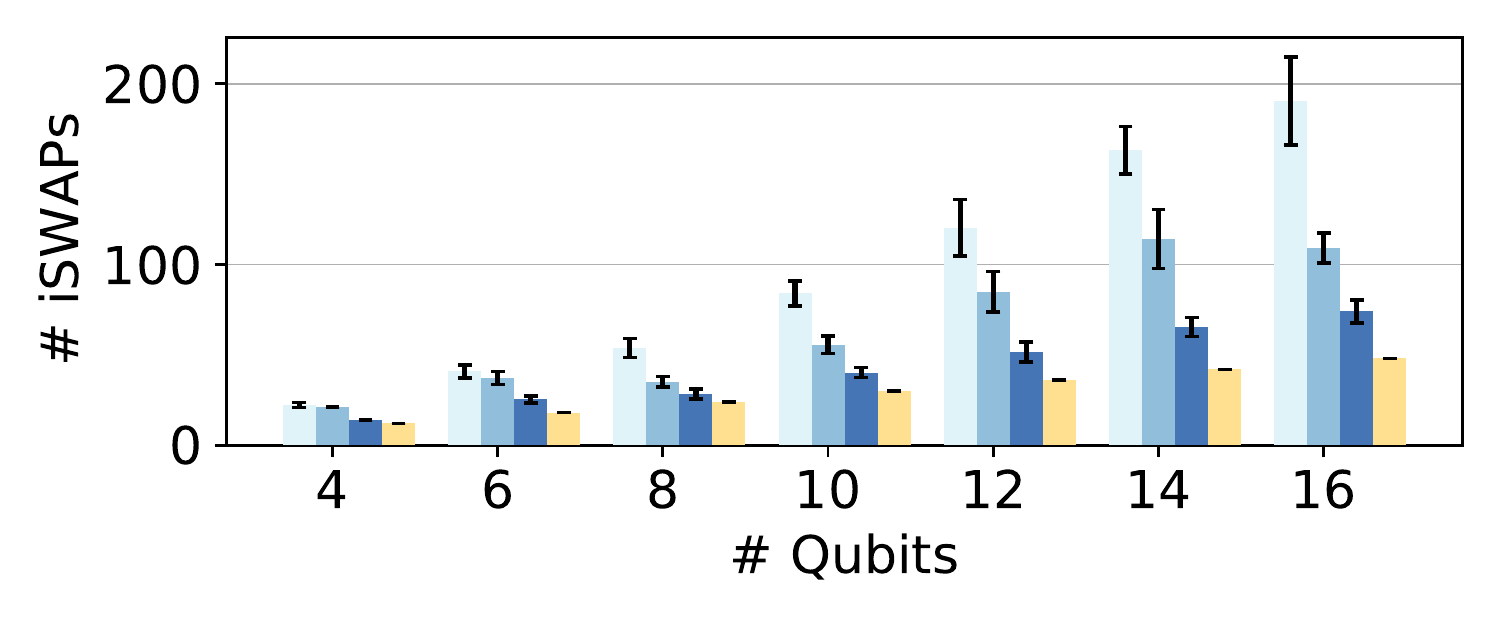}
    \label{fig:aspen_qaoa_iswap}
    }
    \subfloat[iSWAP depth for QAOA-REG-3]{
    \includegraphics[width=0.33\textwidth]{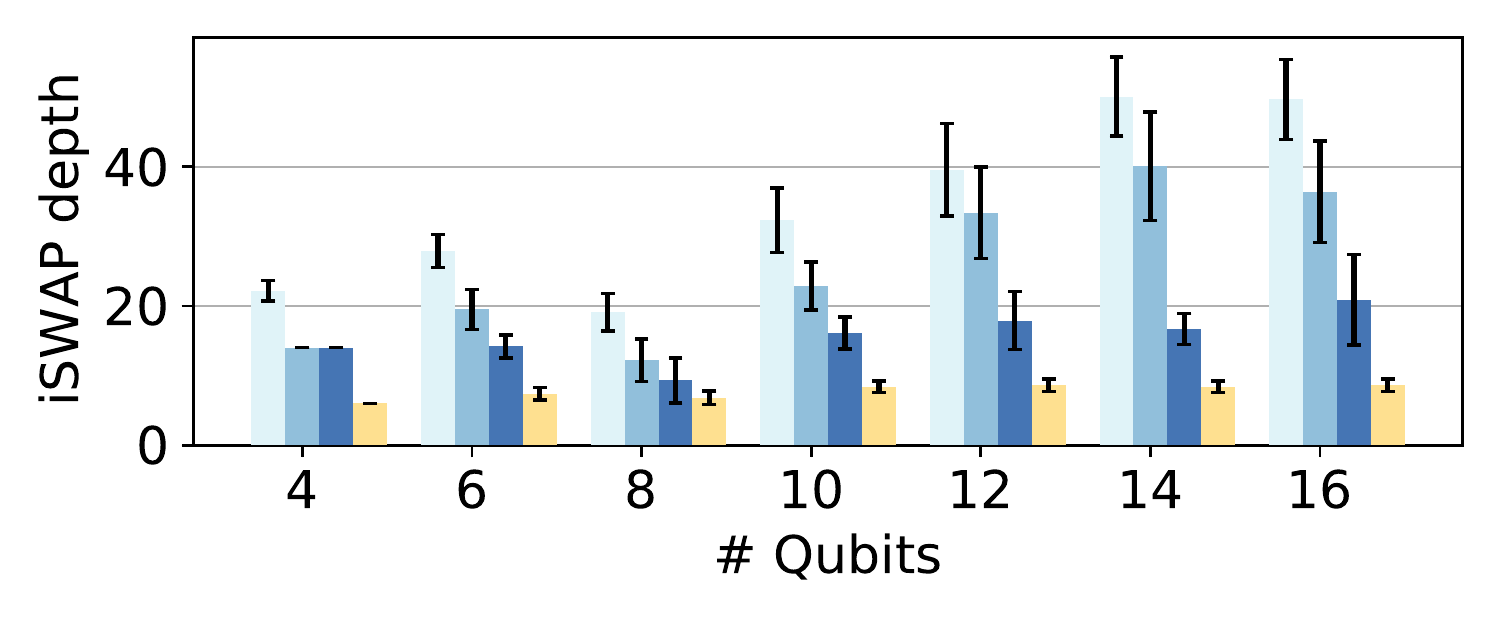}
    \label{fig:aspen_qaoa_iswap_depth}
    }
\caption{Compilation results of the one-layer NNN Heisenberg model, NNN XY, NNN Ising model, and QAOA-REG-3 on the Rigetti Aspen device. The 2QAN compiler has least compilation overhead 
(\# SWAPs, \# iSWAPs, and circuit depth) 
compared to  \tket \cite{tket}, Qiskit \cite{qiskit}.  }
\label{fig:aspen_iswap}
\end{figure*}

\begin{figure*}[tbh!]
 \centering
    \subfloat[SWAP count for NNN Heisenberg model]{
     \includegraphics[width=0.33\textwidth]{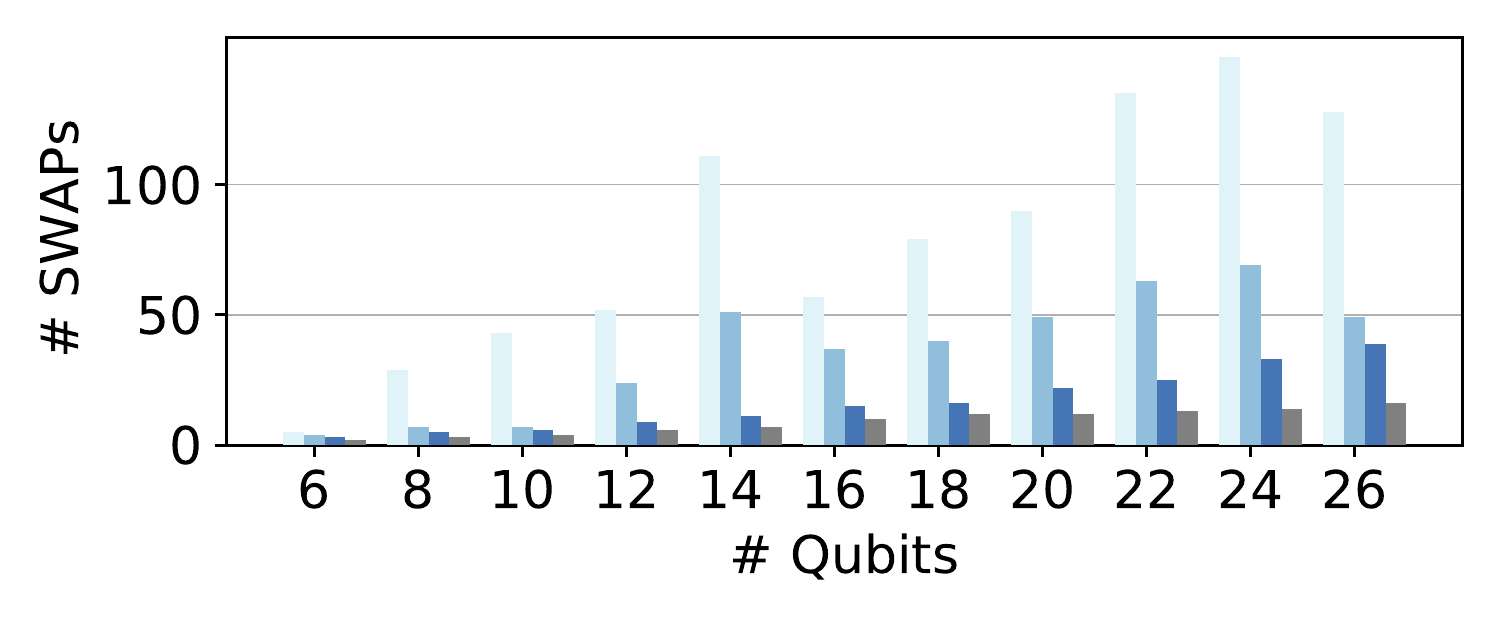}
    \label{fig:montreal_hbxxz_swap}
    }
    \subfloat[CNOT count for NNN Heisenberg model]{
     \includegraphics[width=0.33\textwidth]{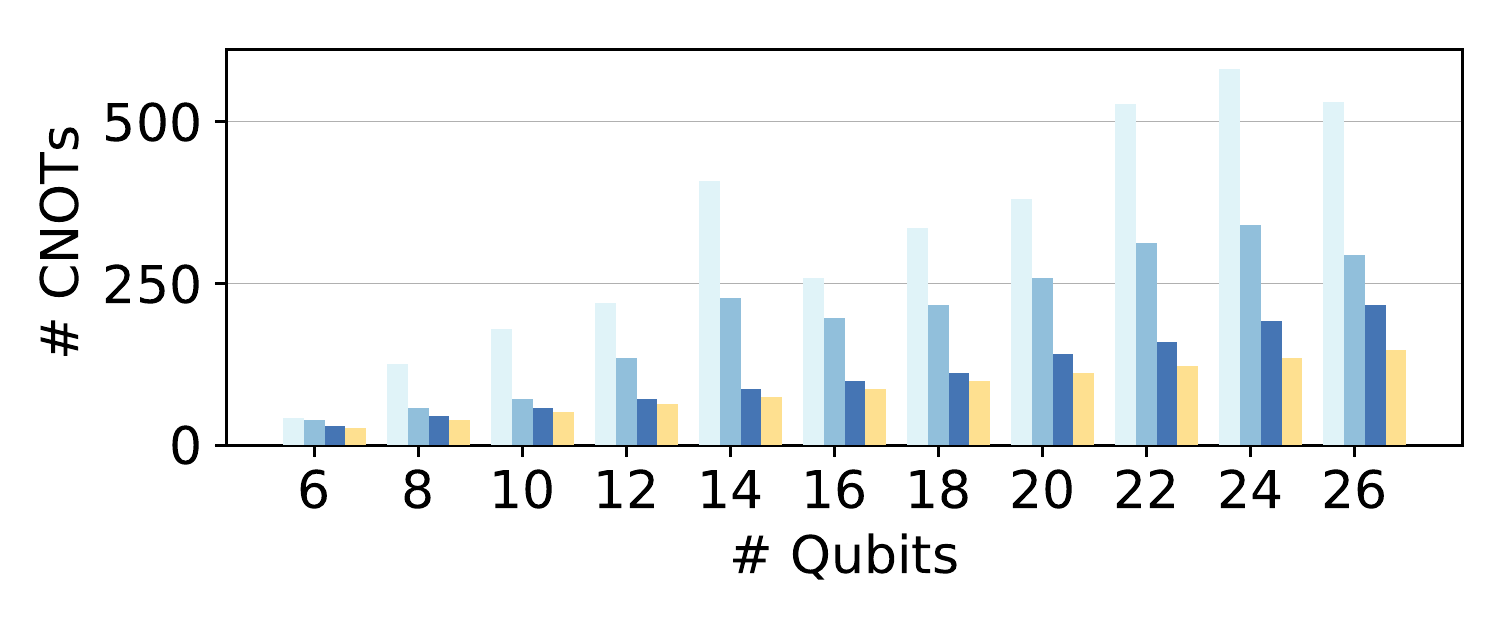}
    \label{fig:montreal_hbxxz_cx}
    }
    \subfloat[CNOT depth for NNN Heisenberg model]{
    \includegraphics[width=0.33\textwidth]{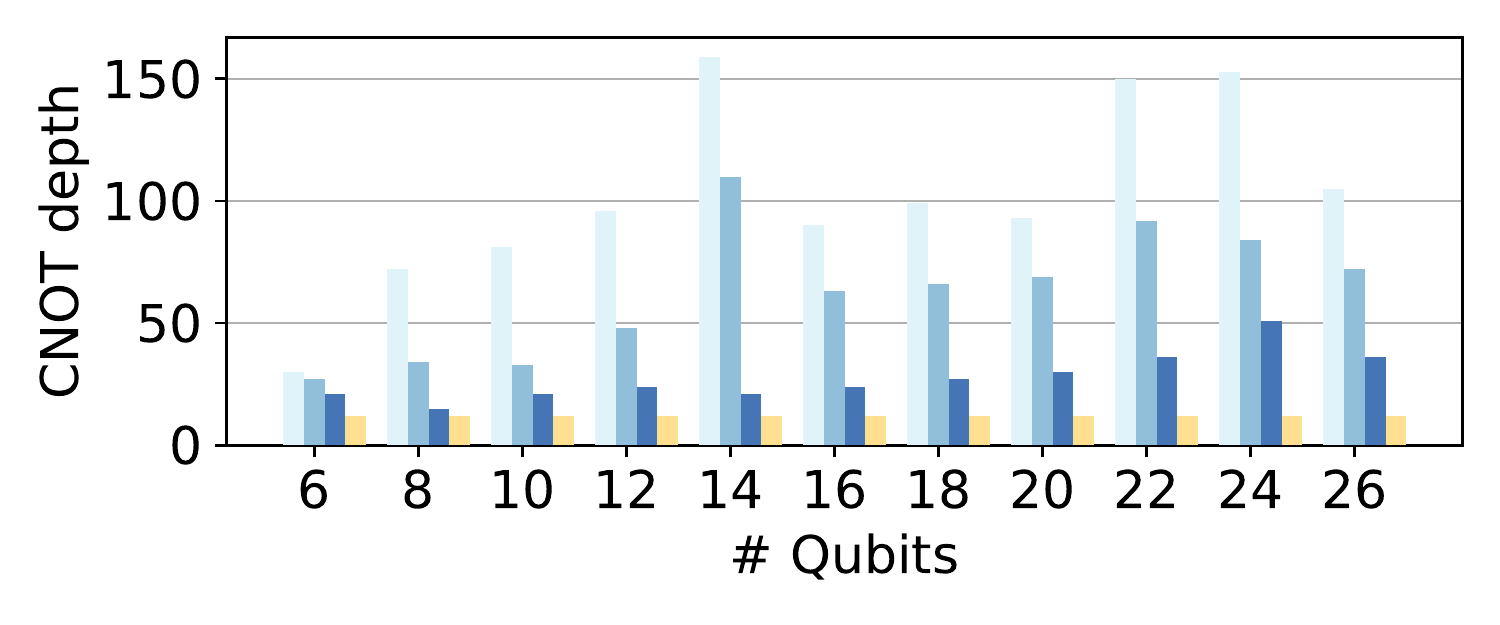}
    \label{fig:montreal_hbxxz_cx_depth}
    }
    
    \subfloat[SWAP count for NNN XY model]{
     \includegraphics[width=0.33\textwidth]{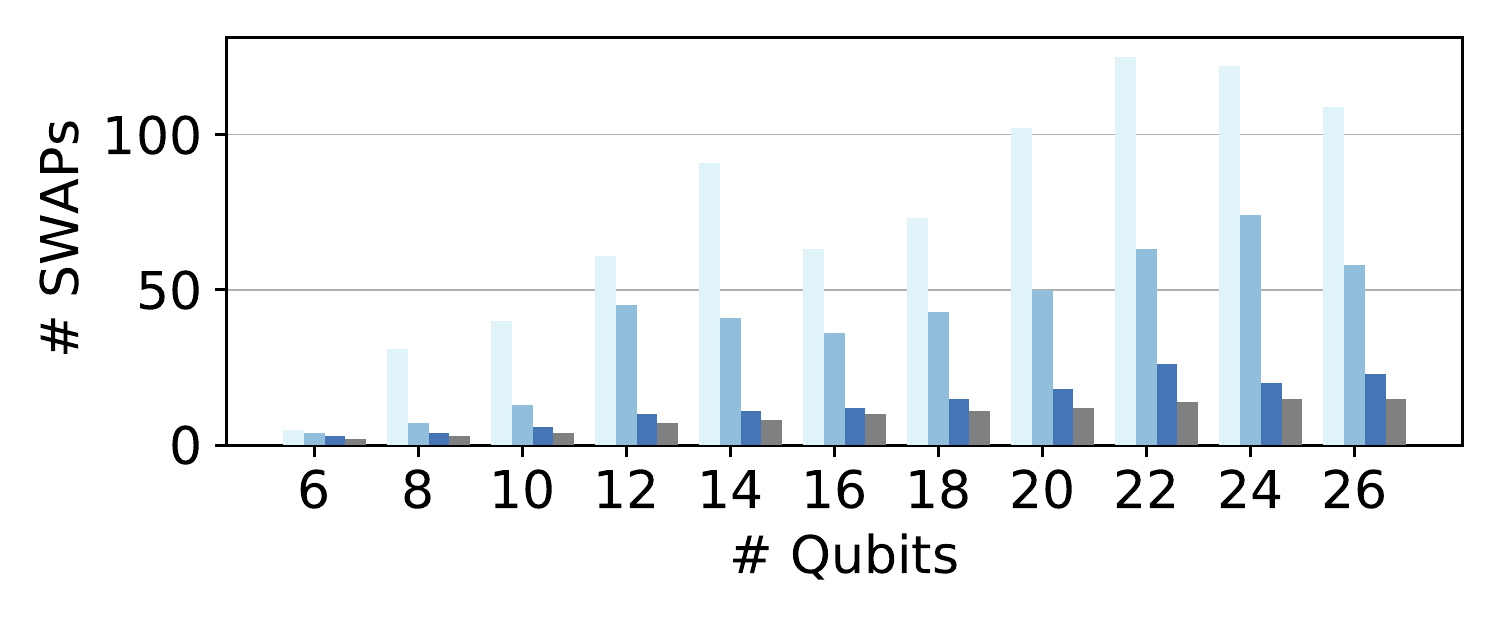}
    \label{fig:montreal_xy_swap}
    }
    \subfloat[CNOT count for NNN XY model]{
     \includegraphics[width=0.33\textwidth]{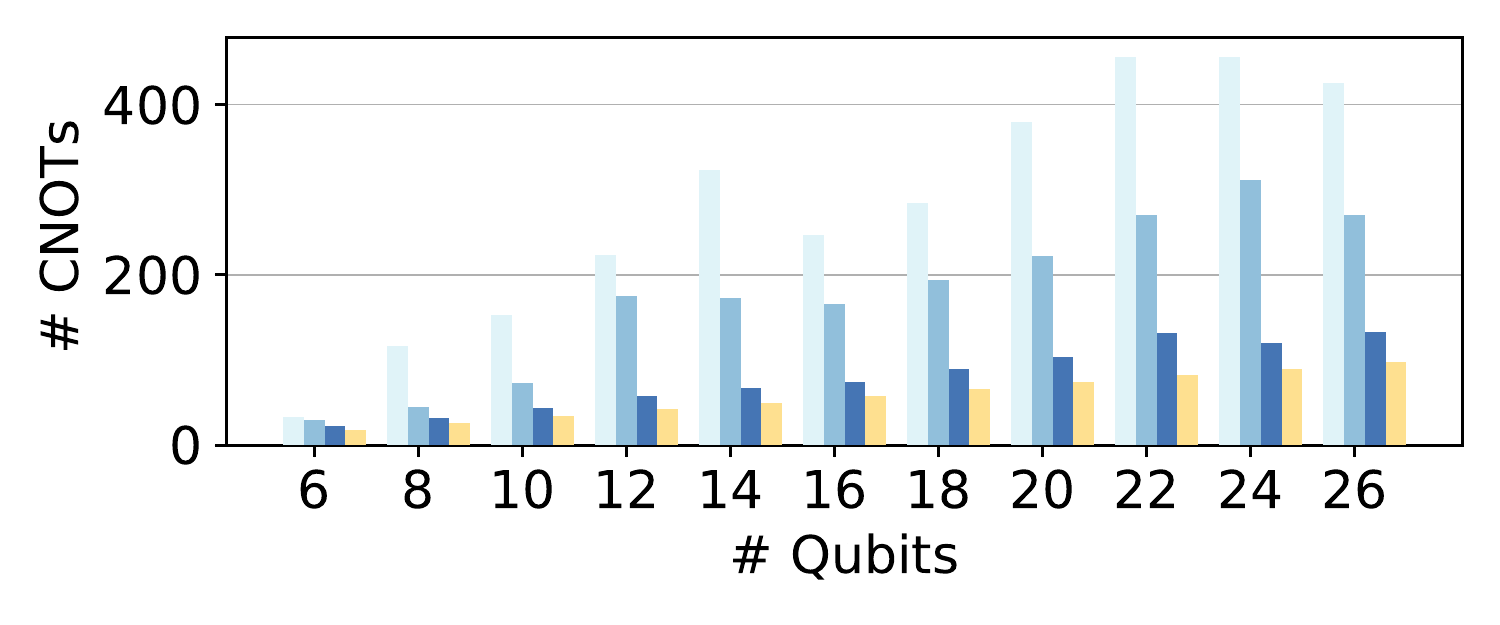}
    \label{fig:montreal_xy_cx}
    }
    \subfloat[CNOT depth for NNN XY model]{
    \includegraphics[width=0.33\textwidth]{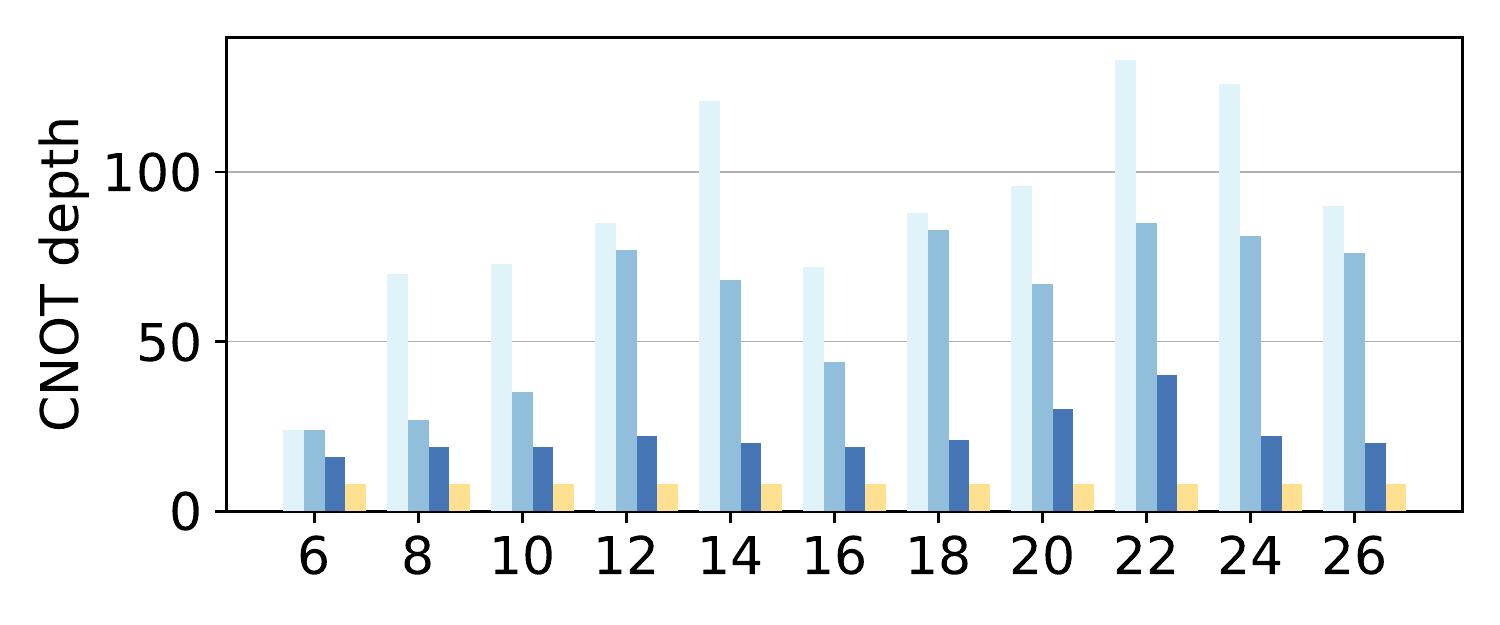}
    \label{fig:montreal_xy_cx_depth}
    }
    
    \subfloat[SWAP count for NNN Ising model]{
     \includegraphics[width=0.33\textwidth]{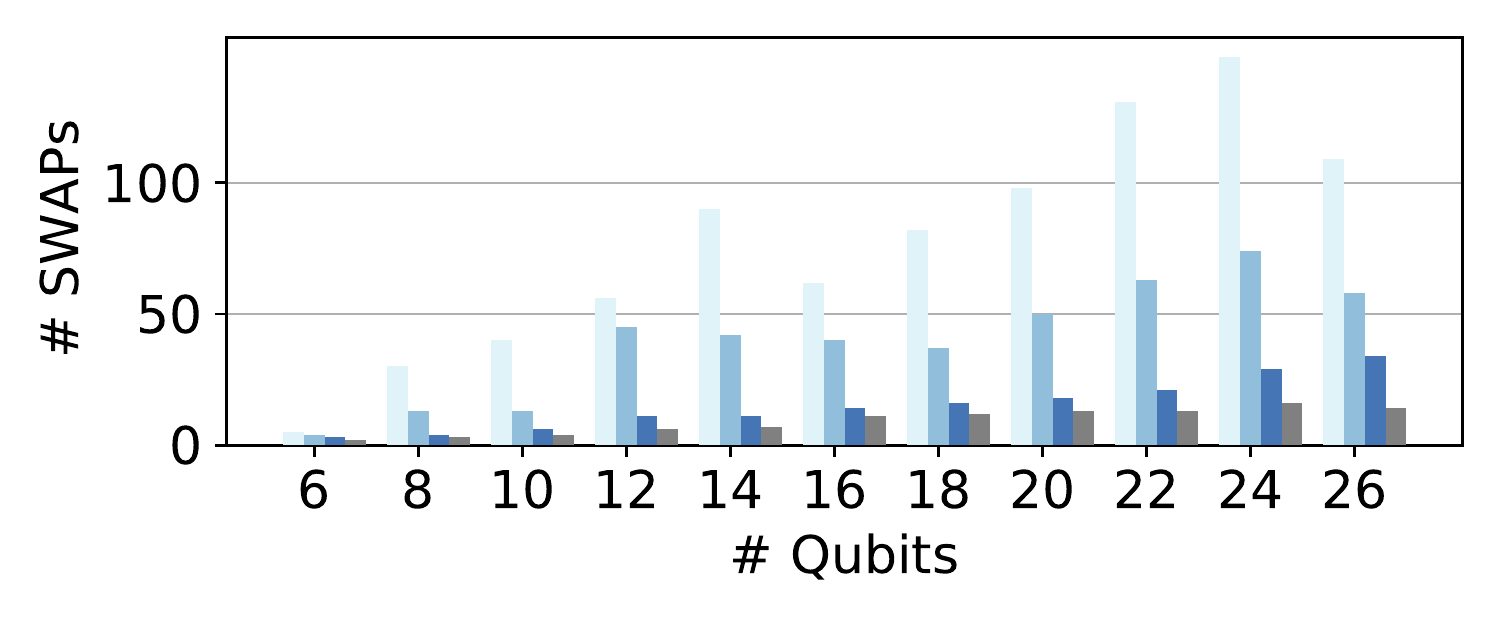}
    \label{fig:montreal_ising_swap}
    }
    \subfloat[CNOT count for NNN Ising model]{
     \includegraphics[width=0.33\textwidth]{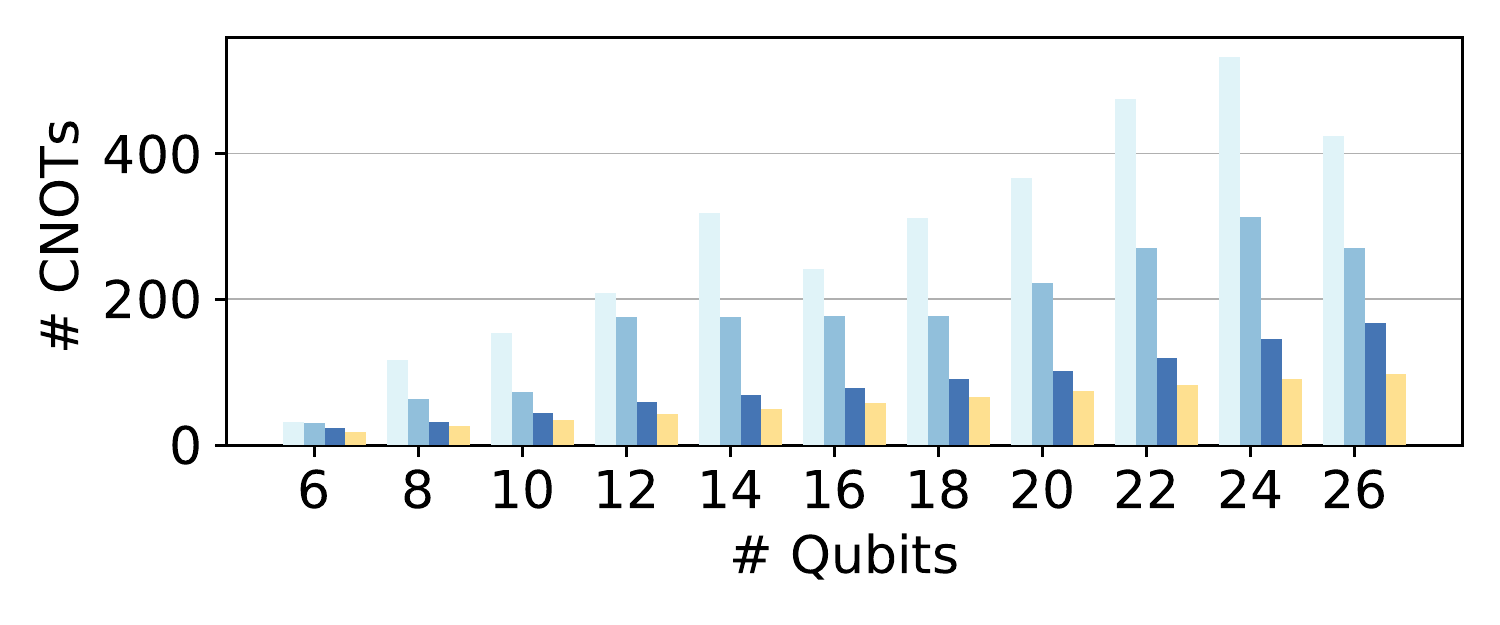}
    \label{fig:montreal_ising_cx}
    }
    \subfloat[CNOT depth for NNN Ising model]{
    \includegraphics[width=0.33\textwidth]{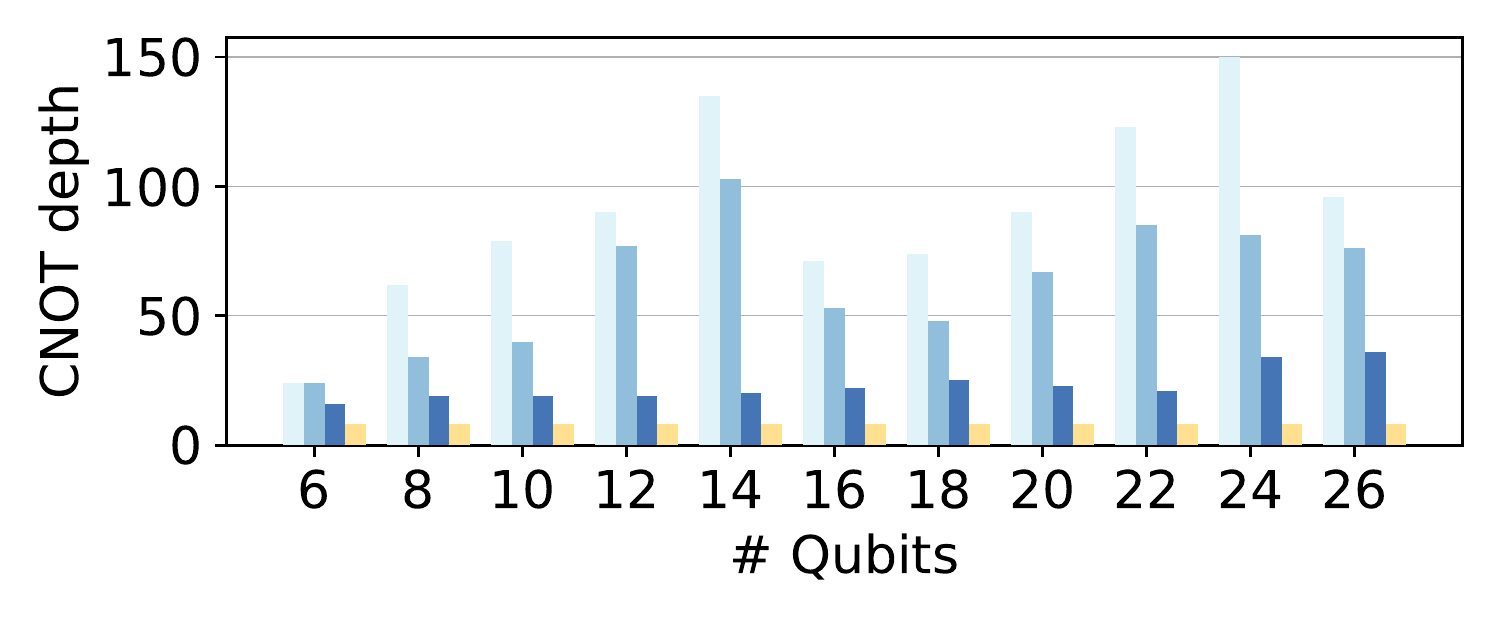}
    \label{fig:montreal_ising_cx_depth}
    }
    
    \subfloat[SWAP count for QAOA-REG-3]{
     \includegraphics[width=0.33\textwidth]{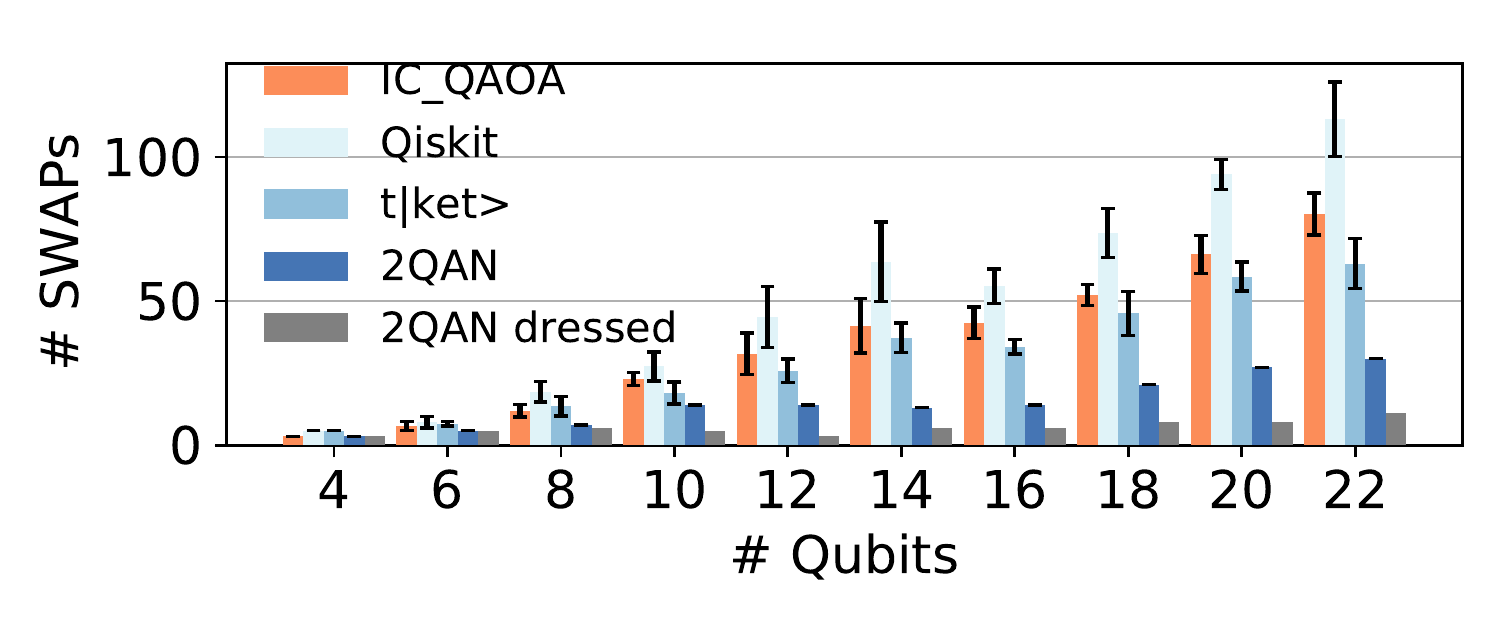}
    \label{fig:montreal_qaoa_swap}
    }
    \subfloat[CNOT count for QAOA-REG-3]{
     \includegraphics[width=0.33\textwidth]{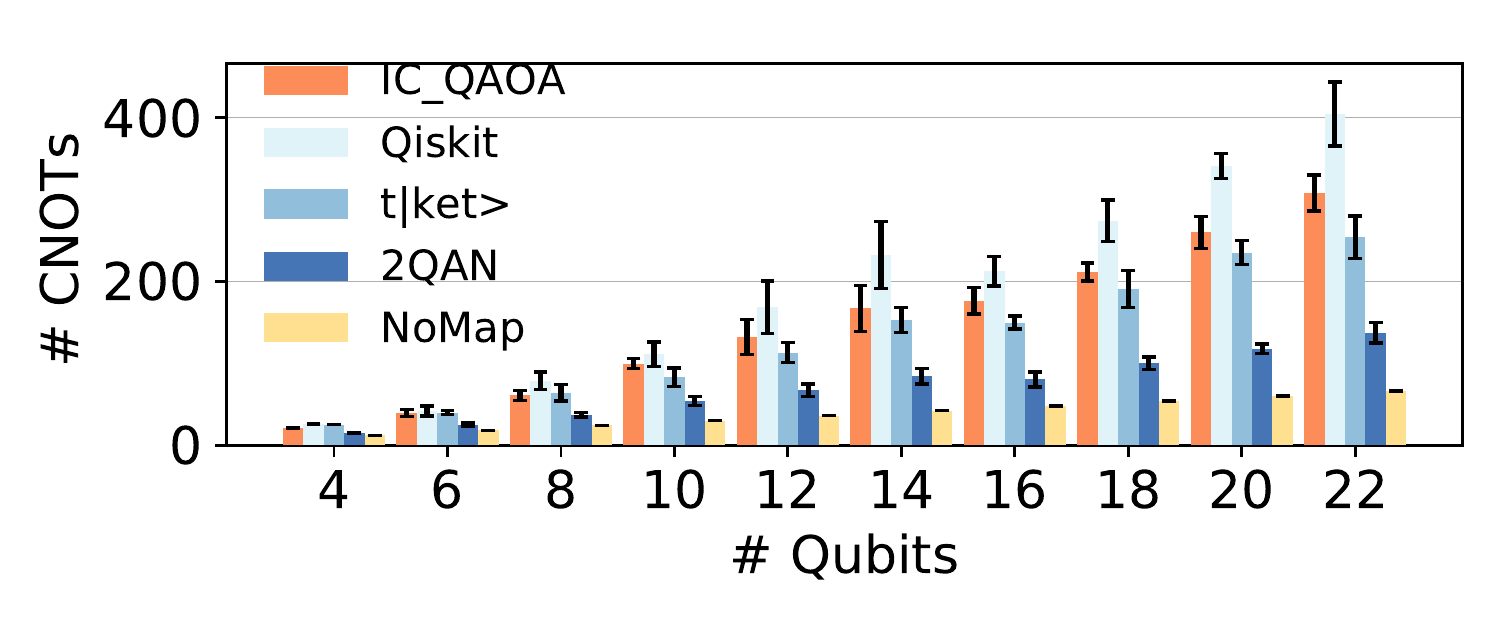}
    \label{fig:montreal_qaoa_cx}
    }
    \subfloat[CNOT depth for QAOA-REG-3]{
    \includegraphics[width=0.33\textwidth]{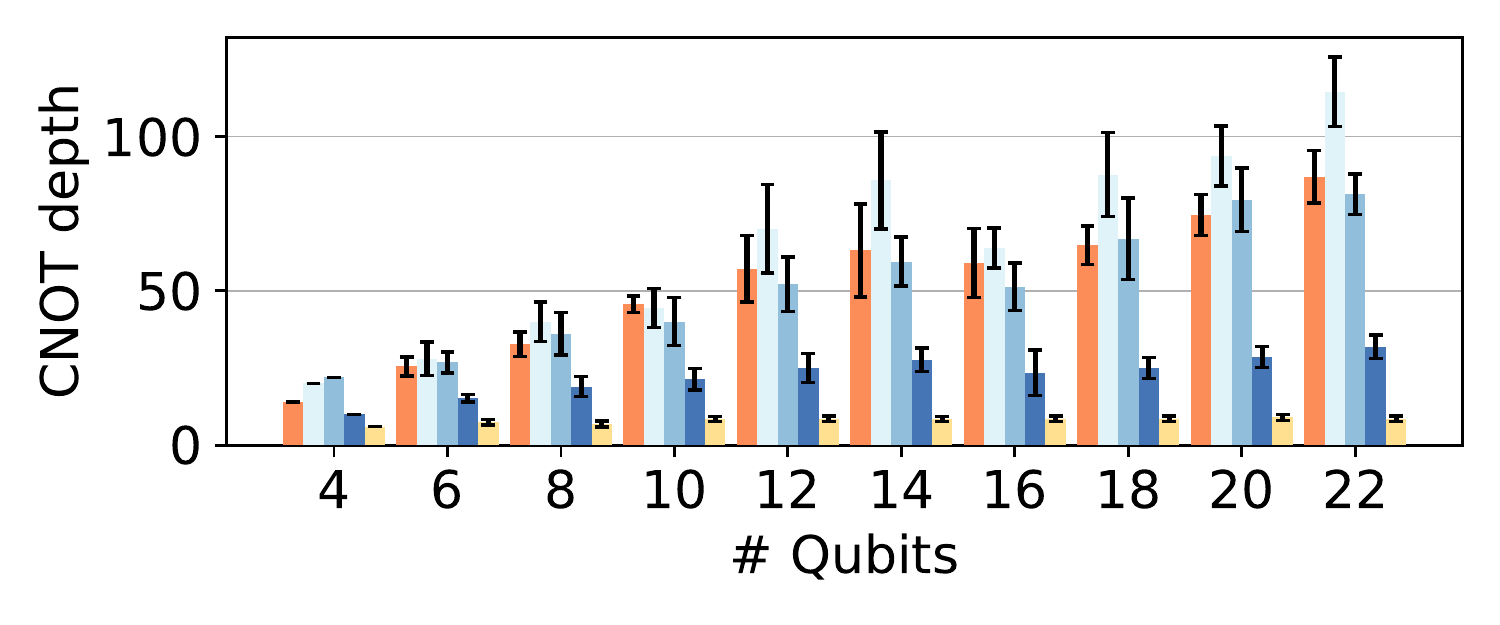}
    \label{fig:montreal_qaoa_cx_depth}
    }
\caption{Compilation results of the one-layer NNN Heisenberg model, NNN XY, NNN Ising model, and QAOA-REG-3 on the IBMQ Montreal device. The 2QAN compiler has least compilation overhead 
(\# SWAPs, \# CNOTs, and circuit depth) 
compared to  \tket \cite{tket}, Qiskit \cite{qiskit}, and the QAOA compiler (IC-QAOA) \cite{alam2020circuit, alam2020efficient, alam2020noise}. }
\label{fig:montreal_cx}
\end{figure*}

\subsection{Reducing overhead compared to generic compilers}
In this section, we compare the compilation overhead of our \ucl~compiler with \tket~and Qiskit. Figures \ref{fig:grid_syc}, \ref{fig:aspen_iswap}, \ref{fig:montreal_cx} show the compilation results on Google Sycamore, Rigetti Aspen, and IBM Montreal, respectively.  Compared to \tket~and Qiskit, \ucl~has least compilation overhead in terms of the number of inserted SWAP gates, the number of hardware two-qubit gates, and the circuit depth. 
We summarize the overhead reduction of \ucl~versus \tket~and Qiskit in Tables \ref{tb:ucl_tket} and \ref{tb:ucl_qiskit} respectively. We define the overhead reduction as the ratio of the overhead of \tket~or Qiskit and the overhead of \ucl. We also compare the compilation results on the CZ gate set for each qubit architecture in Appendix.

Across all benchmarks and quantum computers, \ucl~inserts on average 3.6x fewer SWAPs than \tket~and 9.1x fewer SWAPs than Qiskit. 
This SWAP count reduction will lead to a significant reduction of the hardware two-qubit gate count. 
Moreover, a large percent of SWAPs are unified with circuit gates by the unitary unifiying pass in \ucl~(gray bars in (a), (d), (g) of Figures \ref{fig:grid_syc}-\ref{fig:montreal_cx}), further reducing the gate count.
For example, when compiling to the Sycamore device, \ucl~almost has \textit{no SYC overhead} compared to the baseline `NoMap' for the Heisenberg model. This is because most of the SWAPs can be combined with circuit gates (Figure \ref{fig:grid_hb_swap}) and such an unified gate can be decomposed into similar number of SYCs as a circuit gate $\textup{exp}(it (\alpha XX+\beta YY+\gamma ZZ))$. In contrast, \tket~and Qiskit lead to on average 24.5\% and 92.3\% SYC overhead. 
\ucl~reduces the two-qubit gate overhead by 4x and 10.4x compared to \tket~and Qiskit on average over all evaluated benchmarks.
This gate count reduction helps reduce the circuit depth.
The average depth overhead reduction by using the \ucl~compiler is 2.1x relative to \tket~and 3.4x relative to Qiskit.

\begin{table*}[]
\centering
\resizebox{1.9\columnwidth}{!}{
\textbf{
\begin{tabular}{|l|c|c|c|c|c|c|c|c|c|c|c|c|c|c|c|c|c|c|}
\hline
\multirow{3}{*}{} & \multicolumn{6}{c|}{Sycamore}                                                       & \multicolumn{6}{c|}{Aspen}                                                            & \multicolumn{6}{c|}{Montreal}                                                        \\ \cline{2-19} 
                 Benchmark & \multicolumn{2}{c|}{SWAPs} & \multicolumn{2}{c|}{SYCs} & \multicolumn{2}{c|}{SYC Depth} & \multicolumn{2}{c|}{SWAPs} & \multicolumn{2}{c|}{iSWAPs} & \multicolumn{2}{c|}{iSWAP Depth} & \multicolumn{2}{c|}{SWAPs} & \multicolumn{2}{c|}{CNOTs} & \multicolumn{2}{c|}{CNOT Depth} \\ \cline{2-19} 
                  & avg          & max         & avg         & max         & avg          & max         & avg          & max         & avg          & max          & avg          & max         & avg          & max         & avg         & max          & avg          & max         \\ \hline
NNN Heisenberg                & 1.7x         & 3.9x        & --          & --          & 1.2x         & 2.7x        & 1.1x         & 1.8x        & 3.2x         & 5x           & 1.5x         & 3.2x        & 2.2x         & 4.6x        & 6x          & 12.8x        & 2.4x         & 5.2x        \\ \hline
NNN XY             & 1.7x         & 3.4x        & 6.2x        & 21.1x       & 1.2x         & 2.1x        & 1.1x         & 1.7x        & 1.9x           & 3.8x         & 1.3x         & 2.5x        & 2.8x         & 4.5x        & 5.3x        & 8.3x         & 2.7x         & 3.9x          \\ \hline
NNN Ising             & 1.9x         & 3.9x        & 5.6x        & 10.7x       & 1.3x         & 2.9x        & 1.1x         & 1.8x        & 2x           & 3.2x         & 1.5x         & 3.4x        & 2.7x         & 4.1x        & 4.9x        & 7.8x         & 2.7x         & 4x          \\ \hline
QAOA-REG-3              & 1.8x         & 2.5x        & 3.2x        & 4.4x        & 3.9x         & 9.5x        & 1.8x         & 2.4x        & 3x           & 4.5x         & 2.2x         & 3.8x        & 2x           & 2.4x        & 3x          & 4.3x         & 3x           & 4x          \\ \hline
\end{tabular}
}
}
\caption{The average (avg) and maximum (max) compilation overhead reduction when comparing \ucl~with \tket \cite{tket}. For the cases with blank values `--', the \ucl~compiler has negligible overhead.}
\label{tb:ucl_tket}
\end{table*}

\begin{table*}[]
\centering
\resizebox{1.9\columnwidth}{!}{
\textbf{
\begin{tabular}{|l|c|c|c|c|c|c|c|c|c|c|c|c|c|c|c|c|c|c|}
\hline
\multirow{3}{*}{} & \multicolumn{6}{c|}{Sycamore}                                                       & \multicolumn{6}{c|}{Aspen}                                                            & \multicolumn{6}{c|}{Montreal}                                                        \\ \cline{2-19} 
                Benchmark  & \multicolumn{2}{c|}{SWAPs} & \multicolumn{2}{c|}{SYCs} & \multicolumn{2}{c|}{SYC Depth} & \multicolumn{2}{c|}{SWAPs} & \multicolumn{2}{c|}{iSWAPs} & \multicolumn{2}{c|}{iSWAP Depth} & \multicolumn{2}{c|}{SWAPs} & \multicolumn{2}{c|}{CNOTs} & \multicolumn{2}{c|}{CNOT Depth} \\ \cline{2-19} 
                  & avg         & max          & avg        & max          & avg          & max         & avg          & max         & avg          & max          & avg          & max         & avg         & max          & avg         & max          & avg          & max         \\ \hline
NNN Heisenberg        & 6x          & 9.7x         & --         & --           & 1.8x         & 2.6x        & 3.3x         & 4.3x        & 9.3x         & 13x          & 2.7x         & 3.4x        & 5.1x        & 10.1x        & 14x         & 27.8x        & 3.8x         & 7.6x        \\ \hline
NNN XY             & 6.5x         & 11.1x        & 24.1x        & 68.5x       & 2.1x         & 3.5x        & 3.2x         & 4.1x        & 5.4x           & 8.4x         & 2.5x         & 3.4x        & 5.6x         & 8.3x        & 10.7x        & 16.1x         & 4x         & 6x          \\ \hline
NNN Ising             & 6.7x        & 11.5x        & 19x        & 30.7x        & 2.4x         & 4.1x        & 3.3x         & 4.1x        & 5.6x         & 7.4x         & 3x           & 3.9x        & 5.3x        & 8.2x         & 9.7x        & 15x          & 4x           & 7.4x        \\ \hline
QAOA-REG-3              & 4.7x        & 6.4x         & 8x         & 10.6x        & 6.9x         & 21x         & 3.1x         & 4x          & 5.2x         & 6.9x         & 3.5x         & 5x          & 3x          & 3.9x         & 4.4x        & 5.1x         & 3.7x         & 4.8x        \\ \hline
\end{tabular}
}}
\caption{The average (avg) and maximum (max) compilation overhead reduction when comparing \ucl~with Qiskit \cite{qiskit}. For the cases with blank values `--', the \ucl~compiler almost has negligible overhead.}
\label{tb:ucl_qiskit}
\end{table*}

\subsection{Reducing overhead compared to specialized compilers}
We further compare \ucl~with the QAOA compiler in \cite{alam2020circuit, alam2020efficient, alam2020noise} and the Paulihedral compiler in \cite{li2021paulihedral}.
We will only evaluate the compilations results on IBM quantum computers because these two compilers are restricted to the CNOT or CZ gate set.
Compared to the QAOA compiler, \ucl~reduces SWAP count, CNOT count overhead, and CNOT depth overhead by on average 2.6x, 4x, and 2.8x, respectively (Figure \ref{fig:montreal_qaoa_swap}-\ref{fig:montreal_qaoa_cx_depth}).
The Paulihedral compiler is not open-sourced yet and we directly use the numbers from \cite{li2021paulihedral} as shown in Table \ref{tb:ucl_paulihedral}.
The Heisenberg models were compiled by assuming all-to-all connectivity and the QAOA-REG-m were compiled to the IBMQ Manhattan device that has a dodecagon lattice and CNOT as native two-qubit gate as the Montreal device.
We generate 10 random graphs for each QAOA problem and present the average compilation results of \ucl~in Table \ref{tb:ucl_paulihedral}.
The CNOT count and circuit depth achieved by Paulihedral are on average 1.59x and 1.64x as high as the circuits compiled by \ucl. Therefore, by employing the permutation-aware optimizations in the routing and unitary unifying passes, \ucl~outperforms both application-specific compilers in terms of compilation overhead. 
\begin{table}[]
\centering
\resizebox{0.85\columnwidth}{!}{
\textbf{
\begin{tabular}{|l|c|c|c|c|}
\hline
\multirow{2}{*}{Benchmark} & \multicolumn{2}{c|}{Paulihedral} & \multicolumn{2}{c|}{2QAN} \\ \cline{2-5} 
                  & CNOTs           & Depth           & CNOTs         & Depth      \\ \hline
Heisenberg-1D (30 qubits)     & 87             & 13              & 87           & 13         \\ \hline
Heisenberg-2D (30 qubits)     & 216            & 43              & 147          & 25         \\ \hline
Heisenberg-3D (30 qubits)     & 305            & 65              & 177          & 31         \\ \hline
QAOA-REG-4 (20 qubits)        & 366            & 147             & 177(7)     & 71(4)      \\ \hline
QAOA-REG-8 (20 qubits)       & 539            & 246             & 324(9)        & 160(16)    \\ \hline
QAOA-REG-12 (20 qubits)       & 678            & 319             & 419(9)     & 222(22)    \\ \hline
\end{tabular}
}
}
\caption{Circuit size comparison with Paulihedral \cite{li2021paulihedral}. Numbers in brackets show the standard deviation over 10 instances.}
\label{tb:ucl_paulihedral}
\end{table}

\begin{figure*}[tbh!]
 \centering
    \subfloat[Experimental results of 1-layer QAOA]{
    \includegraphics[width=0.33\textwidth]{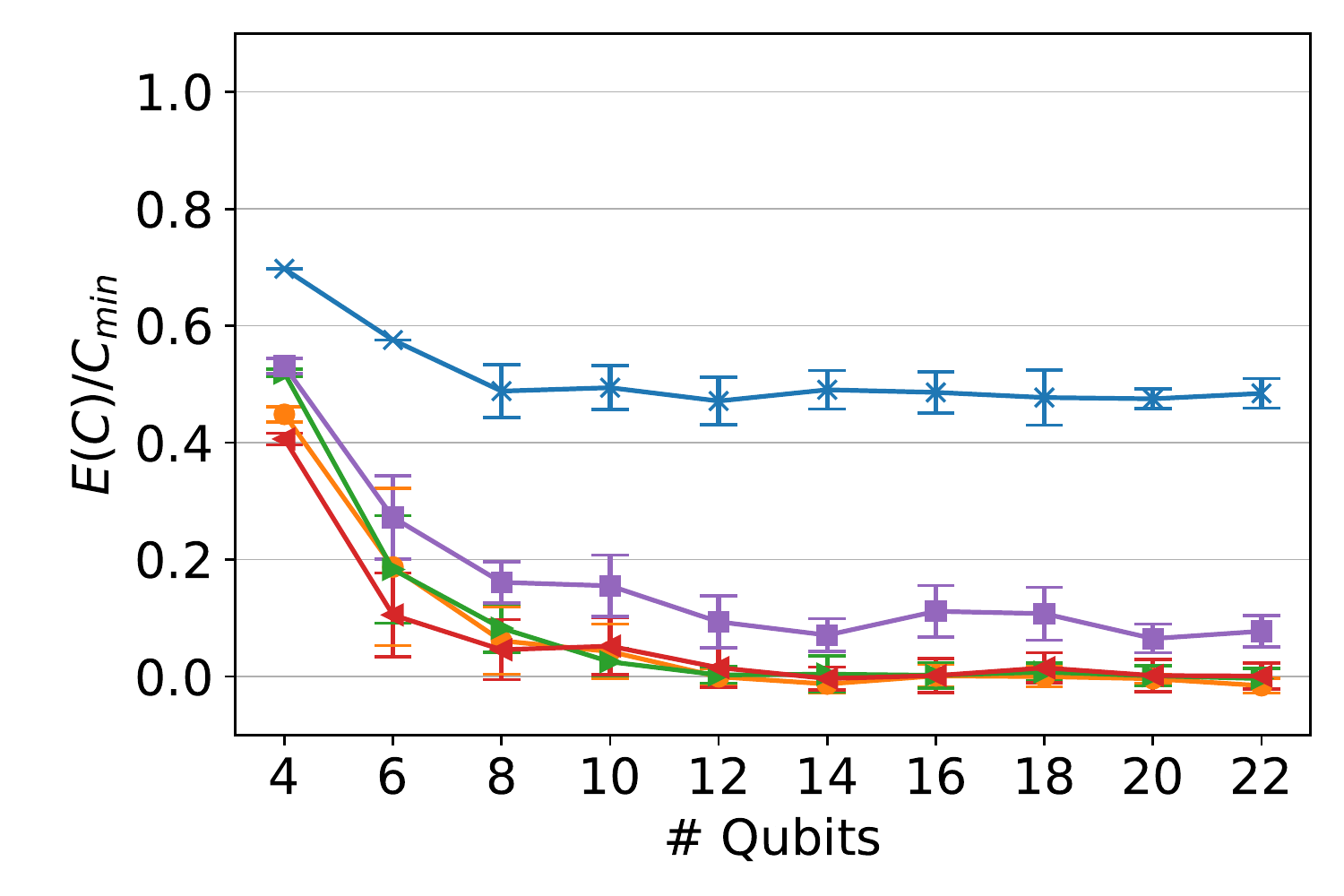}
    \label{fig:montreal_qaoa_cost1}
    }
    \subfloat[Experimental results of 2-layer QAOA]{
     \includegraphics[width=0.33\textwidth]{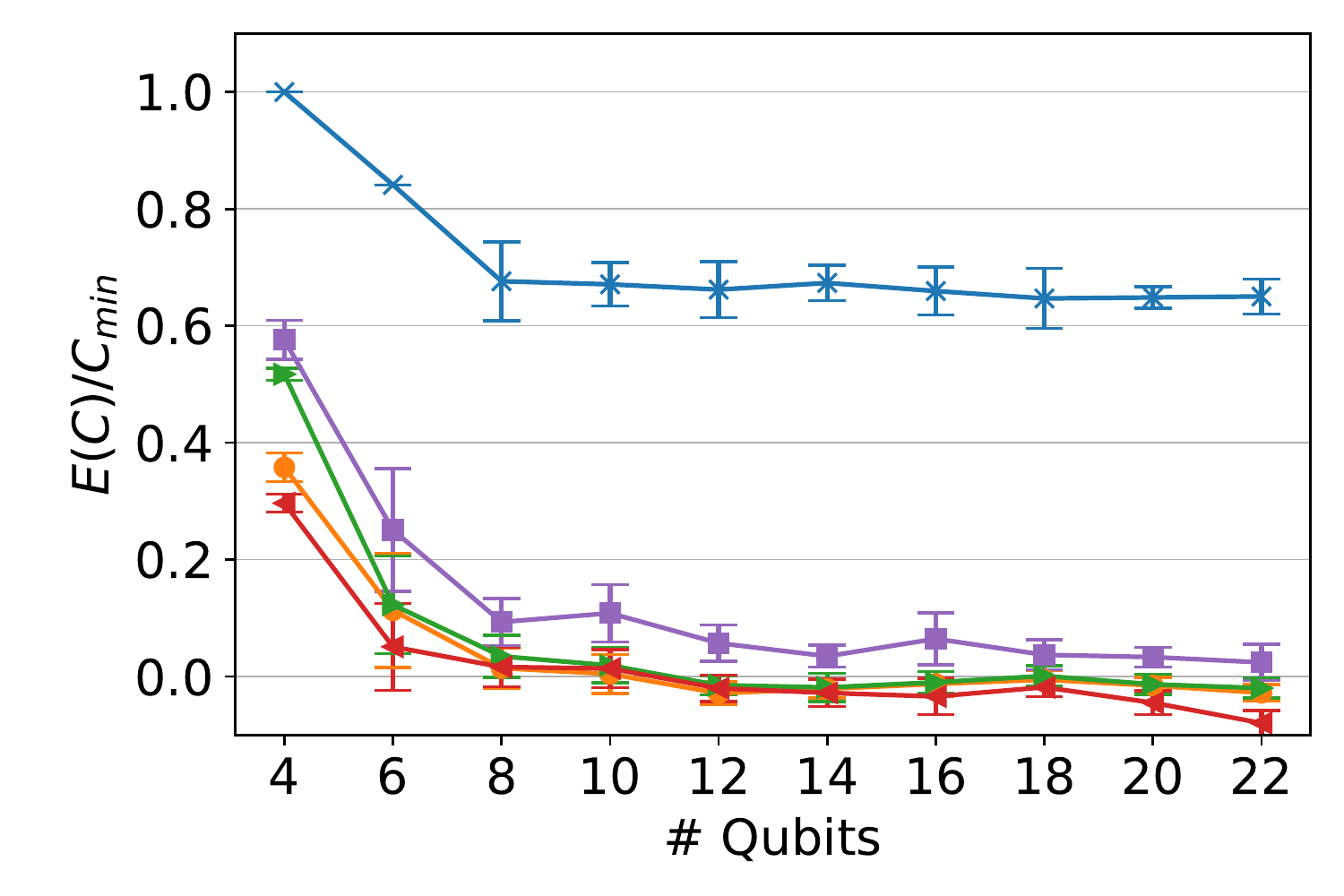}
    \label{fig:montreal_qaoa_cost2}
    }
    \subfloat[Experimental results of 3-layer QAOA]{
     \includegraphics[width=0.33\textwidth]{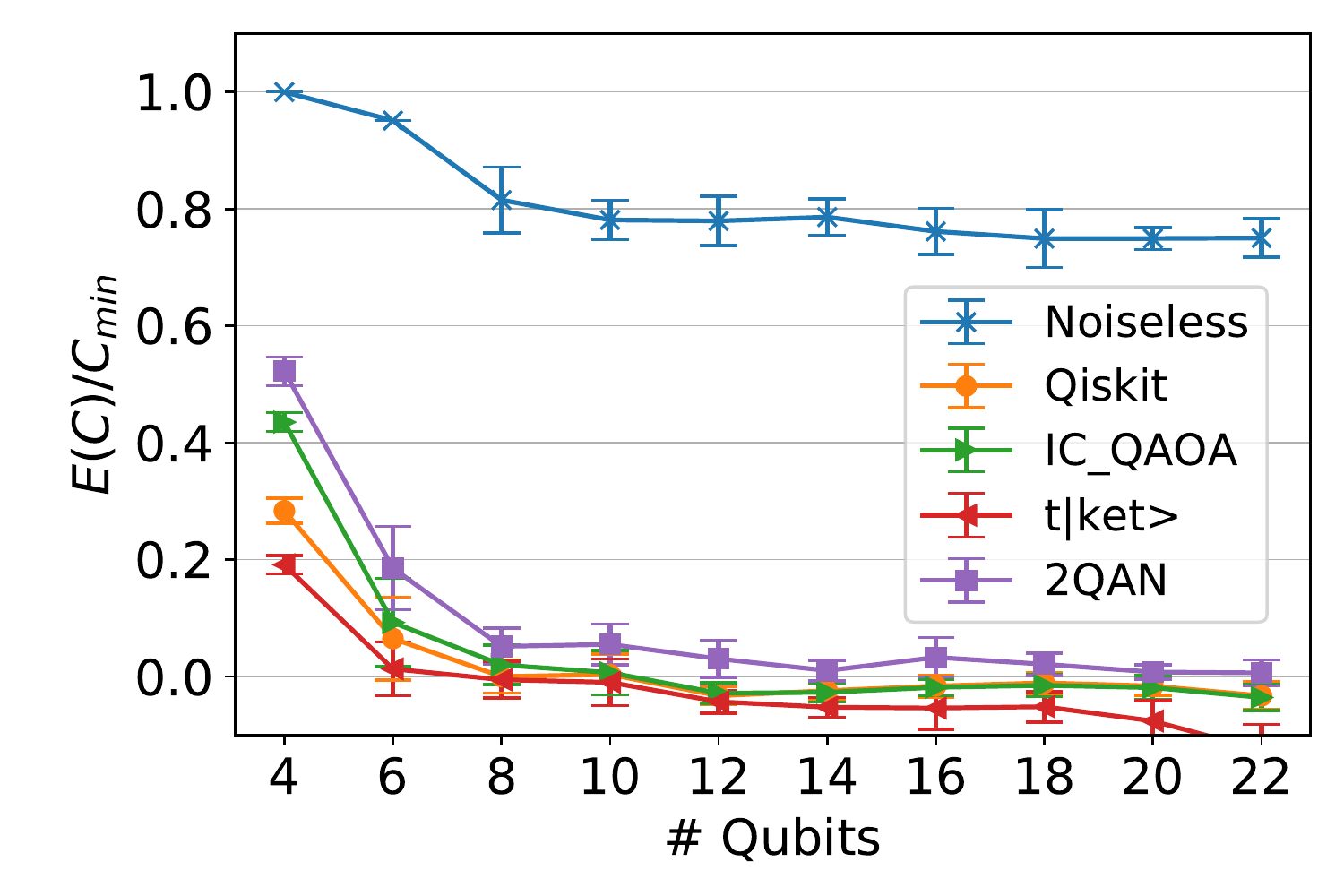}
    \label{fig:montreal_qaoa_cost3}
    }
\caption{Experimental results of running QAOA-REG-3 on the IBMQ Montreal device. The y axis shows the application performance, which is measured by the normalized cost function $ \left \langle  C\right \rangle/C_{\textup{min}}$ (larger is better). Each problem size is averaged over 10 different instances (error bars show the standard deviation). 
For the same layer QAOA, 2QAN always achieves better application performance compared to other compilers.
By using \ucl, the 4-qubit QAOA performance improves when the number of layers is increased from 1 to 2.
}
\label{fig:qaoa_exps}
\end{figure*}

\subsection{Improving application performance}
To study how the compilation overhead reduction impacts application performance, we experimentally implemented the QAOA benchmarks with different numbers of layers on the IBMQ Montreal device.
For the multiple-layer QAOA circuits, the \ucl~compiler only performs compilation for the first layer and obtains a circuit $c_1$. For odd number layers, it directly uses the compiled circuit $c_1$. For even number layers, it simply reverses the two-qubit gate order in $c_1$. In contrast, \tket~and Qiskit compile a multiple-layer QAOA circuit as a whole. 
For all four compilers, the compilation overhead of a $n$-layer QAOA circuit is approximately $n$ times of the overhead of a single-layer circuit (see Figure \ref{fig:montreal_qaoa_l3} in Appendix).
In NISQ computers, the implementation of a quantum application that has fewer hardware gates (reducing gate errors) and shorter circuit depth (reducing decoherence errors) should have better performance (higher fidelity).
This is experimentally demonstrated in Figure \ref{fig:qaoa_exps}, the circuits compiled by our 2QAN compiler have the \textit{best application performance} compared to the results from using \tket, Qiskit, and IC-QAOA for all problem sizes and all QAOA layers. 

For example, as noise accumulates in QAOA circuits they will often converge to the value 0 corresponding to a random guess. Here we see that the 3-layer QAOA benchmarks compiled by \tket~or Qiskit or IC-QAOA already approach zero for problems with 8 qubits. In comparison, \ucl~only comes close to this value for much larger problems (around 20 qubits).
Ideally (without hardware noise), the application performance should improve with the number of QAOA layers. However, in practice there is a trade off, with additional layers also increasing the overall error probabilities, and hence decreasing application performance.
Figure \ref{fig:qaoa_exps} shows that the QAOA performance of all problem sizes decreases when the number of layers increases for all compilers except the 4-qubit QAOA compiled by \ucl~of which performance improves when increasing the layer number from 1 to 2. This implies that efficient compilation techniques can enhance device capacity and potentially pave the way towards practical applications of quantum computing. 

\subsection{Scalability and runtime}
We use the Tabu search algorithm for solving the qubit placement problem. This algorithm is fast for solving small problems, e.g, it takes around 1.6 seconds for the 10-qubit Ising model and 12.2 seconds for the 20-qubit QAOA.
However, it becomes slow for larger problems and takes 330.2 (976.3) seconds for the 40(50)-qubit Heisenberg model. Qubit placement is not our main optimization goal in this work. More efficient algorithms exist for solving QAP problems \cite{burkard1998quadratic} and other qubit mapping techniques \cite{tket,qiskit} could also be applied in future work.
The proposed routing and scheduling heuristics (Algorithm \ref{alg:routing} and Algorithm \ref{alg:scheduling}) scale quadratically with the number of gates in one Trotter step. Moreover, we only perform the compilation for the first Trotter step and simply use this circuit for odd-number steps and reverse the two-qubit gate order for even-number steps. Such an implementation saves compilation time and is similar to the second-order Trotterization in Equation \ref{equ:2order}. The runtime evaluation and scalability analysis give us confidence that the \ucl~compiler will scale up to near-term quantum applications with a considerable number of qubits and gates.

\section{Related work}
\label{sec:priors}
The product formula is the most established approach to simulate the dynamics of quantum systems.
Approximation errors arise when there are anti-commuting terms in the \h. To achieve a desired precision, the time evolution is divided into many small time steps. Assume a simulation circuit has $r$ Trotter steps 
and the operator count for implementing each step is $G$.
Minimizing the circuit size $Gr$ meanwhile maintaining computational accuracy is crucial for a practical implementation.
Besides the high-order approximation approach \cite{suzuki1991general}, many randomization approaches have been proposed to further reduce $Gr$ \cite{childs2019faster, campbell2019random, chen2020quantum, ouyang2020compilation,faehrmann2021randomizing}.
Afterwards, these high-level circuits still need to be decomposed into native hardware gates.
One popular approach for optimizing low-level circuit for Hamiltonian simulation is to first group Pauli terms into commuting sets and then apply simultaneous diagonalization of Pauli exponentials in each set \cite{van2020circuit, cowtan2020generic} or gate cancellation between consecutive Pauli exponentials \cite{hastings2014improving,gui2020term}.
Other works present circuit synthesis method based on the ZX-calculus \cite{cowtan2019phase,de2020architecture}.
However, these low-level circuit optimization approaches are restricted to CNOT or CZ-based circuits and most of them do not consider the qubit connectivity constraint, requiring further compilation.

Many compilation techniques have been proposed to map quantum circuits onto NISQ computers with limited qubit connectivity and native gate sets. Some of them optimize gate count to reduce high probability errors from two-qubit gates or minimize circuit depth to mitigate the effect of decoherence~\cite{zulehner2018efficient,cowtan2019qubit,lao21qmap,childs2019circuit}. Others consider spatial and temporal noise variations and develop noise-aware compilation algorithms to minimize error rates~\cite{li2019tackling,murali2019noise,tannu2019not,murali2020software}. 
Industrial quantum compilers such as \tket~\cite{tket} and Qiskit \cite{qiskit} incorporate multiple  techniques to achieve the best application performance.
These general-purpose compilers typically operate at the gate level and can work for general quantum applications. Nevertheless, they have little knowledge of the mathematical properties at the application level and therefore lack fine-grained optimizations.

Application-specific compilation techniques will be advantageous for NISQ computing. For example, optimized compilers have been developed for variational quantum algorithms \cite{dallaire2016quantum,venturelli2018compiling,shi2019optimized,gokhale2019partial, alam2020circuit, alam2020efficient, alam2020noise,li2021software} and quantum simulation in product formula \cite{li2021paulihedral}.
The optimization techniques in these specialized compilers are restricted to CNOT or CZ gates and are not applicable (or cause higher overhead) for quantum computers with other hardware gates such as the SYC gate in Google Sycamore \cite{arute2019quantum} and the iSWAP gate in Rigetti Aspen \cite{rigetti}.
Furthermore, most of them do not explore the flexible operator permutation property in the product formula approach for quantum simulation problems (no matter whether these operators commute or not).
For instance, the compilers in \cite{venturelli2018compiling,shi2019optimized,gokhale2019partial, alam2020circuit, alam2020efficient, alam2020noise}
only check for commutable gates in the circuit mapping procedure but many terms in the \h~may not commute (e.g., the XY model and Heisenberg model) and changing the order of anti-commuting gates will be prohibited.
The compiler in \cite{li2021paulihedral} considers this flexibility in the term scheduling procedure but lacks optimizations for qubit routing and unitary unifying. 
To address these issues, we exploit the operator permutation flexibility in different compiler passes and present application-specific algorithms to minimize compilation overhead.
Our \ucl~compiler can target different architectures and provide efficient compilation for different gate sets.

\section{Conclusions}
We have developed an application-specific compiler for 2-local qubit \h~simulation problems. The designed \ucl~compiler exploits the flexibility of permuting operators in a \h~and performs optimizations on the qubit routing, gate synthesis, and gate scheduling passes.
Evaluation results showed that \ucl~can significantly reduce compilation overhead compared to two general-purpose compilers and two application-specific compilers across three quantum computers with different topologies and gate sets. 
For some applications, \ucl~even has no gate overhead.
Furthermore, experimental results demonstrated that \ucl~can achieve the highest application fidelity in practice on hardware. We believe application-specific compilation techniques will help enhance the performance of quantum applications on NISQ devices and allow them to explore their maximum capacities.

Future work will perform more optimizations and investigate other research directions.
First, the objective of \ucl~is to minimize the number of SWAP gates and circuit depth. NISQ computers have inhomogeneities and noise-aware compilation techniques can be used to maximize application fidelity \cite{li2019tackling,murali2019noise,tannu2019not,murali2020software}.
Moreover, error mitigation techniques can be applied to further reduce errors \cite{temme2017error,li2017efficient,bravyi21mitigating,tannu2019mitigating}.
In the experimental results reported here, we only compiled the first Trotter step in \h~simulation and simply applied a reverse scheduling for the two-qubit gates in the even-number steps. Prior works prove that randomizing the operator order in each step can reduce the simulation costs \cite{childs2019faster,campbell2019random}. Future work can adapt 2QAN to this randomization and analyze how randomized compiling affects approximation errors and the efficiency of quantum simulation algorithms.
In addition, it is worthwhile to investigate how to generalize these compilation techniques for $k$-local Hamiltonians and other quantum simulation algorithms.  


\section*{Acknowledgements}
We thank Prakash Murali for valuable feedback on the manuscript. We thank Silas Dilkes for help with the \tket~compiler.
We thank César A. Rodríguez Rosario for help with the experimental setup on the IBMQ Montreal device via Strangeworks QC platform.
We acknowledge use of the IBM Q platform for this work. 
We acknowledge funding from the EPSRC Prosperity Partnership in Quantum Software for Modelling and Simulation (Grant No. EP/S005021/1).

\bibliographystyle{unsrt}
\bibliography{references}
\newpage
\appendix
\label{sec:appendix}
In this appendix, we present additional results of comparing 2QAN's compilation performance on the IBM Montreal, Google Sycamore, and Rigetti Aspen devices. 
In addition to the hardware two-qubit gates (SYC and iSWAP) in Figure~\ref{fig:devices}, Sycamore and Aspen also implement CZ gates as native gates.
In this comparison, we use the recommended optimization passes for the CNOT/CZ gates in Qiskit, \tket, and the QAOA compiler. That is, we use Qiskit with optimization level 3, \tket~with the `FullPass' (FullPeepholeOptimise, followed by the default qubit mapping pass, SynthesiseIBM), the QAOA compiler with.
Similarly to results in Section \ref{sec:results}, \ucl~exploits the term permutation flexibility in \h~simulation and has the \textit{least compilation overhead} (the number of inserted SWAPs, the CNOT gate count, and the circuit depth) compared to all three compilers~across all benchmarks and quantum computers.

\begin{figure*}[tbh!]
 \centering
 \subfloat[SWAP count for NNN Heisenberg model]{
     \includegraphics[width=0.33\textwidth]{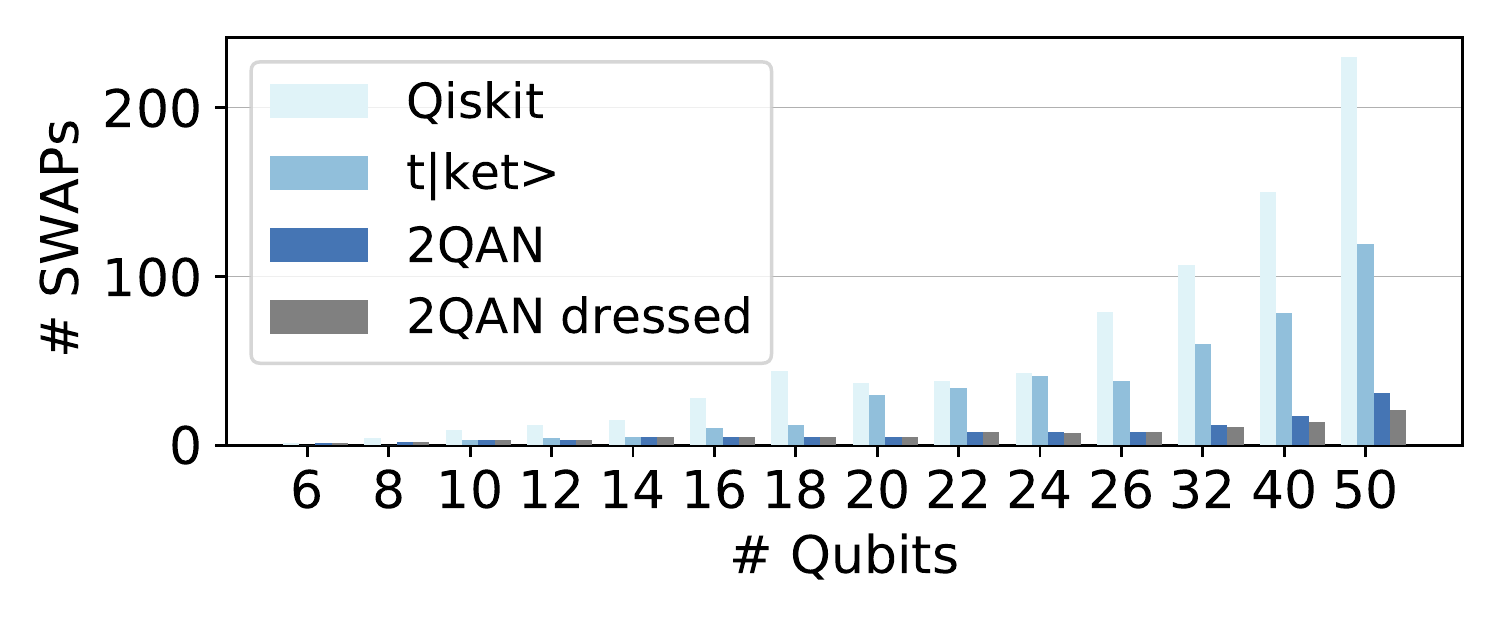}
    \label{fig:grid_hb_cxswap}
    }
    \subfloat[CZ count for NNN Heisenberg model]{
     \includegraphics[width=0.33\textwidth]{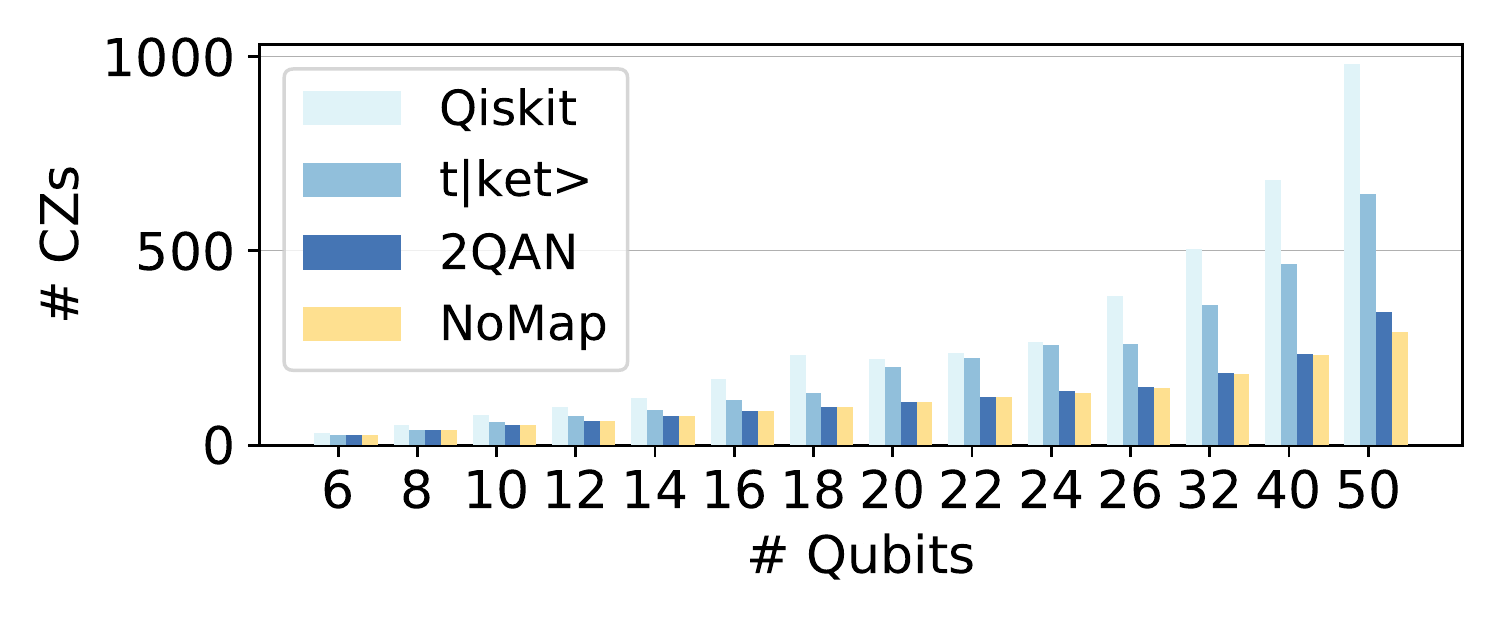}
    \label{fig:grid_hb_cx}
    }
    \subfloat[CZ depth for NNN Heisenberg model]{
    \includegraphics[width=0.33\textwidth]{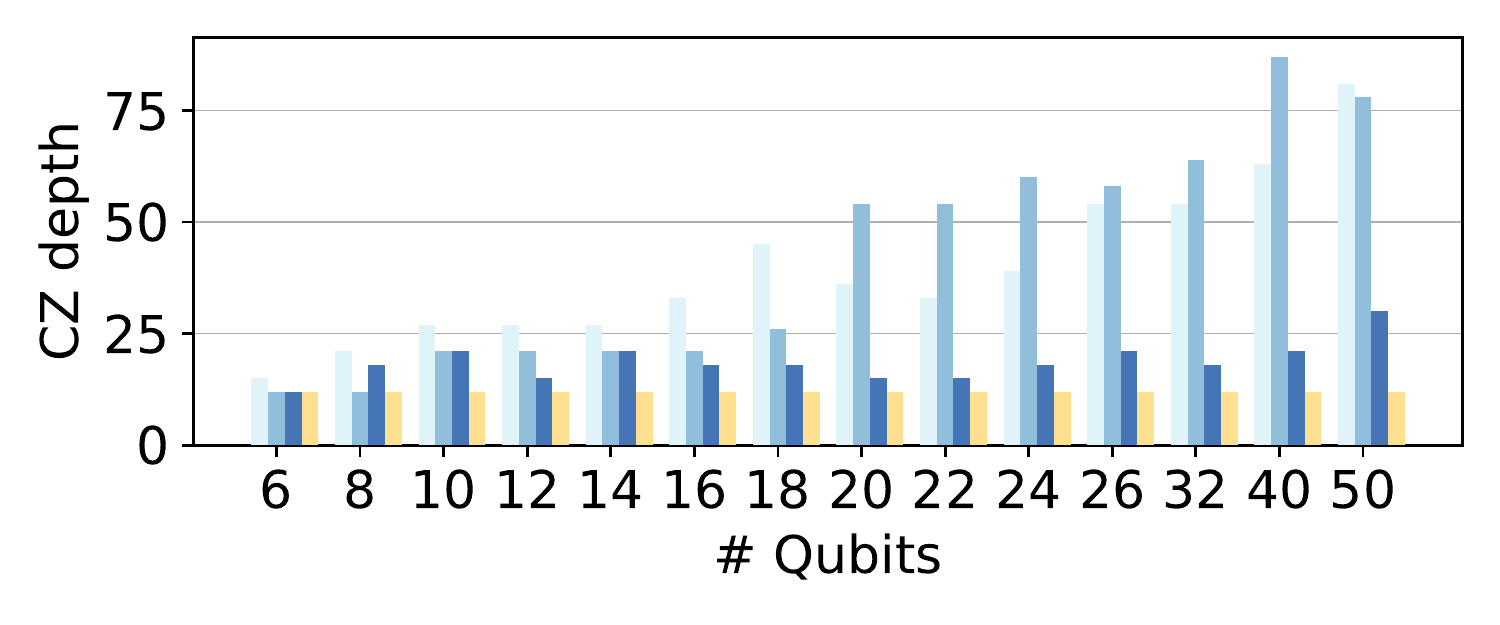}
    \label{fig:grid_hb_cx_depth}
    }
    
    \subfloat[SWAP count for NNN XY model]{
     \includegraphics[width=0.33\textwidth]{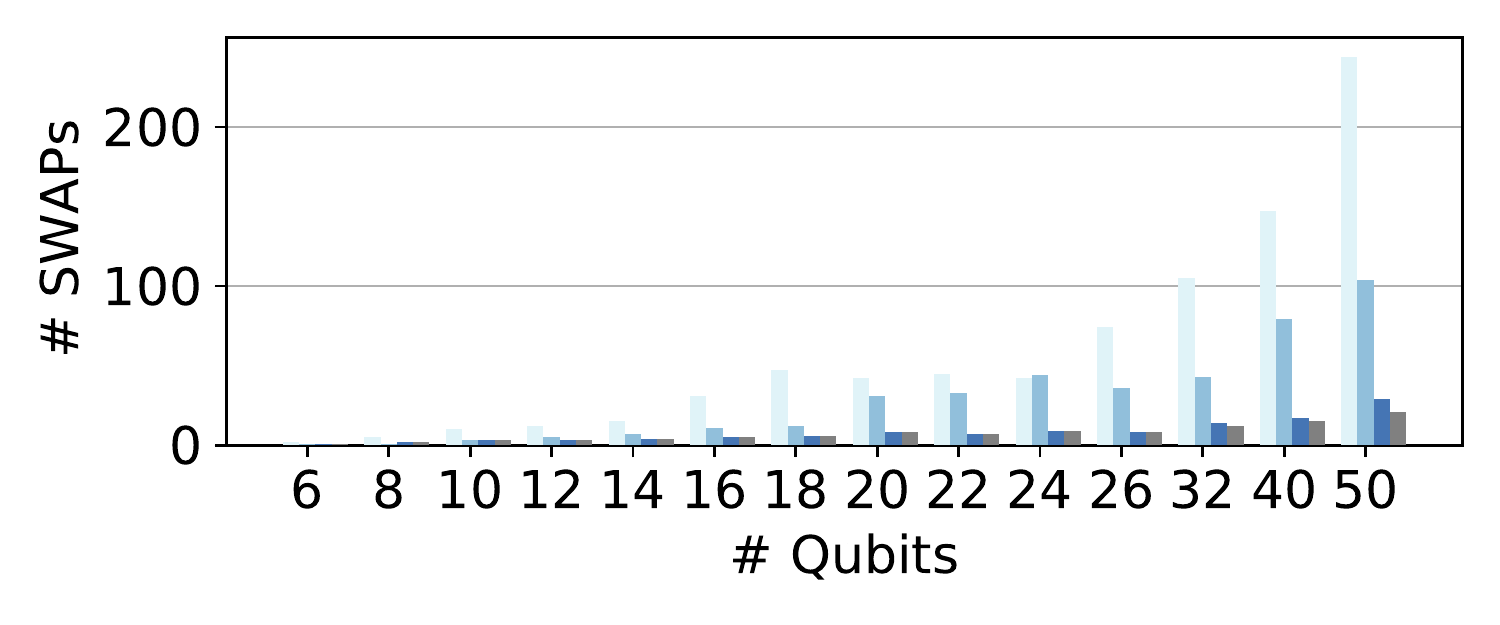}
    \label{fig:grid_xy_cxswap}
    }
    \subfloat[CZ count for NNN XY model]{
     \includegraphics[width=0.33\textwidth]{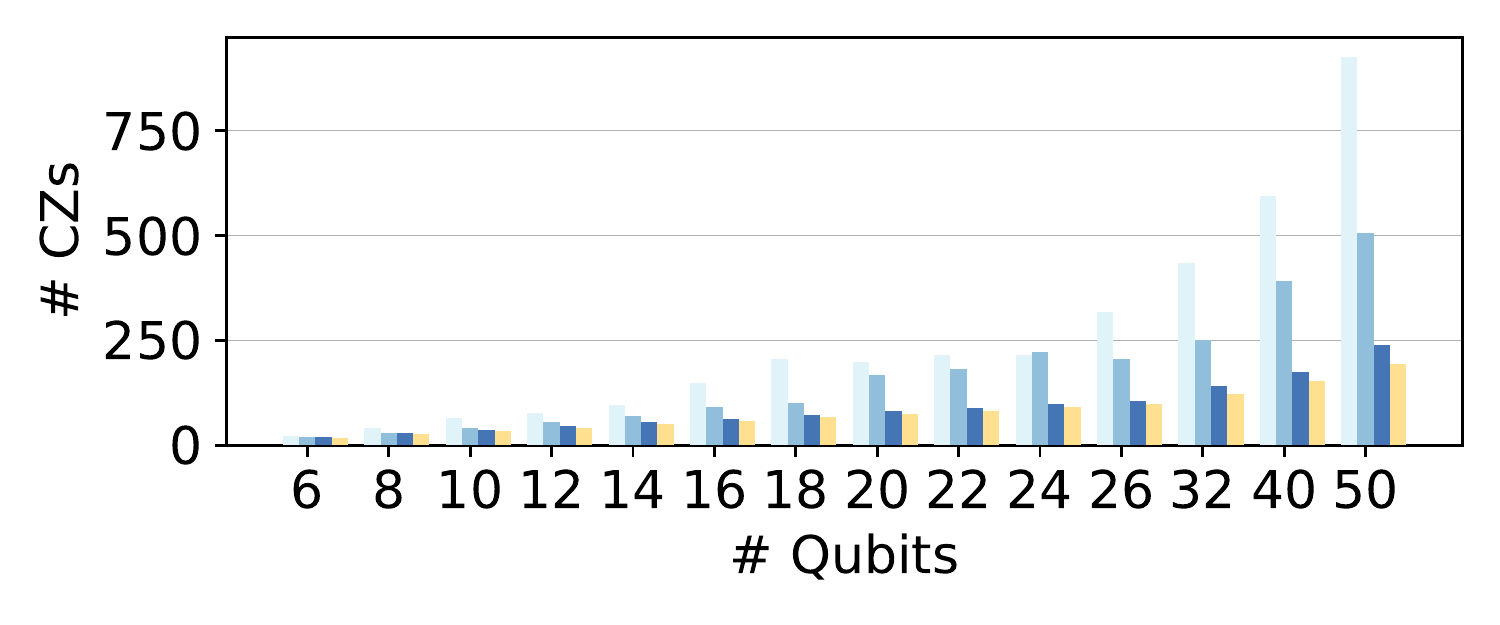}
    \label{fig:grid_xy_cx}
    }
    \subfloat[CZ depth for NNN XY model]{
    \includegraphics[width=0.33\textwidth]{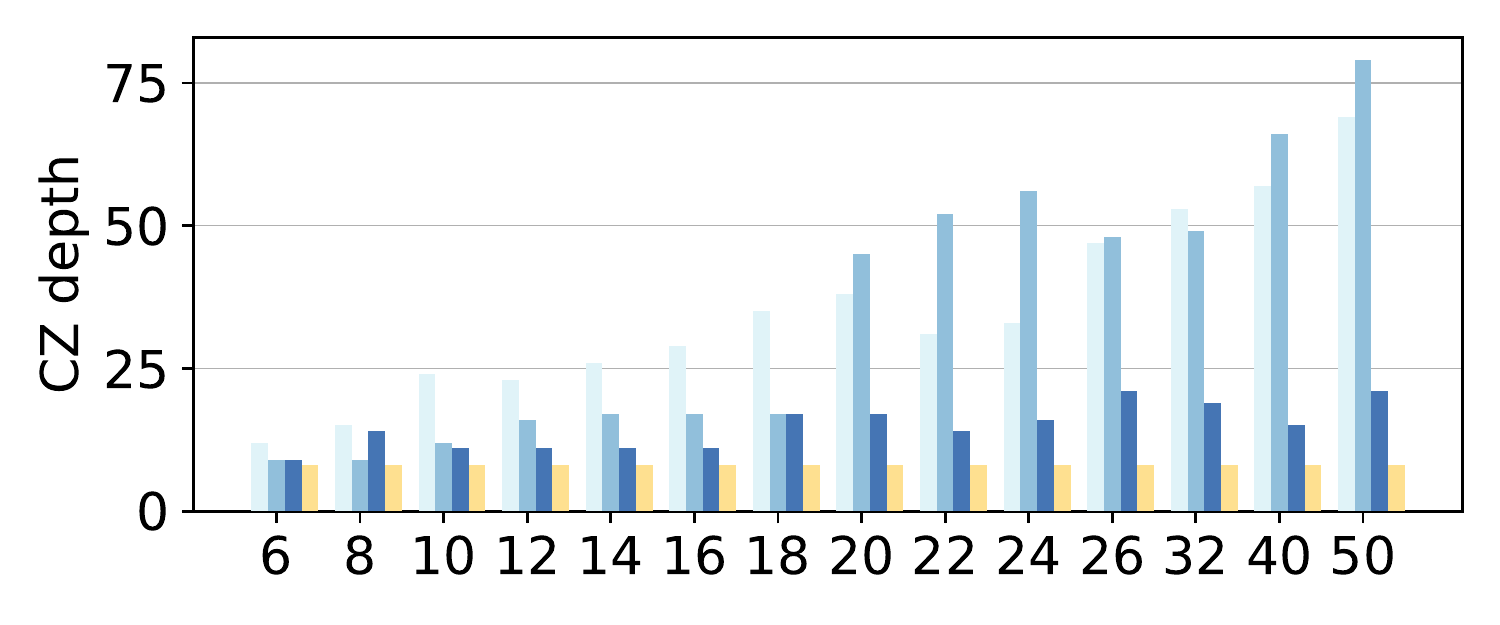}
    \label{fig:grid_xy_cx_depth}
    }
    
    \subfloat[SWAP count for NNN Ising model]{
     \includegraphics[width=0.33\textwidth]{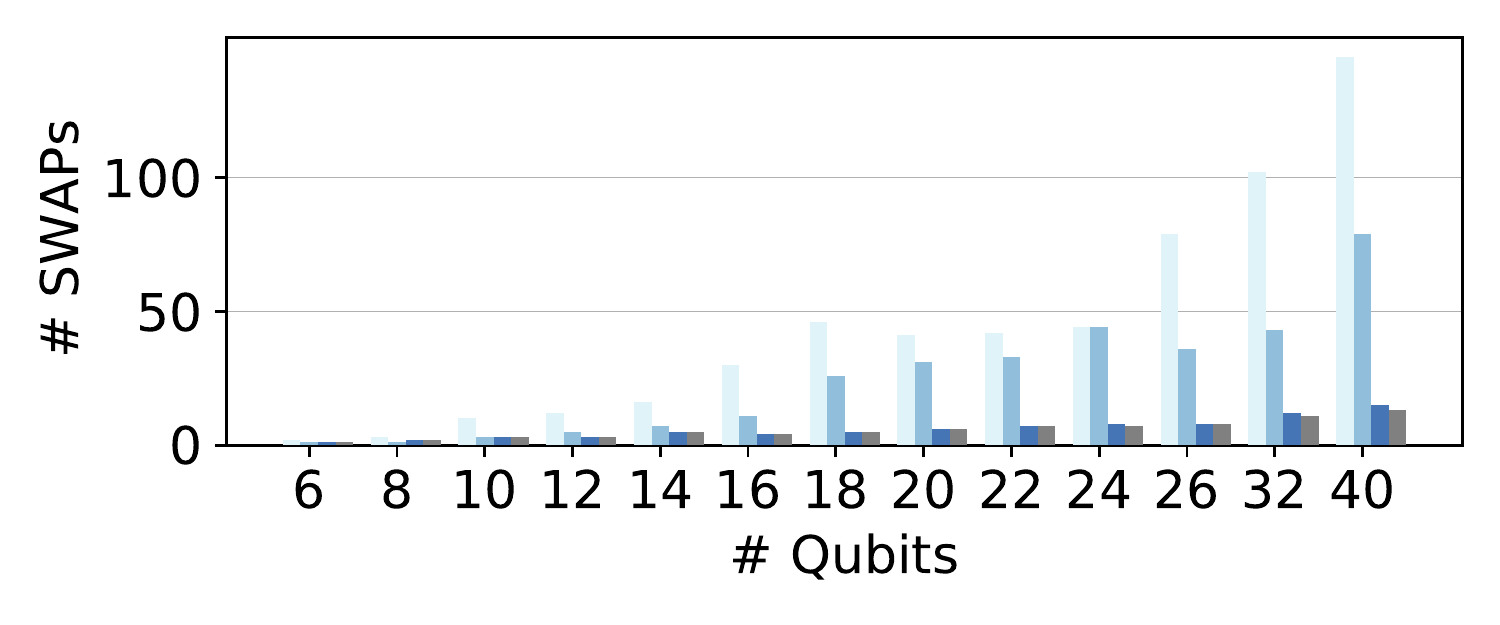}
    \label{fig:grid_ising_cxswap}
    }
    \subfloat[CZ count for NNN Ising model]{
     \includegraphics[width=0.33\textwidth]{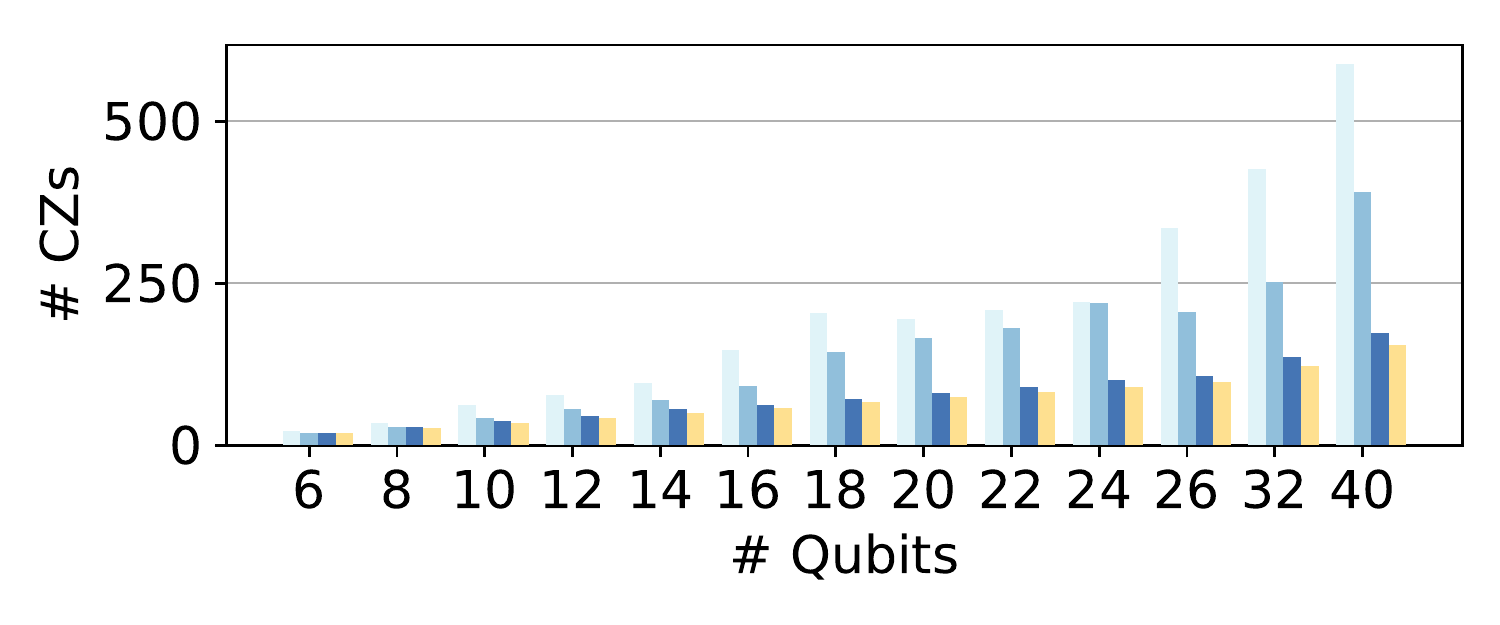}
    \label{fig:grid_ising_cx}
    }
    \subfloat[CZ depth for NNN Ising model]{
    \includegraphics[width=0.33\textwidth]{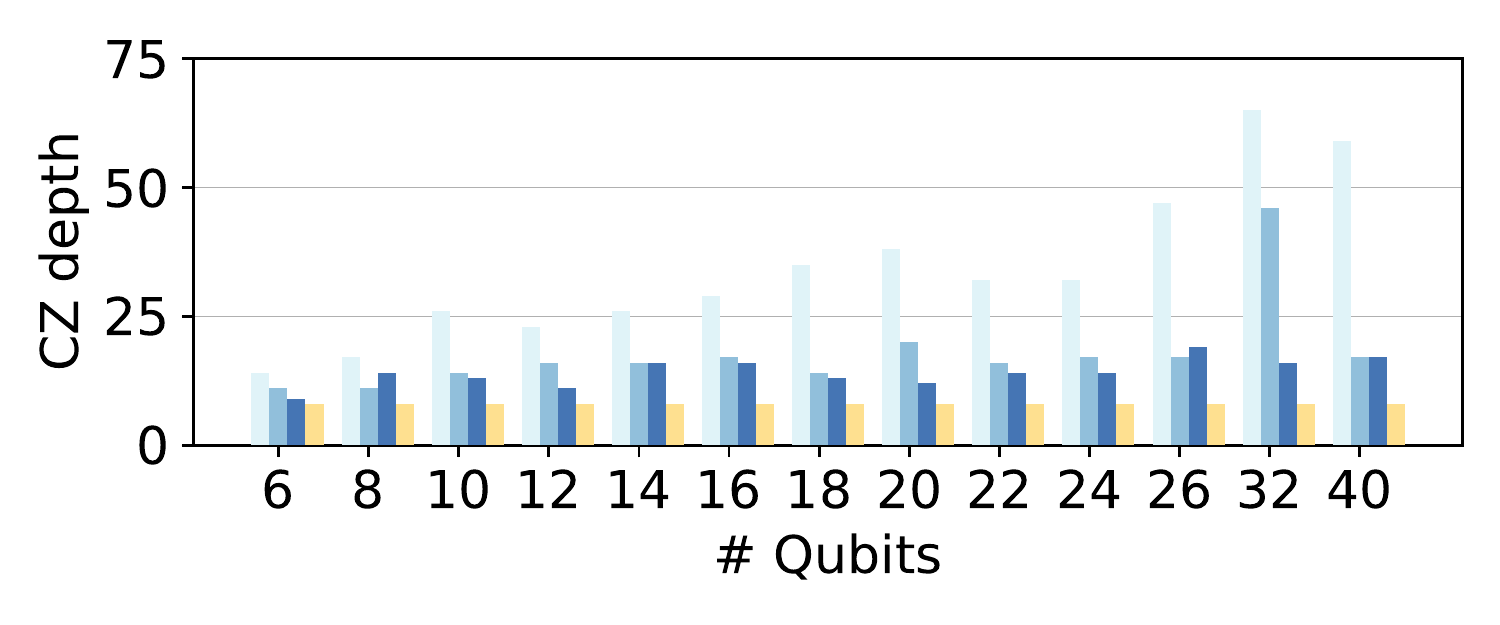}
    \label{fig:grid_ising_cx_depth}
    }
    
    \subfloat[SWAP count for QAOA-REG-3]{
     \includegraphics[width=0.33\textwidth]{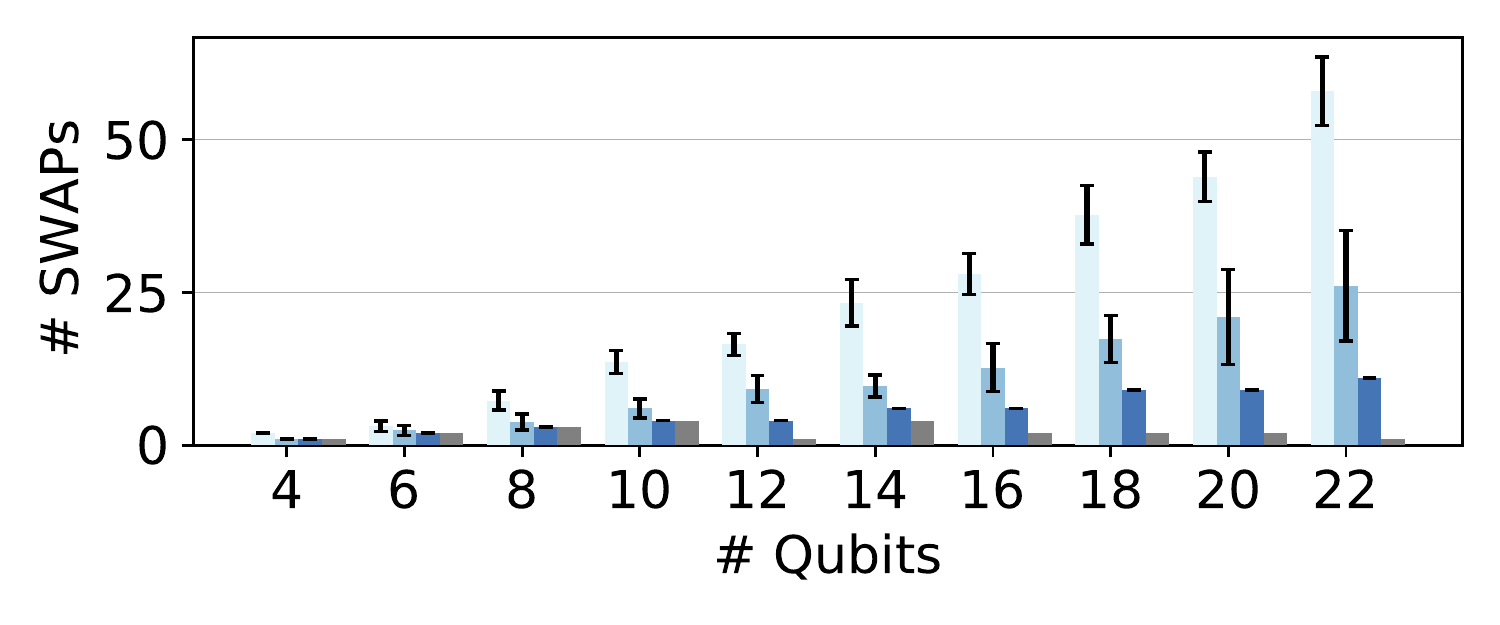}
    \label{fig:grid_qaoa_cxswap}
    }
    \subfloat[CZ count for QAOA-REG-3]{
     \includegraphics[width=0.33\textwidth]{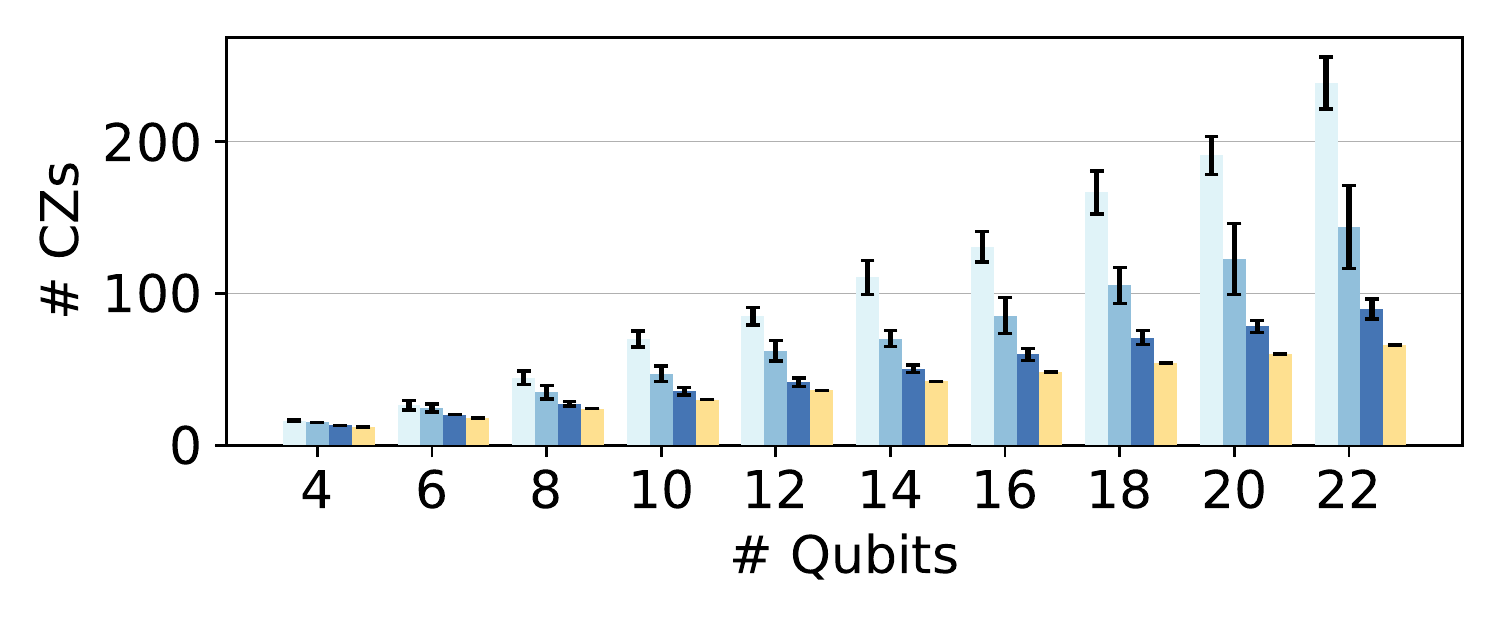}
    \label{fig:grid_qaoa_cx}
    }
    \subfloat[CZ depth for QAOA-REG-3]{
    \includegraphics[width=0.33\textwidth]{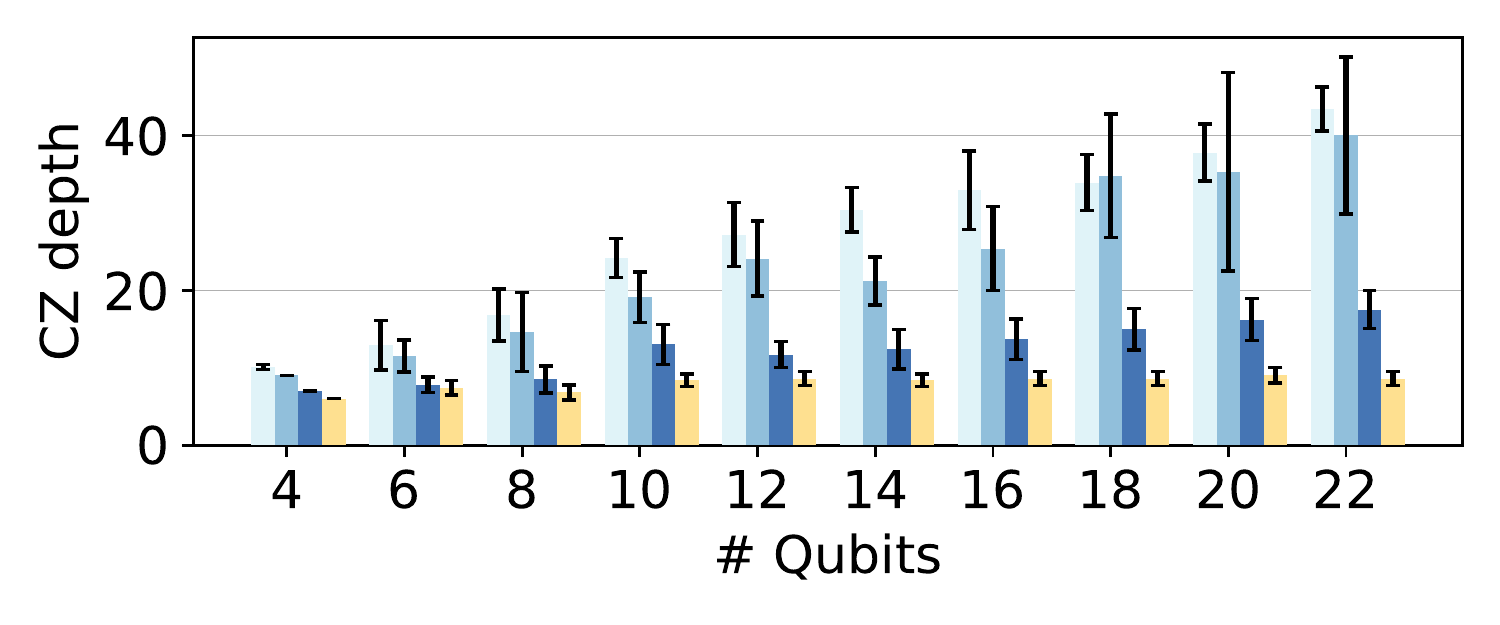}
    \label{fig:grid_qaoa_cxdepth}
    }
\caption{Compilation results of the one-layer NNN Heisenberg model, NNN Ising model, and QAOA on the Google Sycamore architecture with CZ as hardware gate. The 2QAN compiler has least compilation overhead compared to Qiskit and \tket. }
\label{fig:grid_cx}
\end{figure*}

Figure \ref{fig:grid_cx} and Table \ref{tb:grid_cx} show the compilation results on the Sycamore architecture with CZ as the hardware two-qubit gate.
\ucl~almost has \textit{no CZ overhead} compared to `NoMap' for the Heisenberg model. This is because most of the SWAPs can be combined with circuit gates (Figure \ref{fig:grid_hb_cxswap}) and such an unified gate can be decomposed into the same number (3) of CZs as a circuit gate $\textup{exp}(it (\alpha XX+\beta YY+\gamma ZZ))$. In contrast, \tket~and Qiskit lead to on average 55.6\% and 107\% CZ overhead, respectively. Similarly, \ucl~almost has no overhead for the XY model and only has on average 8.7\% CZ overhead for the Ising model (the circuit unitary $\textup{exp}(i\theta ZZ)$ requires 2 CZs).
The maximum depth reduction over four benchmarks is 10.3x relative to \tket~and 13.7x compared to Qiskit.

\begin{figure*}[tbh!]
 \centering
    \subfloat[SWAP count for NNN Heisenberg model]{
     \includegraphics[width=0.33\textwidth]{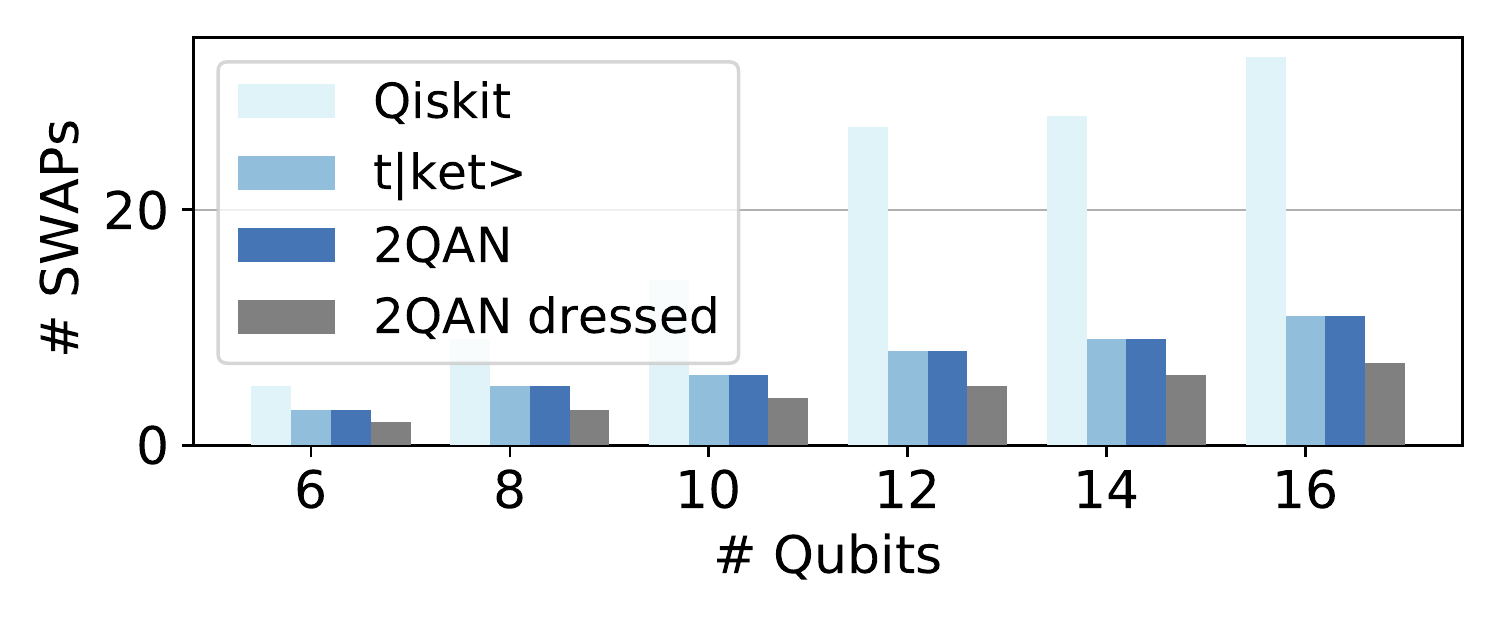}
    \label{fig:aspen_hbxxz_cxswap}
    }
    \subfloat[CZ count for NNN Heisenberg model]{
     \includegraphics[width=0.33\textwidth]{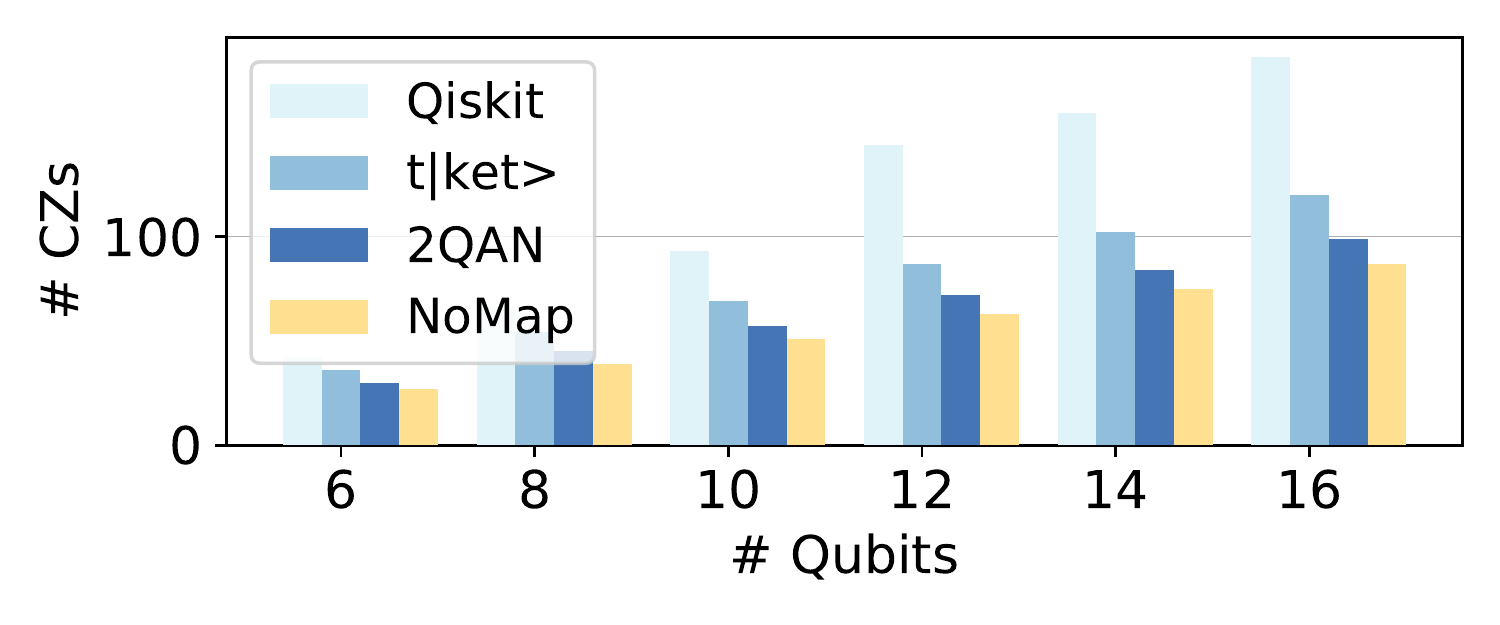}
    \label{fig:aspen_hbxxz_cx}
    }
    \subfloat[CZ depth for NNN Heisenberg model]{
    \includegraphics[width=0.33\textwidth]{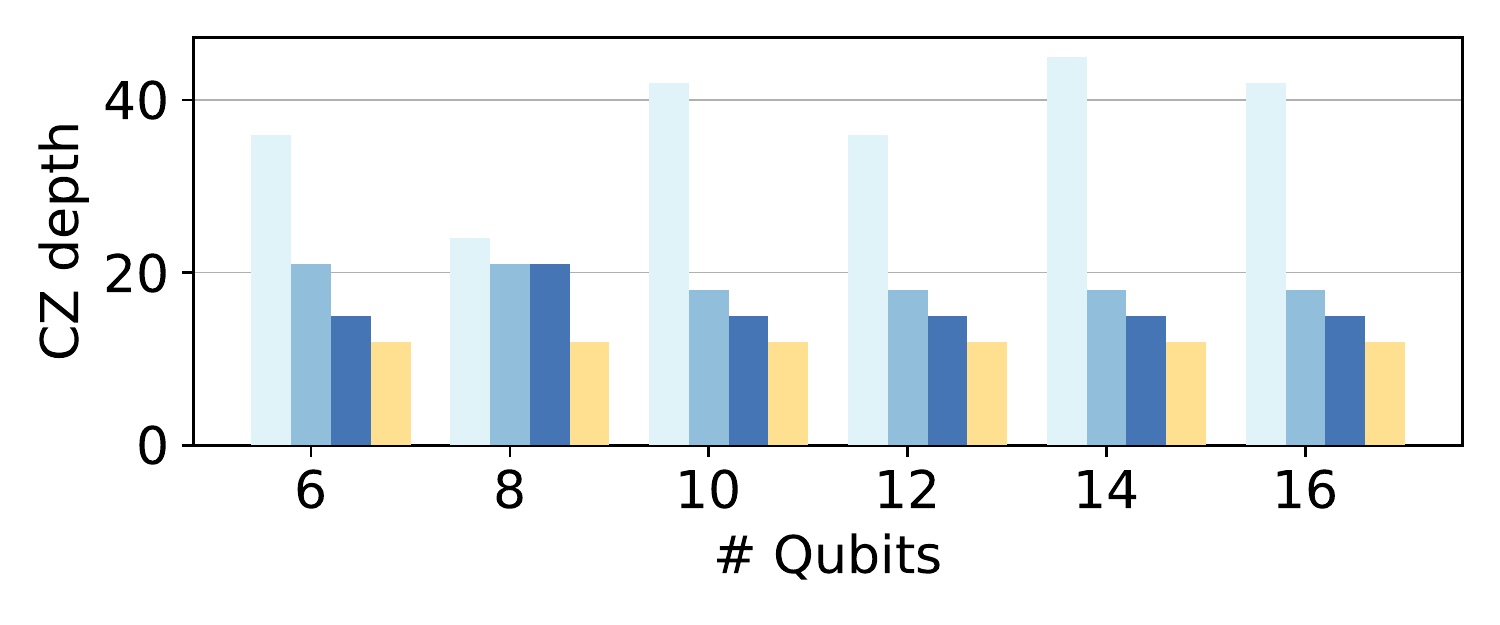}
    \label{fig:aspen_hbxxz_cx_depth}
    }
    
    \subfloat[SWAP count for NNN XY model]{
     \includegraphics[width=0.33\textwidth]{aspen_nnn_xy_cx_swap.pdf}
    \label{fig:aspen_xy_cxswap}
    }
    \subfloat[CZ count for NNN XY model]{
     \includegraphics[width=0.33\textwidth]{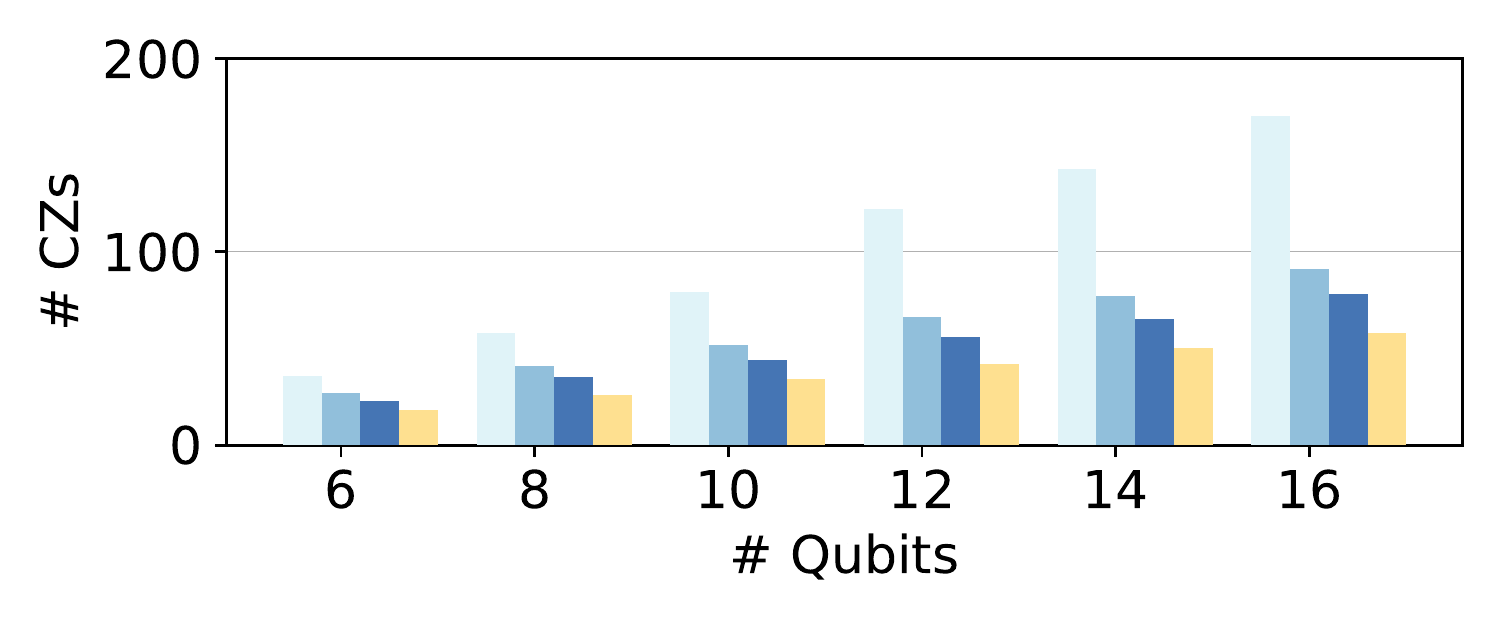}
    \label{fig:aspen_xy_cx}
    }
    \subfloat[CZ depth for NNN XY model]{
    \includegraphics[width=0.33\textwidth]{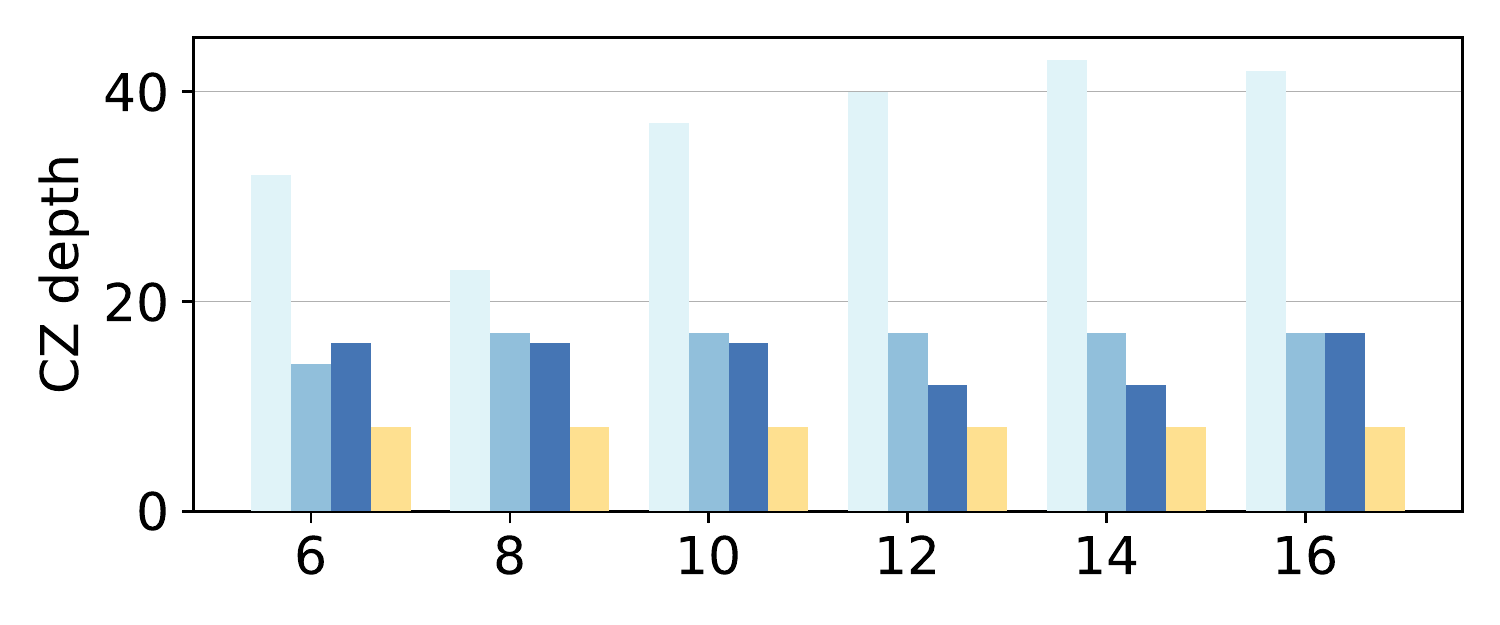}
    \label{fig:aspen_xy_cx_depth}
    }
    
    \subfloat[SWAP count for NNN Ising model]{
     \includegraphics[width=0.33\textwidth]{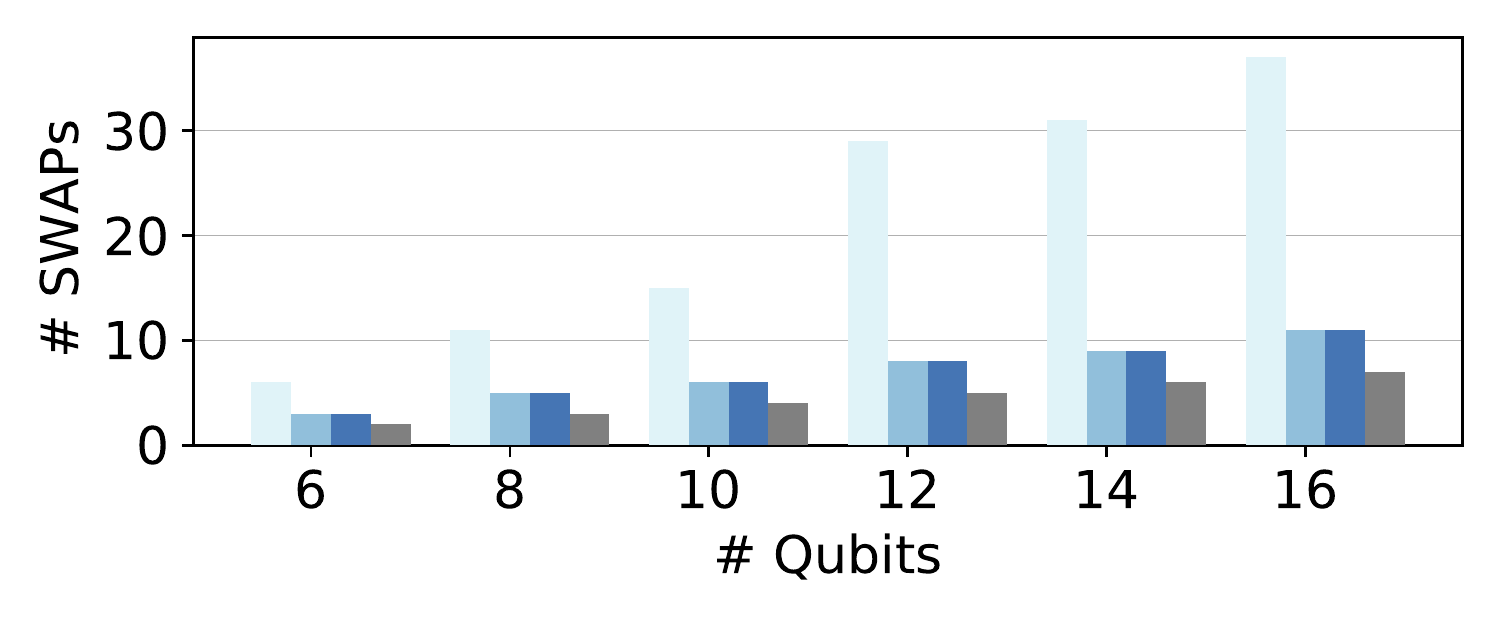}
    \label{fig:aspen_ising_cxswap}
    }
    \subfloat[CZ count for NNN Ising model]{
     \includegraphics[width=0.33\textwidth]{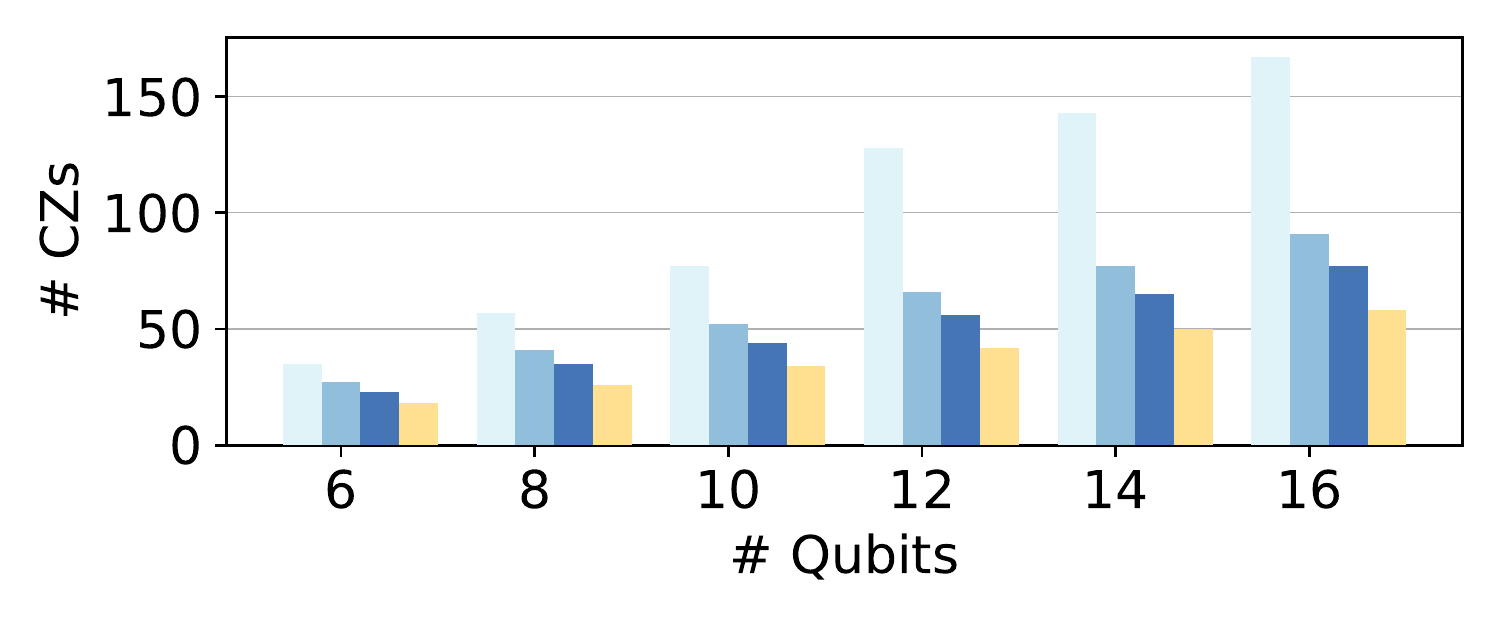}
    \label{fig:aspen_ising_cx}
    }
    \subfloat[CZ depth for NNN Ising model]{
    \includegraphics[width=0.33\textwidth]{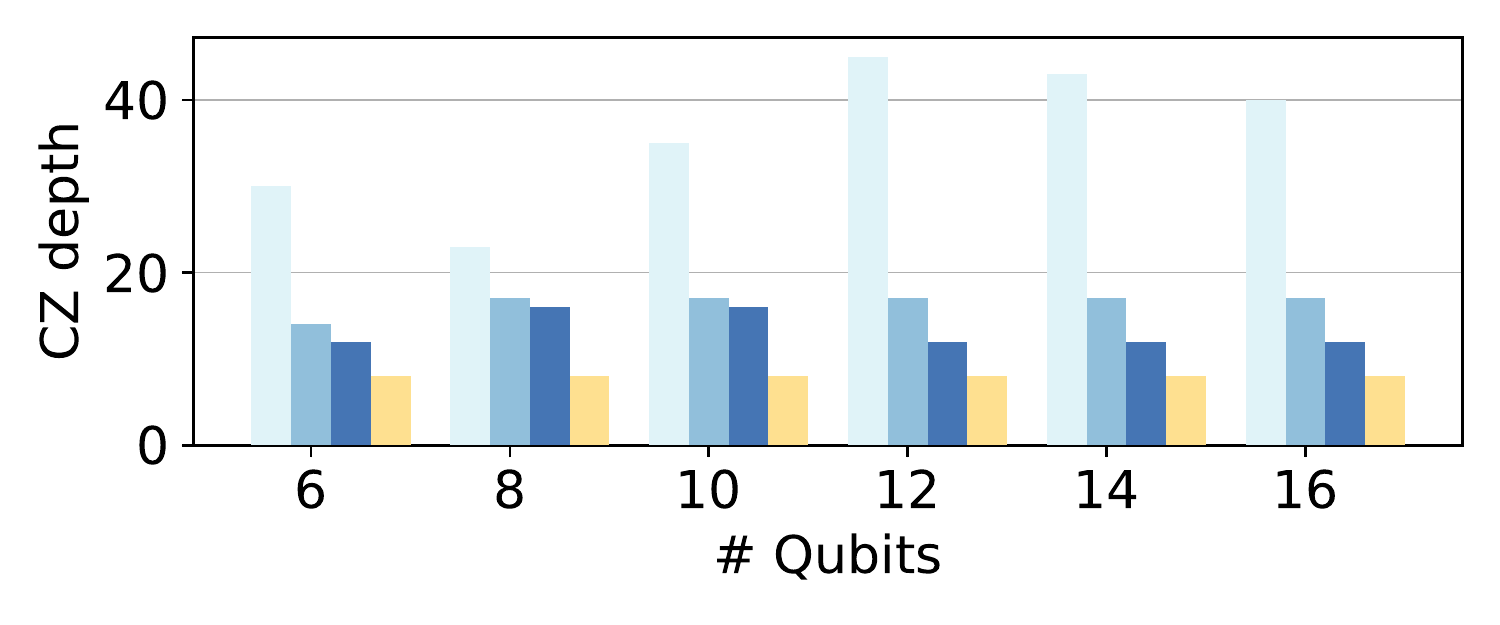}
    \label{fig:aspen_ising_cx_depth}
    }
    
    \subfloat[SWAP count for QAOA-REG-3]{
     \includegraphics[width=0.33\textwidth]{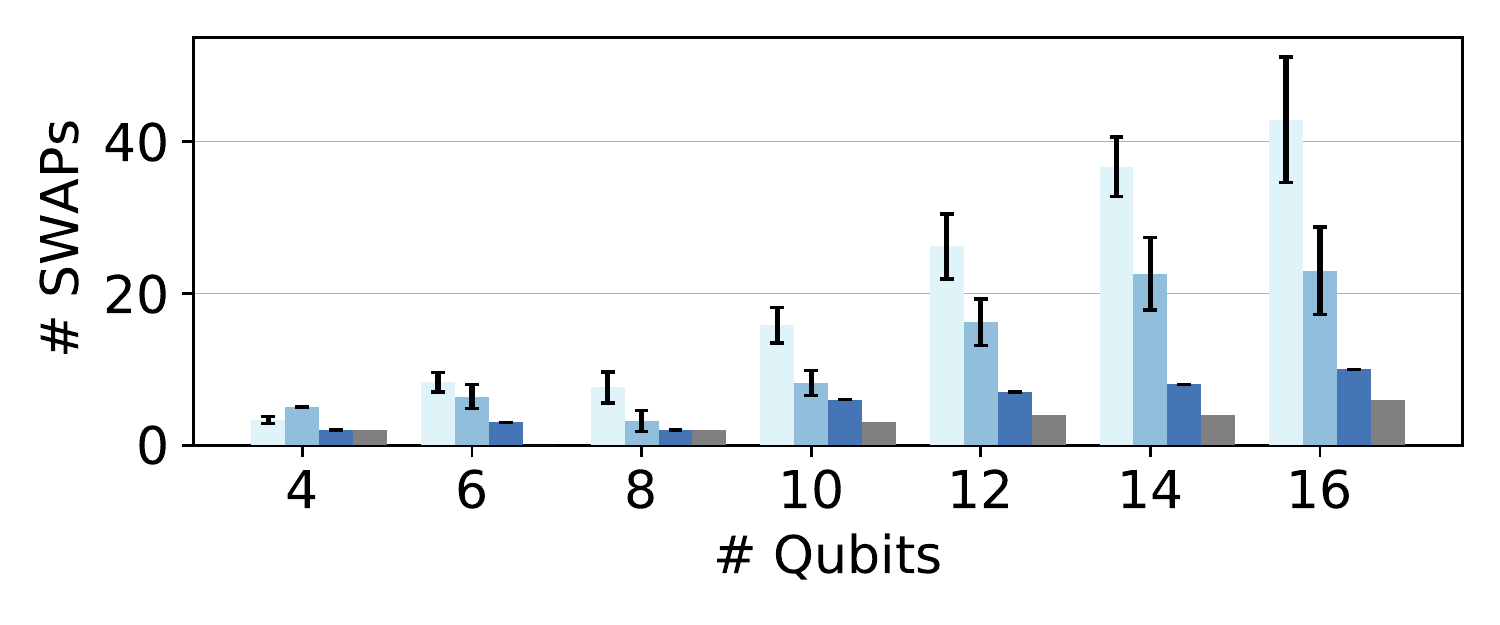}
    \label{fig:aspen_qaoa_cx_swap}
    }
    \subfloat[CZ count for QAOA-REG-3]{
     \includegraphics[width=0.33\textwidth]{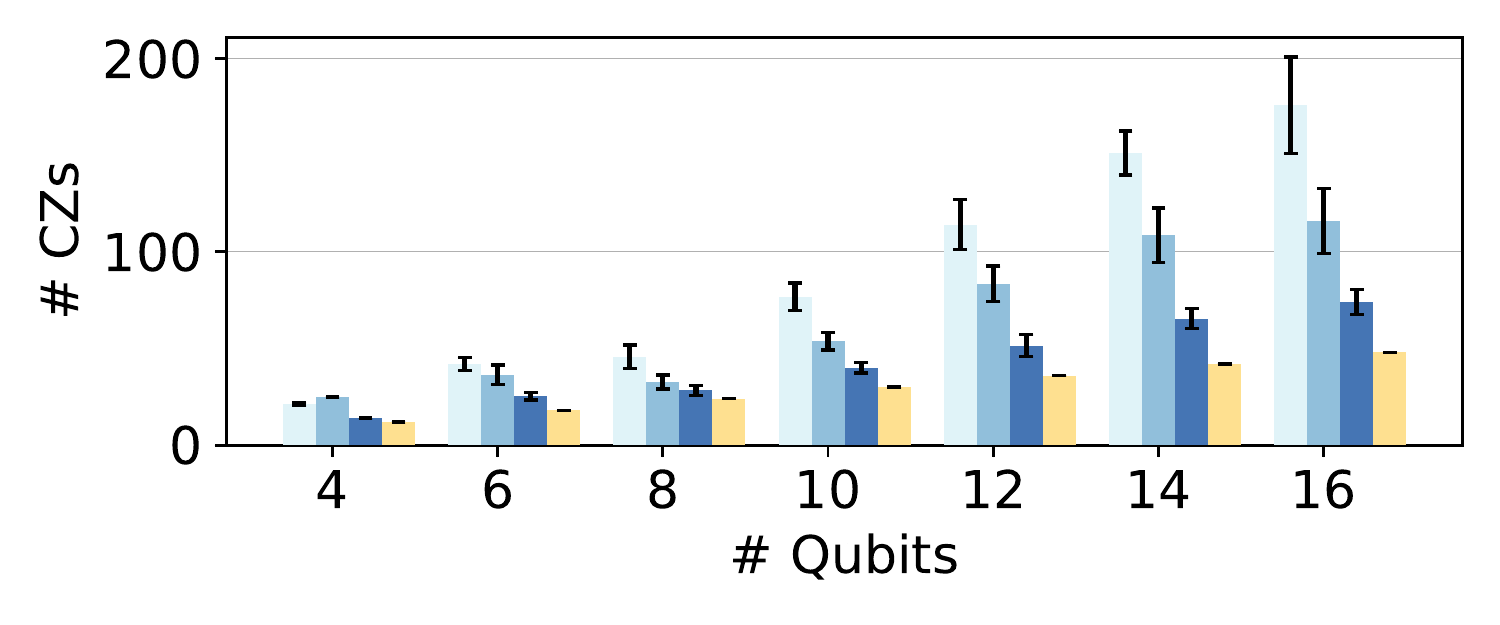}
    \label{fig:aspen_qaoa_cx}
    }
    \subfloat[CZ depth for QAOA-REG-3]{
    \includegraphics[width=0.33\textwidth]{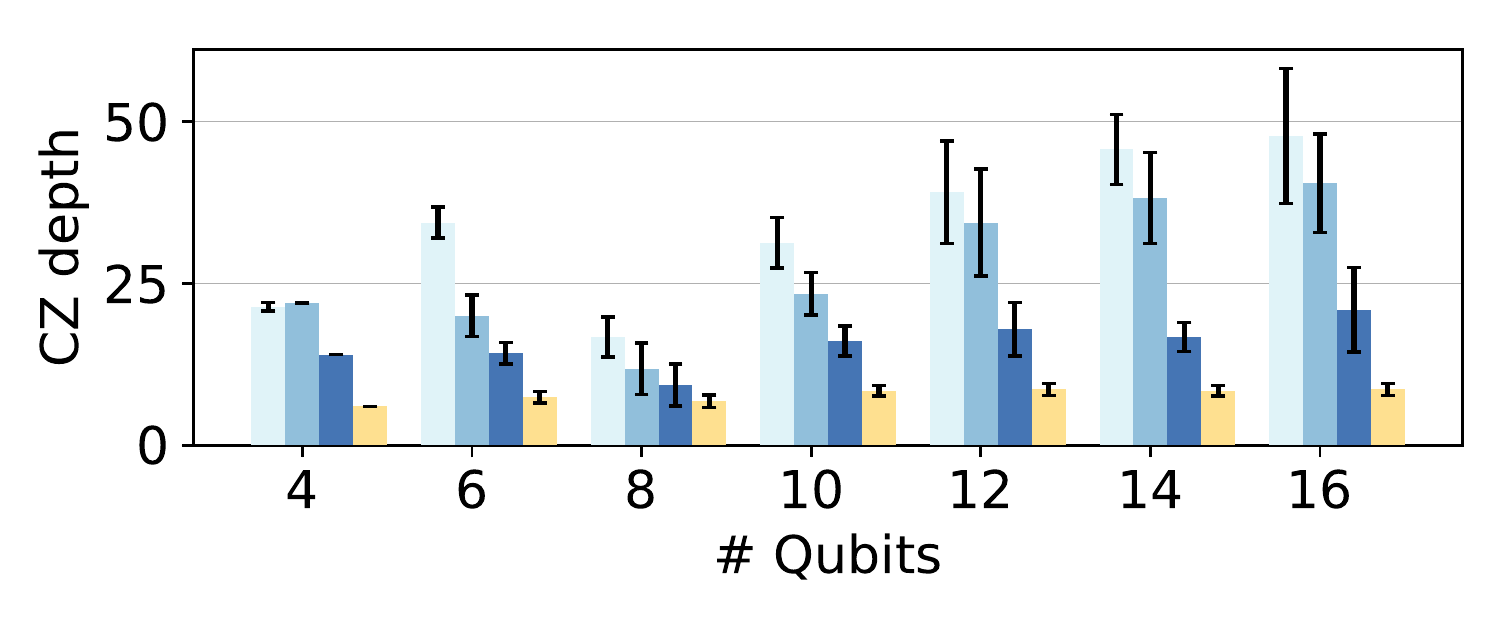}
    \label{fig:aspen_qaoa_cx_depth}
    }
\caption{Compilation results of the one-layer NNN Heisenberg model, NNN Ising model, and QAOA on the Rigetti Aspen architecture with CZ as hardware gate. The 2QAN compiler has least compilation overhead compared to Qiskit and \tket. }
\label{fig:aspen_cx}
\end{figure*}

Figure \ref{fig:aspen_cx} and Table \ref{tb:aspen_cx} show the results of compilation on the Aspen architecture.
\ucl~inserts the same number of SWAPs as \tket~for all three models. More than 60\% of these SWAPs can be combined with circuit gates, which will help reduce hardware gate count. For example, \ucl~reduces CZ overhead of the Heisenberg model by on average 2.8x (up to 3.0x) compared to \tket~and 7.2x (up to 9.3x) with respect to Qiskit. 
The average depth reduction over three benchmarks is 1.6x and 2.9x compared to \tket~and Qiskit, respectively.

\begin{figure*}[htb!]
 \centering
    \subfloat[SWAP count for 3-layer QAOA-REG-3]{
     \includegraphics[width=0.33\textwidth]{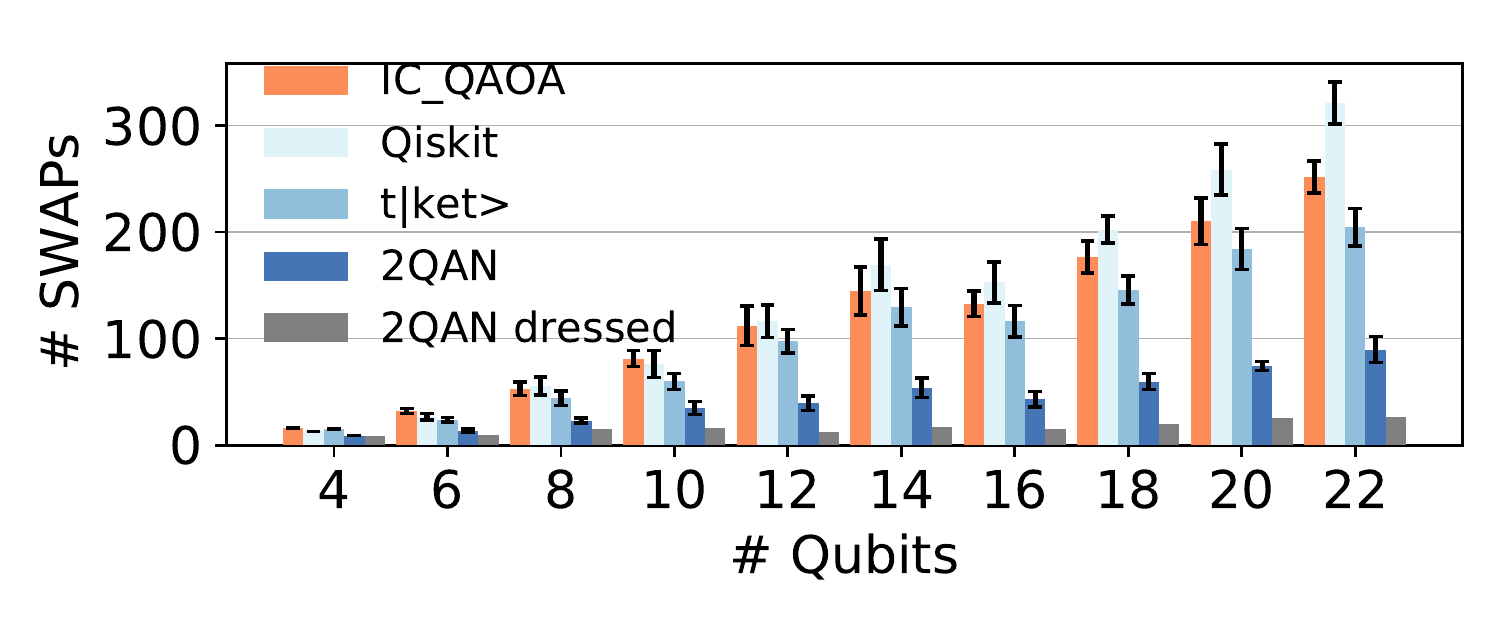}
    \label{fig:montreal_qaoa_swap3}
    }
    \subfloat[CNOT count for 3-layer QAOA-REG-3]{
     \includegraphics[width=0.33\textwidth]{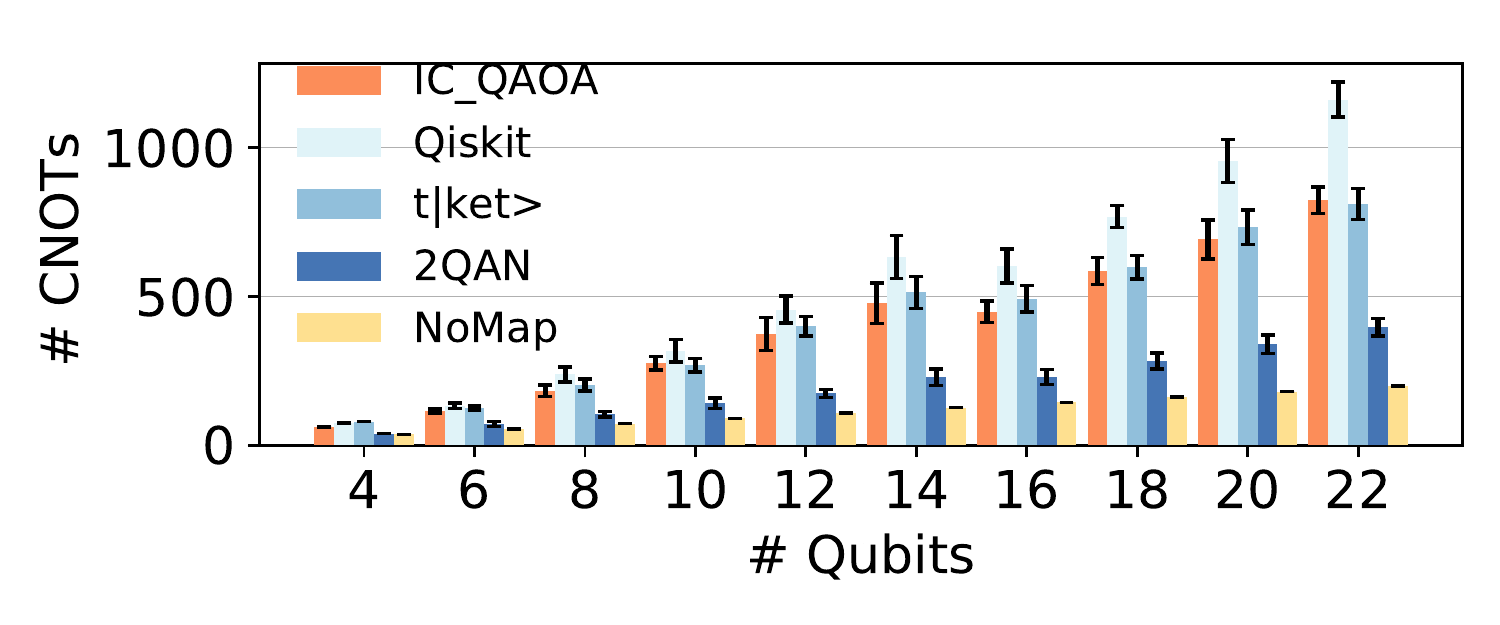}
    \label{fig:montreal_qaoa_cx3}
    }
    \subfloat[CNOT depth for 3-layer QAOA-REG-3]{
    \includegraphics[width=0.33\textwidth]{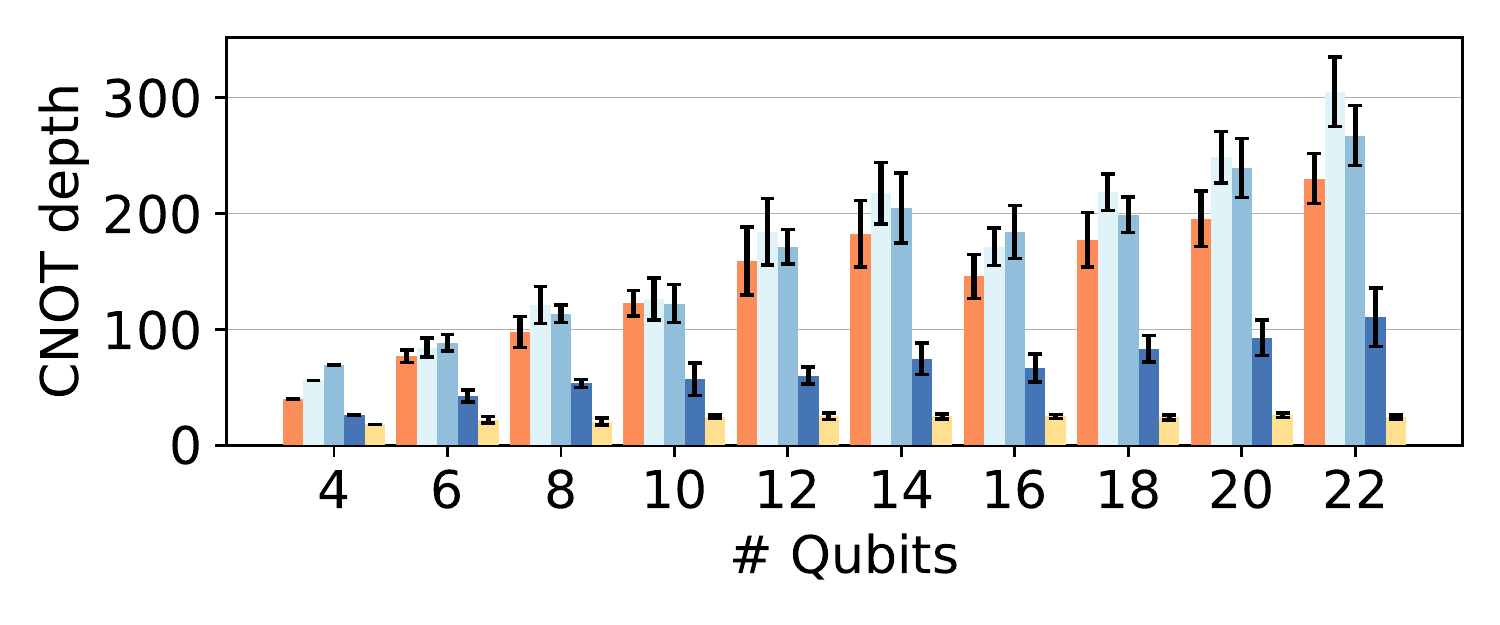}
    \label{fig:montreal_qaoa_cx_depth3}
    }
\caption{Compilation results of 3-layer QAOA-REG-3 on the IBMQ Montreal device. 
Each QAOA problem size is averaged over 10 different instances (error bars show the standard deviation) and operator parameters of QAOA circuits were chosen at their theoretically optimal values. `2QAN dressed' shows the number of SWAPs that were merged with circuit gates by \ucl, which helps to reduce hardware gate count. `NoMap' represents the compilation results when assuming all-to-all qubit connectivity.
The overhead of 3-layer QAOA circuits is approximately 3 times of the overhead of 1-layer QAOA circuits. 
The 2QAN compiler has least compilation overhead compared to Qiskit, \tket, and the QAOA compiler (IC-QAOA). }
\label{fig:montreal_qaoa_l3}
\end{figure*}

Figure \ref{fig:montreal_qaoa_l3} shows the compilation results of 3-layer QAOA on the Montreal device. Since QAOA has periodic circuit structures, the 3-layer QAOA circuits have approximately 3x compilation overhead compared to 1-layer QAOA circuits in Figure \ref{fig:montreal_cx}.

\begin{table*}[]
\centering
\resizebox{1.5\columnwidth}{!}{
\textbf{
\begin{tabular}{|l|c|c|c|c|c|c|c|c|c|c|c|c|}
\hline
\multirow{3}{*}{} & \multicolumn{6}{c|}{\tket}                                                          & \multicolumn{6}{c|}{Qiskit}                                                        \\ \cline{2-13} 
                  & \multicolumn{2}{c|}{SWAPs} & \multicolumn{2}{c|}{CZs} & \multicolumn{2}{c|}{Depth} & \multicolumn{2}{c|}{SWAPs} & \multicolumn{2}{c|}{CZs} & \multicolumn{2}{c|}{Depth} \\ \cline{2-13} 
                  & avg          & max         & avg        & max         & avg         & max          & avg          & max         & avg         & max        & avg         & max          \\ \hline
NNN Heisenberg                & 3.3x         & 5x          & --         & --          & 2.2x        & 4.1x         & 5.1x         & 9x          & --          & --         & 2.1x        & 3x           \\ \hline
NNN XY             & 2.8x         & 4.9x          & 7.5x       & 14.6x       & 2.2x        & 4.4x         & 5.7x         & 9.2x        & 15.4x       & 27.5x      & 2.3x        & 3.8x         \\ \hline 
NNN Ising             & 2.6x         & 5x          & 8.7x       & 15.6x       & 1.3x        & 2.9x         & 5.2x         & 8.9x        & 16.4x       & 29.6x      & 2.4x        & 4.1x         \\ \hline
QAOA-REG-3              & 1.8x         & 2.4x        & 3.4x       & 4.8x        & 4.3x        & 10.3x        & 3.6x         & 5.3x        & 6.7x        & 9x         & 5.5x        & 13.7x        \\ \hline
\end{tabular}
}}
\caption{The average (avg) and maximum (max) compilation overhead reduction when comparing \ucl~with \tket~and Qiskit on the Sycamore architecture with CZ as hardware gate. For the cases with blank values `--', \ucl~has negligible CZ overhead.}
\label{tb:grid_cx}
\end{table*}

\begin{table*}[]
\centering
\resizebox{1.5\columnwidth}{!}{
\textbf{
\begin{tabular}{|l|c|c|c|c|c|c|c|c|c|c|c|c|}
\hline
\multirow{3}{*}{} & \multicolumn{6}{c|}{\tket}                                                          & \multicolumn{6}{c|}{Qiskit}                                                        \\ \cline{2-13} 
                  & \multicolumn{2}{c|}{SWAPs} & \multicolumn{2}{c|}{CZs} & \multicolumn{2}{c|}{Depth} & \multicolumn{2}{c|}{SWAPs} & \multicolumn{2}{c|}{CZs} & \multicolumn{2}{c|}{Depth} \\ \cline{2-13} 
                  & avg          & max         & avg         & max        & avg          & max         & avg          & max         & avg         & max        & avg          & max         \\ \hline
NNN Heisenberg                & 1x           & 1x          & 2.8x        & 3x         & 1.2x         & 1.4x        & 2.7x         & 3.5x        & 7.2x        & 9.3x       & 2.4x         & 3x          \\ \hline
NNN XY             & 0.99x         & 1x          & 1.7x       & 1.8x       & 1.1x        & 1.4x         & 2.8x         & 3.4x        & 4.8x       & 6.2x      & 2.5x        & 3.6x         \\ \hline
NNN Ising             & 1x           & 1x          & 1.7x        & 1.8x       & 1.2x         & 1.4x        & 2.7x         & 3.5x        & 4.9x        & 6.2x       & 2.8x         & 3.7x        \\ \hline
QAOA-REG-3              & 2.1x         & 2.8x        & 3.1x        & 6.5x       & 2.4x         & 3.6x        & 3.3x         & 4.6x        & 4.6x        & 5x         & 3.4x         & 4.6x        \\ \hline
\end{tabular}
}}
\caption{The average (avg) and maximum (max) compilation overhead reduction when comparing \ucl~with \tket~and Qiskit on the Aspen architecture with CZ as hardware gate.}
\label{tb:aspen_cx}
\end{table*}

\end{document}